\documentclass[%
 aip,
 amsmath,amssymb,
 reprint,%
]{revtex4-1}

\usepackage{graphicx}
\usepackage{dcolumn}
\usepackage{bm}

\usepackage[utf8]{inputenc}
\usepackage[T1]{fontenc}
\usepackage{mathptmx}
\usepackage{hyperref}

\begin{document}

\preprint{AIP/123-QED}

\title[Global and Local Reduced Models for Interacting, Heterogeneous Agents]{Global and Local Reduced Models for Interacting, Heterogeneous Agents}

\author{Thomas N. Thiem}
\affiliation{Department of Chemical and Biological Engineering, Princeton University, USA}

\author{Felix P. Kemeth}
\affiliation{Department of Chemical and Biomolecular Engineering, Whiting School of Engineering, Johns Hopkins University, USA}

\author{Tom Bertalan}
\affiliation{Department of Mechanical Engineering, Massachusetts Institute of Technology, USA}

\author{Carlo R. Laing}
\affiliation{School of Natural and Computational Sciences, Massey University, NZ}

\author{Ioannis G. Kevrekidis}
\affiliation{Department of Applied Mathematics and Statistics, Johns Hopkins University, USA}
\altaffiliation[Also at ]{Department of Chemical and Biomolecular Engineering, Whiting School of Engineering, Johns Hopkins University, USA}

\date{\today}

\textbf{Author to whom correspondence should be addressed:} yannisk@jhu.edu

\begin{abstract}
Large collections of coupled, heterogeneous agents can manifest complex dynamical behavior presenting difficulties for simulation and analysis. However, if the collective dynamics lie on a low-dimensional manifold then the original agent-based model may be approximated with a simplified surrogate model on and near the low-dimensional space where the dynamics live. Analytically identifying such simplified models can be challenging or impossible, but here we present a data-driven coarse-graining methodology for discovering such reduced models. We consider two types of reduced models: globally-based models which use global information and predict dynamics using information from the whole ensemble, and locally-based models that use local information, that is, information from just a subset of agents close (close {\em in heterogeneity space}, not physical space) to an agent, to predict the dynamics of an agent. For both approaches we are able to learn laws governing the behavior of the reduced system on the low-dimensional manifold directly from time series of states from the agent-based system. These laws take the form of either a system of ordinary differential equations (ODEs), for the globally-based approach, or a partial differential equation (PDE) in the locally-based case. For each technique we employ a specialized artificial neural network integrator that has been templated on an Euler time stepper (i.e. a ResNet) to learn the laws of the reduced model. As part of our methodology, we utilize the proper orthogonal decomposition (POD) to identify the low-dimensional space of the dynamics. Our globally-based technique uses the resulting POD basis to define a set of coordinates for the agent states in this space, and then seeks to learn the time evolution of these coordinates as a system of ODEs. For the locally-based technique, we propose a methodology for learning a partial differential equation representation of the agents; the PDE law depends on the state variables and partial derivatives of the state variables with respect to the model heterogeneities. We require that the state variables are smooth with respect to the model heterogeneities, which permits us to cast the discrete agent-based problem as a continuous one in heterogeneity space. The agents in such a representation bear similarity to the discretization points used in typical finite element/volume methods. As an illustration of the efficacy of our techniques, we consider a simplified coupled neuron model for rhythmic oscillations in the pre-B\"{o}tzinger complex and demonstrate how our data-driven surrogate models are able to produce dynamics comparable to the dynamics of the full system. A nontrivial conclusion is that the dynamics can be equally well reproduced by an all-to-all coupled as well as by a locally coupled model of the same agents.
\end{abstract}

\maketitle

\begin{quotation}
Interacting agents can be used to model a diverse class of dynamic behaviors; they are typically comprised of a collection of individual agents, a set of rules or laws governing the interactions of the agents, and an interaction topology that links the agents together. The interactions may be deterministic or stochastic in nature and could be based on simple heuristics or detailed rules. The versatility of agent-based models has contributed to their broad appeal across disciplines such as disease outbreaks and response \cite{carley2006biowar, kattis2016modeling, hoertel2020stochastic, silva2020covid}, city parking \cite{benenson2008parkagent}, the modeling of urban sprawl \cite{zou2012accelerating, torrens2013simple}, the assessment of the impacts of shared autonomous vehicles \cite{fagnant2014travel, martinez2017assessing}, finance \cite{siettos2012equation, liu2015equation}, social interactions \cite{tsoumanis2010equation, zou2012model}, and the evaluation of changes in flood risk caused by climate change \cite{haer2020safe} to name a few. This paper discusses alternative, data-driven ways to obtain reduced models of the agent behavior.
\end{quotation}

\section{Introduction}

It is well-known that even simplistic agents can, through their interactions engender complex behaviors; the process is sometimes referred to as emergence, as the behaviors ``emerge'' from the interactions among the agents \cite{anderson1972more}. The high-dimensionality of the agent system can impede simulation and analysis. Historically, it has been a challenge to find reduced descriptions of these complex systems, a process known as coarse-graining, and this remains an open area of research in complex system analysis. Coarse-grained descriptions strive to find new variables with which to summarize the collective dynamics of the agents, such as the momentum and density fields used in fluid mechanics or concentrations, temperatures and pressures in thermodynamics. These coarse variables can then be used to construct simplified reduced order surrogate models of the dynamics, typically as systems of ordinary or partial differential equations, to facilitate simulation and analysis. The coarse variables themselves may be derived analytically, such as in kinetic theory and statistical mechanics, defined empirically \cite{kuramoto1984chemical} or found through a variety of data-driven techniques including proper orthogonal decomposition (POD) \cite{berkooz1993proper, kerschen2005method, hinze2005proper, kunisch2002galerkin}, POD-Galerkin \cite{sirovich1987coherent, deane1991low, shvartsman1998low, shvartsman1998nonlinear, shvartsman2000order}, deep/manifold Galerkin \cite{lee2020model}, and manifold learning techniques \cite{kemeth2018emergent, holiday2019manifold, thiem2020emergent}.

In the case of coupled oscillator models, the coarse variables may take the form of a set of order parameters which provides a global or macroscopic description of the dynamics \cite{kuramoto1984chemical, watanabe1993integrability, o2017oscillators}, for instance the level of clustering of the oscillators. If an order parameter description is insufficient, one may instead consider a description in terms of continuous fields, such as the oscillator density distribution \cite{strogatz2000kuramoto}. If the dynamics exhibit an underlying low-dimensional structure, such as an inertial manifold, then it may be possible to derive the evolution equations of the order parameter(s) or the field(s) analytically \cite{ott2008low, tyulkina2018dynamics}. A major drawback of the analytical approach is that it is typically limited to special cases and requires considerable effort and insight, if it is possible at all (see \cite{bick2020understanding} for a review of analytical mean-field reductions). As an alternative to the analytical approach, we instead propose a data-driven methodology for identifying approximations of the temporal evolution of the reduced dynamics with machine learning (ML) techniques, specifically artificial neural networks. Such a methodology offers a generalized approach to coarse-graining agent-based systems that could be utilized when analytical techniques are infeasible or unavailable.

Data-driven nonlinear system identification is an established field \cite{kumpati1990identification,rico1992discrete,rico1994continuous, brunton2020machine} that is currently experiencing a resurgence of interest in the literature \cite{raissi2018multistep, chen2018neural, vlachas2018data, lu2019deeponet, nardini2020learning} particularly with regards to the data-driven identification \cite{long2018pde, raissi2018forward, arbabi2020linking, arbabi2020particles, linot2020deep} and solution of partial differential equations \cite{lu2019deepxde, raissi2019physics, bhattacharya2020model}. With regards to agent-based systems, data-driven techniques for learning both interaction laws \cite{lu2019nonparametric, lu2021learning} as well as order parameters from system trajectories have been demonstrated \cite{thiem2020emergent}. In this paper we endeavor to extend these data-driven model reduction techniques to coupled agent-based systems with a focus on coupled oscillator systems in particular. We consider systems with regions of low-dimensional dynamics, for example inertial manifolds, and seek to identify reduced order surrogate models of the dynamics in these regions in a data-driven way. For this purpose we propose two approaches, one global in nature and the other local.

Both of our approaches take advantage (one crucially, the other only for filtering) of the proper orthogonal decomposition technique (POD) to define the low-dimensional manifold on which our surrogate models operate from time series of agent states. Our global approach uses the POD basis to construct a set of coordinates for this low-dimensional manifold and then learns the system of ordinary differential equations governing the temporal evolution of these coordinates with a neural network integrator (in effect, a ResNet \cite{he2016deep}). This type of model reduction bears a resemblance to our previous coarse-graining work with polynomial chaos basis functions \cite{choi2016dimension, bertalan2017coarse, rajendran2016modeling} (instead of PODs, for the spatial description).

For our local approach we treat the state variable(s) of the agents as continuous field(s) defined over the agents' heterogeneities; here the agents act as discretization points of the field(s). We learn the temporal evolution of the field(s) as a partial differential equation written in terms of the partial derivatives of the field(s) with respect to the heterogeneities. Here, we use the POD basis to define a filter for the high wavenumber components of the solution, in order to prevent the learned dynamics from leaving the low-dimensional space in which the true dynamics live. During integration of the learned PDE this filter provides stability to the solution and can be viewed as analogous to the concept of hyperviscosity used in computational fluid dynamics simulations \cite{smith1996crossover, lamorgese2005direct, cook2005hyperviscosity, frisch2008hyperviscosity}. For each of these approaches the reduced order surrogate model describes the dynamics on and near the underlying low-dimensional manifold of the original system dynamics, which permits the reduced models to be used for coarse-grained simulation and analysis tasks, potentially providing a new understanding of the original system and its dynamics.

The remainder of this paper is organized as follows. In Section~\ref{sec:hh_model} we introduce a simplified model of coupled Hodgkin-Huxley neurons which serves as our example agent-based system with which we demonstrate our model reduction methodology. In Section~\ref{sec:neural_network_integrator} we describe our neural network integrator architecture, followed by the method of proper orthogonal decomposition in Section~\ref{sec:POD}. Our globally-based model reduction technique is presented in Section~\ref{sec:global_approximation_technique} with Section~\ref{sec:reconstructing_POD_modes} discussing techniques for reconstructing the dynamics of higher POD modes. Section~\ref{sec:local_approximation_technique} focuses on our locally-based model reduction technique where we learn an effective partial differential equation representation of a coupled oscillator model. We show how such a model can be constructed from subsampled data in Section~\ref{sec:domain_subsampling}. Finally, we conclude with a summary of our results and a brief discussion in Section~\ref{sec:discussion}.

\section{Background}

The goal of this paper is to provide a data-driven framework for both globally and locally-based model reductions of interacting agent-based problems. In this section, we introduce our example system and describe the data-driven techniques that are employed for the model reduction methodology.

\subsection{A Simplified Hodgkin-Huxley Model} \label{sec:hh_model}

We consider a simplified model for the rhythmic oscillations of the bursting neurons found in the pre-B\"{o}tzinger complex
\cite{butera1999models, laing2012managing, rubin2002synchronized}. This particular model is a cutdown version of the realistic model proposed by Butera et al. and describes the slow transitions between the windows of bursting and quiescence. It involves a membrane potential $V_{i}$, and a channel state $h_{i}$, related to the inactivation of persistent sodium, for each neuron $i$. The time evolution of each neuron is governed by the following ordinary differential equations
\begin{equation}
\begin{split}
    C\frac{dV_{i}}{dt}&=-g_{Na}m(V_{i})h_{i}(V_{i}-V_{Na})-g_{l}(V_{i}-V_{l})+I_{syn}^{i}+I_{app}^{i},\\
    \frac{dh_{i}}{dt}&=\frac{h_{\infty}(V_{i})-h_{i}}{\tau(V_{i})},
\label{eq:model_equations}
\end{split}
\end{equation}
for $i=1,\dots,N$. The neurons are coupled through a synaptic current $I_{syn}^{i}$ with a symmetric adjacency matrix $A$,
\begin{equation}
    I_{syn}^{i}=\frac{g_{syn}(V_{syn}-V_{i})}{N}\sum_{j=1}^{N}A_{ij}s(V_{j}).
\label{eq:synaptic_current}
\end{equation}
The membrane potential $V_{i}$ incorporates both a persistent sodium current, with parameters $g_{Na}$ and $V_{Na}$, and a passive leakage current, parametrized by $g_{l}$ and $V_{l}$, represented by the first and second terms of the right hand side of the membrane potential equation Eq.~\ref{eq:model_equations}, respectively. The synaptic communication acts through the function $s$, while the functions $\tau$, $h_{\infty}$, and $m$ are standard functions used in the Hodgkin-Huxley formalism:

\begin{align}
    s(V)&=\frac{1}{1+\exp(-(V+40)/5)}, \\
    \tau(V)&=\frac{1}{\epsilon \cosh((V+44)/12)}, \\
    h_{\infty}(V)&=\frac{1}{1+\exp((V+44)/6)}, \\
    m(V)&=\frac{1}{1+\exp(-(V+37)/6)}.
\label{eq:additional_model_functions}
\end{align}

\begin{figure*}
    \centering
    \includegraphics[width=\textwidth]{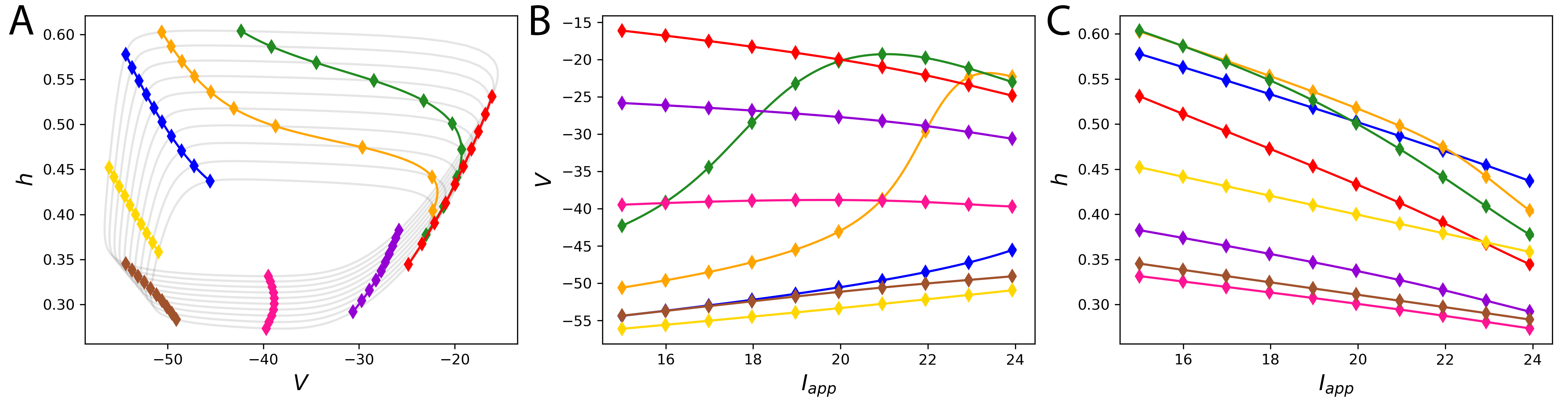}
    \caption{Plots of temporal snapshots of selected $(V, h)$ profiles consisting of 128 oscillators over the limit cycle solution of the system. Each of the snapshots of the system states is colored and the colors are shared between the figures to highlight the behavior over time. The profiles are presented three ways: (\textbf{A}) as profiles in $(V, h)$ space with a subset of the oscillators highlighted with diamonds and their corresponding trajectories included in grey, (\textbf{B}) as $V$ profiles over the $I_{app}^{i}$ domain, (\textbf{C}) and as $h$ profiles over the $I_{app}^{i}$ domain.}
    \label{fig:demonstration_profiles}
\end{figure*}

\begin{figure}
    \centering
    \includegraphics[width=0.48\textwidth]{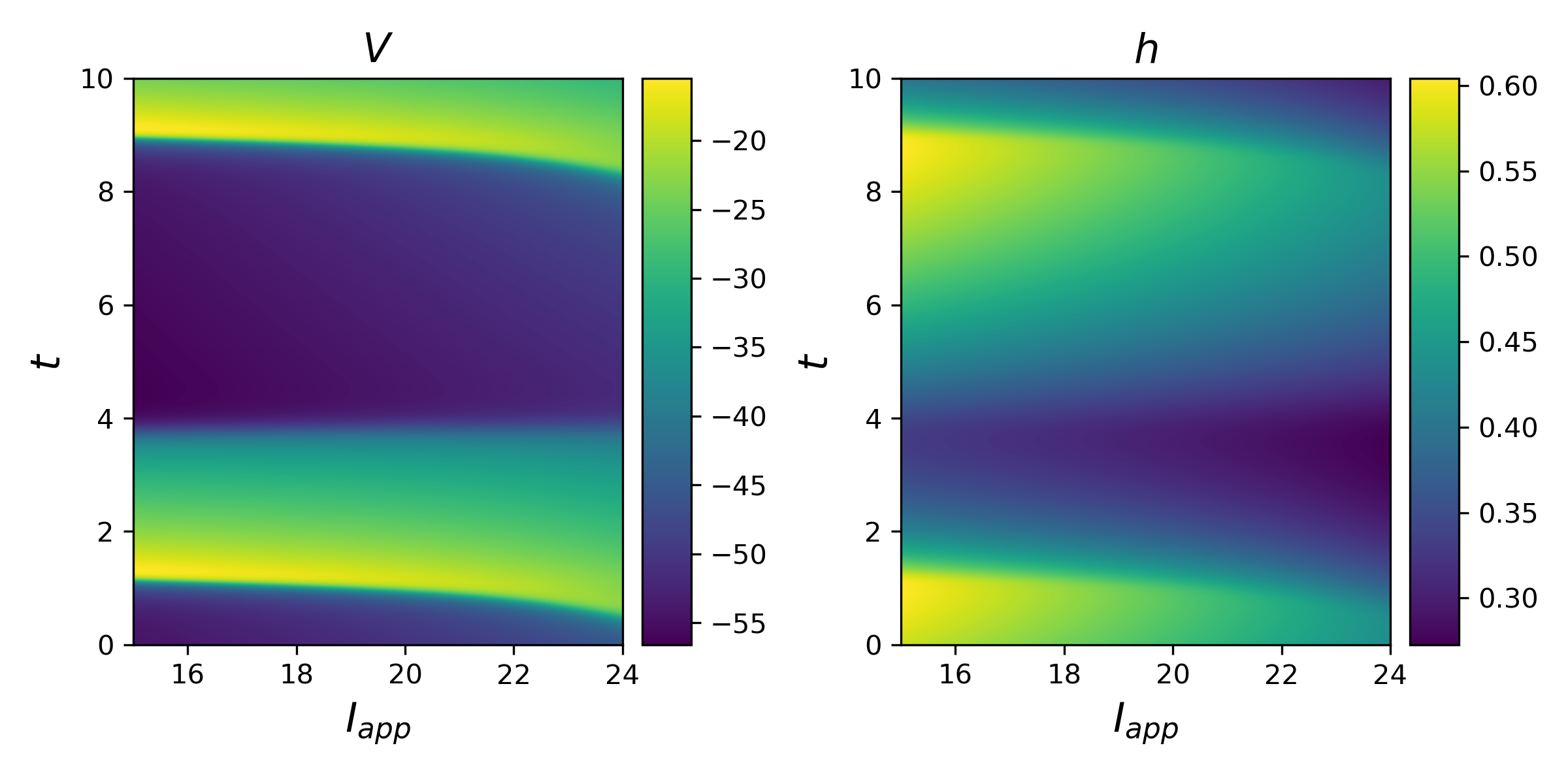}
    \caption{Spacetime plots of the state variables $(V, h)$ over the $I_{app}^{i}$ domain, colored by the respective state variable.}
    \label{fig:demonstration_spacetime}
\end{figure}

We consider an all-to-all coupled network of $N=128$ oscillators (although the particular network topology is not crucial for the success of our methodology) that includes self coupling, i.e. $A_{ij}=1 \ \forall i,j$, and select model parameters of: $C=0.21$, $g_{Na}=2.8$, $V_{Na}=50$, $g_{l}=2.4$, $V_{l}=-65$, $g_{syn}=0.3$, $V_{syn}=0$, and $\epsilon=0.1$. It has been observed that with these parameters, this coupling, and the applied currents $I_{app}^{i}$ uniformly distributed over the interval $[15, 24]$ \cite{laing2012managing, rubin2002synchronized}, the oscillators synchronize, and eventually settle on a limit cycle solution after an initial transient. However, due to the heterogeneity present in $I_{app}^{i}$, the oscillators trace out unique orbits in $(V, h)$ phase space (Fig.~\ref{fig:demonstration_profiles}(a)). Furthermore, on the limit cycle the values of $V_{i}$ and $h_{i}$ smoothly vary with the heterogeneity $I_{app}^{i}$ as depicted by Fig.~\ref{fig:demonstration_profiles}(b, c). This allows us to consider $V$ and $h$ as functions of the heterogeneity on and near the limit cycle (Fig.~\ref{fig:demonstration_spacetime}). As we will show, the smoothness of the state variables with respect to the heterogeneity combined with the low-dimensionality of the limit cycle behavior allows us to achieve both global and local model reductions.

\subsection{Neural Network Based Integrator}\label{sec:neural_network_integrator}

Renowned for their expressiveness and generality, artificial neural networks have become a standard part of nonlinear modeling with widespread use throughout areas as varied as medical \cite{li2014medical, liang2018combining, hou2016patch}, hyperspectral \cite{mou2017deep}, and general \cite{ciregan2012multi} image classification, regression \cite{specht1991general, kolehmainen2001neural}, sentence modeling \cite{kalchbrenner2014convolutional}, social media bot detection \cite{kudugunta2018deep, mohammad2019bot}, and hurricane trajectory and intensity modeling \cite{alemany2019predicting, ghosh2018improvements}. On the dynamical systems front, neural networks have been utilized for the accurate approximation of functions and their derivatives \cite{cardaliaguet1992approximation}, system approximation \cite{funahashi1993approximation}, identification and control \cite{kumpati1990identification}, and modeling \cite{rico1994continuous}. Myriad neural network architectures have been studied including feedforward networks \cite{rico1992discrete}, recurrent networks \cite{funahashi1993approximation, wang1998runge}, multistep networks \cite{raissi2018multistep}, long short-term memory networks \cite{vlachas2018data}, and physics informed/guided networks \cite{raissi2019physics, karpatne2017physics} along with specialized training techniques, such as gradient norm clipping \cite{pascanu2013difficulty}.

\begin{figure*}
    \centering
    \includegraphics[width=\textwidth]{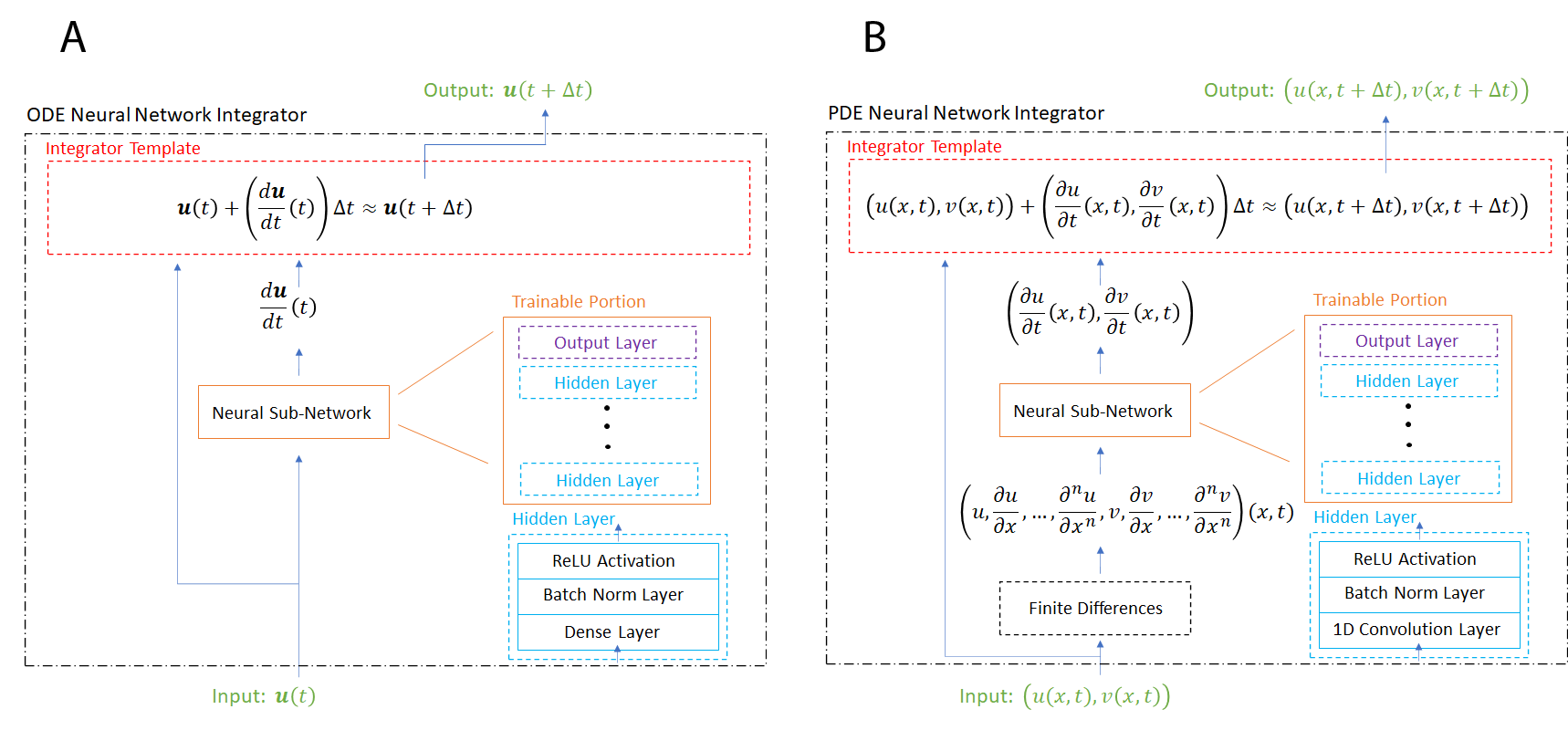}
    \caption{Neural network integrators based on an Euler scheme for: (\textbf{A}) ODEs, and (\textbf{B}) PDEs. In each case, the integrator contains a neural sub-network that approximates the time derivative of the respective system, which is subsequently used as part of the selected template to predict a future state from the current state input. After training the integrator, the sub-network can be extracted to serve as an approximation of the time derivative of the system suitable for use with any time stepper algorithm.}
    \label{fig:neural_network_integrator}
\end{figure*}

We utilize a neural network inspired by (templated on) an explicit integration scheme architecture to deduce the system dynamic evolution laws from discrete time series data. This approach has been successfully applied to both ordinary differential equations (ODEs) \cite{rico1992discrete, krischer1993model, rico1993continuous} and partial differential equations (PDEs) that have been discretized through a method of lines approach \cite{gonzalez1998identification}. This method works by approximating the time derivative of the system with a feedforward neural network, which is then incorporated into a larger network templated on an integrator. An example of this approach with an Euler scheme is illustrated in Fig.~\ref{fig:neural_network_integrator}(a, b) for ODEs and PDEs respectively. These neural network integrators are trained by providing input-output pairs, consisting of points $(\mathbf{u}(t), \mathbf{u}(t+\Delta t))$ for ODEs or profiles $((u(x,t),v(x,t)),(u(x,t+\Delta t),v(x,t+\Delta t)))$ for PDEs, where $x$ is the spatial coordinate(s) of the time-dependent profiles (as we will later show the heterogeneities of the model form the ``space''). By training an integrator to accurately produce the output as a function of the input, part of the network, the neural sub-network, learns an approximation of the time derivative of the system. This sub-network is a surrogate model which can be easily evaluated on out-of-sample states and, after training, can be provided to typical time stepper algorithms to produce an approximation of the system flow that closely matches the ground truth. Careful consideration of both the template and the time step $\Delta t$ of the neural network integrator is important to ensure a meaningful approximation. The length of the time step directly affects the accuracy of the learned time derivative, and should ideally be selected short enough to provide an accurate approximation of the dynamics, while the choice of the integration template determines the accuracy order. This is because the sub-network of the integrator learns a time derivative that matches the training flow when integrated using the integration scheme of the selected template. As such, when possible, a scheme should be selected based on its suitability for the particular dynamics of the problem.

While the ODE and PDE neural network integrators have many similarities, they also have two main differences, the first of which is the type of inputs on which they operate. In the ODE case, Fig.~\ref{fig:neural_network_integrator}(a), the inputs take the form of a vector of states $\mathbf{u}(t)$, while for the PDE version, Fig.~\ref{fig:neural_network_integrator}(b), the input states are profiles over the spatial domain $u(x,t)$ (here one-dimensional for the single heterogeneity $I_{app}^{i}$) and their spatial partial derivatives $u_x(x,t)$, $u_{xx}(x,t)$, etc. These spatial partial derivatives can be estimated by any standard technique, such as finite differences or spectral methods, with the choice ultimately being problem dependent. One of the challenges of the use of spatial partial derivatives for the PDE model is the selection of the features (i.e. how many spatial derivatives should be included in the model). Fortunately, there are a variety of feature selection techniques available for this purpose, such as Gaussian process regression with automatic relevance determination (ARD), and diffusion map based methods \cite{lee2020coarse}.

The second difference lies in the hidden layers and output layers used for the sub-network contained in each neural network integrator. The objective of the ODE sub-network is to predict the time derivatives as a function of all of the states (ODE variables), a multivariate regression problem that fits the classical description of a fully connected neural network. Thus, the hidden layers for the ODE integrator are composed of a dense layer followed by a batch normalization layer to aid training, and a ReLU activation. For the output layer we simply use a dense layer with a linear activation. The PDE sub-network has a similar objective to the ODE network, but vastly different inputs. Here the inputs consist of profiles of the state variables and their respective spatial partial derivatives. Because the goal of the PDE network is to learn a PDE that is valid over the entire domain, the sub-network must learn the same function for each input profile and for each point of the profile (the output of the function can vary over the different inputs, but the function itself must be the same). This problem exhibits similarities to image recognition problems, which use convolutional kernels to encode translationally invariant features. Building on this idea, we construct the hidden layers of the PDE integrator with a one-dimensional convolutional layer with a kernel size of one followed by a batch normalization and a ReLU activation. This construction builds the translational invariance requirement of the PDE into the model while taking advantage of highly optimized image processing libraries (we note that this construction is formally equivalent to looping over the input points with a dense network). For the output layer we use a final one-dimensional convolutional layer with a kernel size of one and a linear activation. To the best of our knowledge this technique was first used for approximating nonlinear PDEs by our group in 1998 \cite{gonzalez1998identification}, following the original shift-invariant neural network pattern recognition architectures \cite{zhang1988shift} and before the term ``convolutional neural networks" became established \cite{lecun1995convolutional}. The expressiveness of the ODE (PDE) integrator can be increased through the addition of more hidden layers and/or additional neurons (filters) in each dense (convolutional) layer. In the following sections, we utilize these neural network integrators to learn ODE and PDE approximations of the dynamics of our coupled neuron model Eq.~\ref{eq:model_equations} near and on its limit cycle.

\subsection{Proper Orthogonal Decomposition (POD) and Filtering}\label{sec:POD}

Invented by Karl Pearson in 1901 \cite{pearson1901liii} and further refined by Hotelling in 1933 \cite{hotelling1933analysis}, the principal component analysis method and its analogs go by many names, such as the proper orthogonal decomposition method, Karhunen-Lo\`eve transform/decomposition, and singular value decomposition \cite{liang2002proper, van1983matrix}. The principal component analysis method defines an orthogonal linear transformation that results in a new coordinate system such that the new coordinates are uncorrelated and ranked in order of decreasing variance. This method is particularly valuable when most of the variation of the transformed data is due to variations in only a few of the new coordinates. When this is the case, the dimensionality of the original data set can be reduced by keeping only a limited number of the new coordinates.

\subsubsection{Covariance Method}

One method for computing the PCA transformation is through the eigendecomposition of the covariance matrix of the data. Consider a vector $\mathbf{X}\in\mathbb{R}^p$ consisting of $p$ coordinates $(x_1,\dots,x_p)$, where each coordinate is a feature. For $N$ such samples we define the $p\times N$ observation matrix as $[\textbf{X}_1 \cdots \textbf{X}_N]$. In order to prepare the data for principal component analysis, the observation matrix must be put into mean-deviation form. The sample mean of the observation vectors is defined as
\begin{equation}
    \textbf{M}=\frac{1}{N}(\textbf{X}_1+\cdots+\textbf{X}_N).
\end{equation}
Subtracting the sample mean from each observation vector centers the data around the origin and produces an observation matrix with zero sample mean. For $i=1,\dots,N$ let
\begin{equation}
    \hat{\textbf{X}}_i=\textbf{X}_i-\textbf{M}.
\end{equation}
Then the $p\times N$ matrix $B$ is the mean-deviation form of the observation matrix,
\begin{equation}
    B=[\hat{\textbf{X}}_1 \cdots \hat{\textbf{X}}_N].
\end{equation}
We define the $p\times p$ covariance matrix $S$ as
\begin{equation}
    S=\frac{1}{N-1}BB^{T}.
\end{equation}
The normalized eigenvectors of $S$ are the principal components of the data and define the orthogonal transformation according to
\begin{equation}
    Y=P^{T}\hat{X},
\label{eq:PCA_transform}
\end{equation}
where $Y$ is the $p\times n$ matrix of new coordinates and $P=[\textbf{u}_1 \cdots \textbf{u}_p]$ is the matrix of eigenvectors $\textbf{u}_1,\dots,\textbf{u}_p$ (the principal components) of $S$. The PCA transformation preserves the variance of the data set with the variance of $y_i$ given by the corresponding eigenvalue $\lambda_i$. The quotient $\lambda_i/ \sum_j \lambda_j$ measures the fraction of the total variance explained by $y_i$ and provides a method for determining the important coordinates for dimensionality reduction.

\subsubsection{POD Filtering}

The dimensionality reduction properties of the PCA transform make it amenable to use as a data filter. We begin by considering a data set $\hat{X}$ in mean-deviation form for which we have the principal components $P$. We transform the data set into the PCA/POD coordinates $Y$ according to transformation Eq.~\ref{eq:PCA_transform}. We then select a reduced set of the PCA coordinates $y_1,\dots,y_j$, $j<p$, to retain (the first $j$ columns of $Y$), as one would do for dimensionality reduction, and zero out the values of the remaining coordinates $y_{j+1},\dots,y_{p}$ (columns of $Y$), producing $Y_{reduced}$. Finally, we transform the reduced data data back to the original coordinates with the inverse transform and define the result of this process as the POD filtered data $\bar{X}$,
\begin{equation}
    \bar{X}=PY_{reduced}.
\end{equation}
The POD filter removes the contributions of the higher POD/PCA modes from the data leaving low-dimensional data in the original data space. In the following sections we make use of this filter to confine dynamical data to a low-dimensional subspace.

\section{A Global ML Approach to Spatiotemporal Modeling} \label{sec:global_approximation_technique}

In this section we introduce a methodology for what we refer to as a globally-based model reduction technique. This technique utilizes the proper orthogonal decomposition method to derive a set of POD modes $\{\mathbf{u}_j(I_{app}^{i})\in\mathbb{R}^2\, | \, i=1, \dots, N, \, j=1, \dots, 2N\}$, which is used to transform the original system of $N$ coupled oscillators (indexed by $i$) each with 2 states $(V, h)$ into an equivalent system of the $2N$ coefficients $c_j(t)$ of the POD mode representation of the states,

\begin{equation}
\label{eq:POD_mode_decomposition}
    (V(I_{app}^{i}, t), h(I_{app}^{i}, t)) = \sum_{j=1}^{2N} c_j(t) \mathbf{u}_j(I_{app}^{i}).
\end{equation}

The behavior of this new system is described by a, possibly nonlinear, system of ODEs
\begin{equation}
\label{eq:POD_ODEs}
    \frac{d\mathbf{c}}{dt}=\mathbf{f}(\mathbf{c}(t)).
\end{equation}

This system governs the time evolution of the POD coordinates $\mathbf{c}(t)$, the coefficients of the POD decomposition of the oscillator states, which can be learned through the artificial neural network integrator technique described in Section~\ref{sec:neural_network_integrator}. To the best of our knowledge, the first time this approach was used to identify spatially distributed dynamical systems from (experimental) spatiotemporal data was by our group in 1993 \cite{krischer1993model, rico1995nonlinear}.

If the original system features low-dimensional dynamics, then the dimensionality of the transformed system can be reduced by discarding all but a select few of the POD coordinates, where the exact number to retain depends on the dimensionality of the dynamics and the desired level of accuracy of the reduced model. This technique acts globally in the sense that the resulting laws of the reduced system depend on the entire ensemble of oscillators, more precisely on the POD mode decomposition of their profiles, and thus the evaluation of a future state requires global information about the current state.

There is an important distinction to make. If the number of POD modes retained is sufficient to completely span the spatiotemporal data, then this approach ``learns" the POD-Galerkin equation right-hand-side. If, however, the number of POD modes retained is {\em not enough to accurately span the data}, but enough to parametrize the system's inertial manifold (in the spirit of what Foias and coworkers called ``determining modes" \cite{foias1991determining}), then what is learned is an {\em approximate inertial form}. This has been discussed, for example, in \cite{shvartsman2000order}.

\subsection{An Illustrative Example: Coupled Neuron Model}

\begin{figure}
    \centering
    \includegraphics[width=0.48\textwidth]{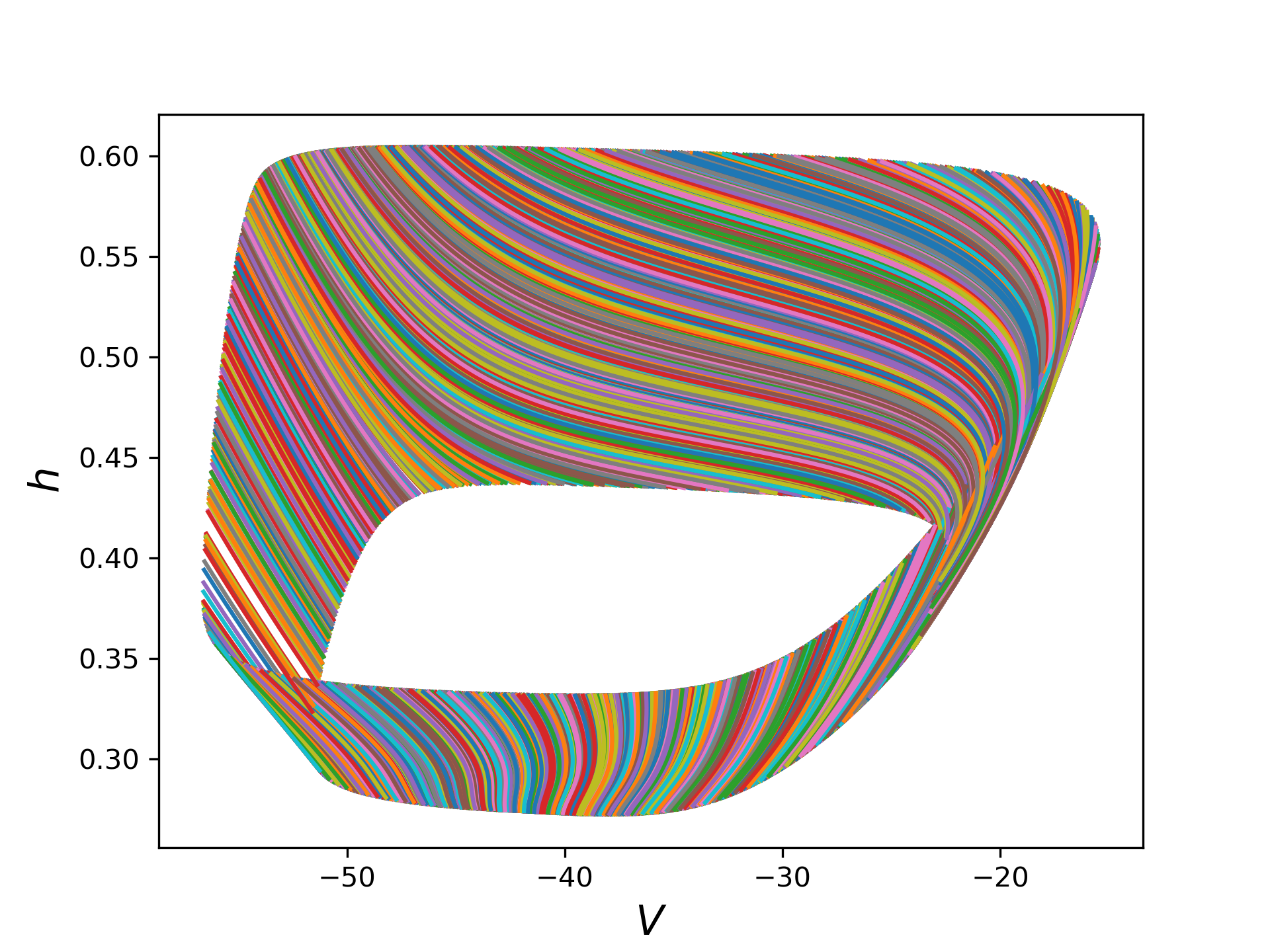}
    \caption{The 5000 snapshots of the system dynamics plotted in $(V, h)$ space. Due to the intrinsic heterogeneity $I_{app}^{i}$ the oscillators each trace out slightly different orbits, leading to one-dimensional snapshots at each time step (see the color coding). Because of the large number of samples, colors have been repeated. These snapshots define the low-dimensional dynamics of the system.}
    \label{fig:starting_points}
\end{figure}

We consider the simplified coupled neuron model described in Section~\ref{sec:hh_model} as an example of a biological coupled-agent model with low-dimensional behavior and use it to showcase our model reduction methodology described above. We study a system of $N=128$ oscillators with heterogeneity $I_{app}$ uniformly distributed over $[15, 24]$, and the coupling and model parameters described in Section~\ref{sec:hh_model}. This system is integrated in time with Scipy's Runge-Kutta integrator (the RK45 solver of the \textit{integrate.solve\_ivp} library) \cite{2020SciPy-NMeth} until it converges to the limit cycle solution depicted in Fig.~\ref{fig:demonstration_profiles}, at which point we collect 5000 time snapshots of the dynamics. Each snapshot is an element of $\mathbb{R}^{256}$ and is formed by concatenating the 128 $V_{i}$-values of the oscillators to the corresponding 128 $h_{i}$-values at a given time. We select the times such that the snapshots cover the entire limit cycle (Fig.~\ref{fig:starting_points}). As these snapshots serve to define the low-dimensional dynamics of the system it is important to have enough to adequately sample the low-dimensional manifold.

\subsection{Finding the POD Modes}

\begin{figure*}
    \centering
    \includegraphics[width=\textwidth]{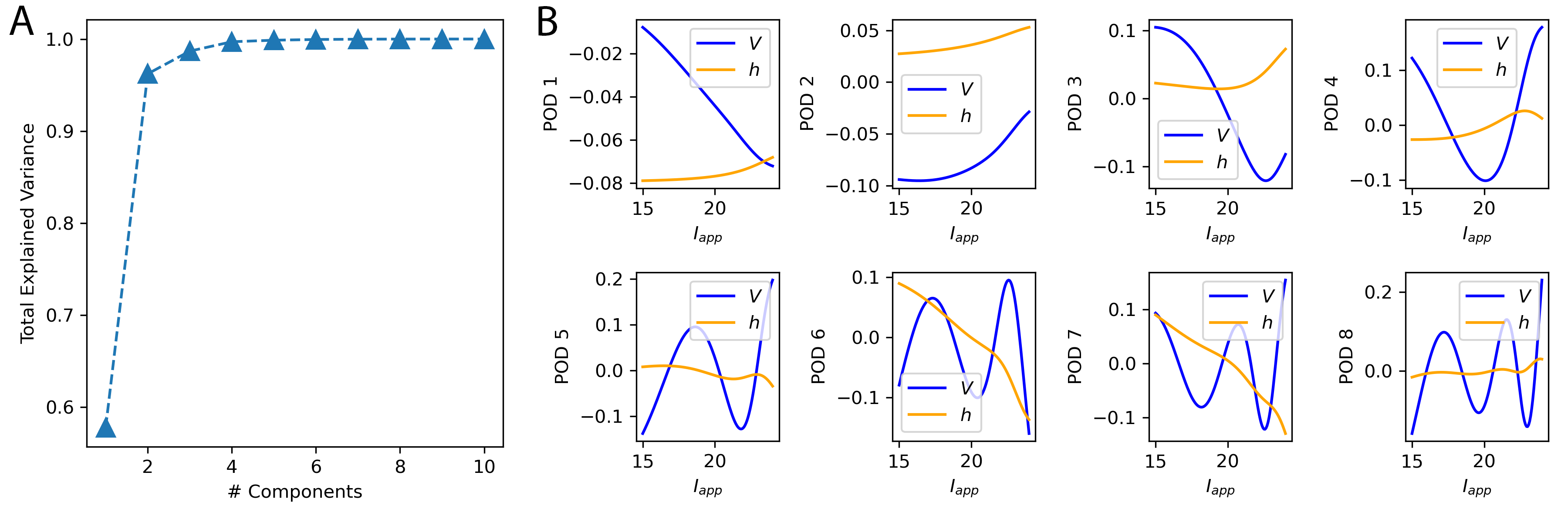}
    \caption{(\textbf{A}) The total variance explained by the POD modes as a function of the number of modes retained. The amount of additional variance captured by each mode greatly decreases after the second mode and reaches a plateau by the fourth. (\textbf{B}) The first 8 POD modes computed from the time snapshot data. The modes are separated into a $V$ (colored in blue) and $h$ (colored in orange) part and are plotted with respect to the oscillator heterogeneity $I_{app}$.}
    \label{fig:POD_modes_variance}
\end{figure*}

Before applying the proper orthogonal decomposition transformation, we normalize the snapshot data on a per dimension basis by subtracting the mean and dividing by the standard deviation in order to remove the influence of the differences in scale between the two state variables (see, for example, the axes in Fig.~\ref{fig:starting_points}). The POD transformation produces an equivalent system defined in terms of the coordinates of the POD modes, which together form a basis for the ambient 256-dimensional space. Because this high-dimensional system is intrinsically low-dimensional in the vicinity of the limit cycle its behavior can be adequately approximated with only a few of the POD modes. As Fig.~\ref{fig:POD_modes_variance}(a) depicts, the fraction of the total variance of the data explained by a reduced set of POD modes increases rapidly with additional modes and plateaus at 4 modes, which are sufficient to explain 99\% of the variance of the original data. The first 8~POD modes themselves are plotted over the oscillator heterogeneity $I_{app}$ in Fig.~\ref{fig:POD_modes_variance}(b), which illustrates that higher modes correspond to higher wavenumber dynamics.

\begin{figure*}
    \centering
    \includegraphics[width=\textwidth]{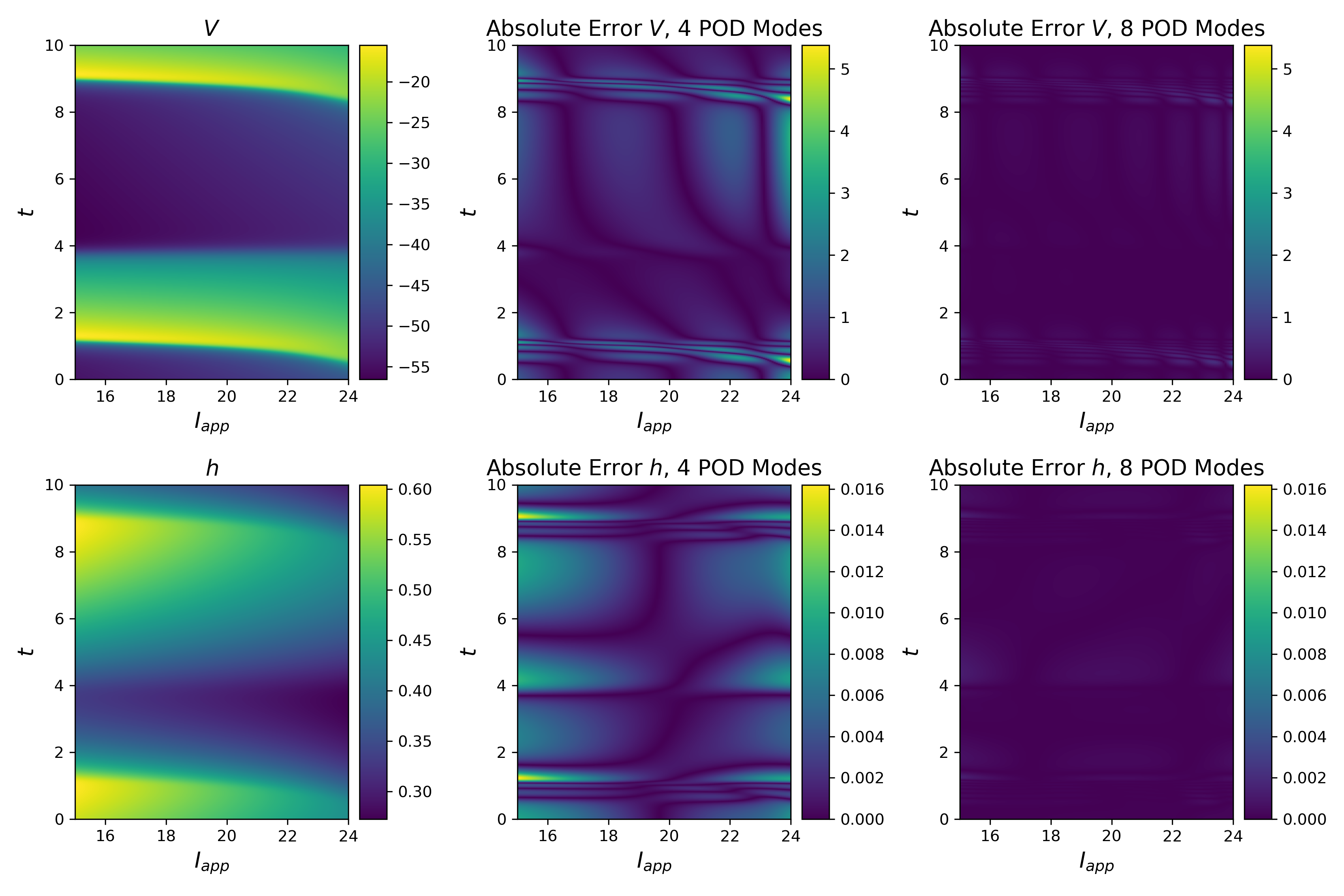}
    \caption{A comparison between different filtered versions of the limit cycle states. The absolute error between the data filtered to 4 POD modes and the truth exhibits noticeable deviations as well as fluctuations that are not present in the truth. Filtering with 8 POD modes greatly diminishes the magnitude of the errors.}
    \label{fig:3d_POD_states}
\end{figure*}

While the high wavenumber modes only account for a small fraction of the variance of the data, they are nonetheless required for a highly accurate representation of the system states. As we illustrate in Fig.~\ref{fig:3d_POD_states}, 8~POD modes are needed for an accurate reproduction of states on the limit cycle. Nevertheless, this transformation furnishes an 8-dimensional reduction of the original 256-dimensional system.

\subsection{Learning the POD ODEs}

We consider 8 POD modes as sufficient for our desired accuracy and seek to learn laws governing their time evolution. These laws will take the form of a system of 8 ODEs with one ODE for each POD coordinate. To this end we construct an artificial neural network integrator based on an Euler time stepper (Fig.~\ref{fig:neural_network_integrator}(a)). This neural network takes states $\mathbf{c}(t)$ at time $t$ and produces new states $\hat{\mathbf{c}}(t+\Delta t)$ at time $t+\Delta t$. By training this network to accurately reproduce a state at a later time $\hat{\mathbf{c}}(t+\Delta t)$ from an initial state $\mathbf{c}(t)$, part of the network learns an approximation for the time derivative of the system, and hence the law governing the time evolution of the system. We seek to apply this method to the POD system, but need input-output pairs of data $(\mathbf{c}(t), \mathbf{c}(t+\Delta t))$ to train the neural network. As we desire stability not only on, but also near the limit cycle, we perturb points off the limit cycle and collect pairs of data from long trajectories as they are attracted back to the limit cycle under the true system dynamics (Eq.~\ref{eq:model_equations}).

\begin{figure}
    \centering
    \includegraphics[width=0.48\textwidth]{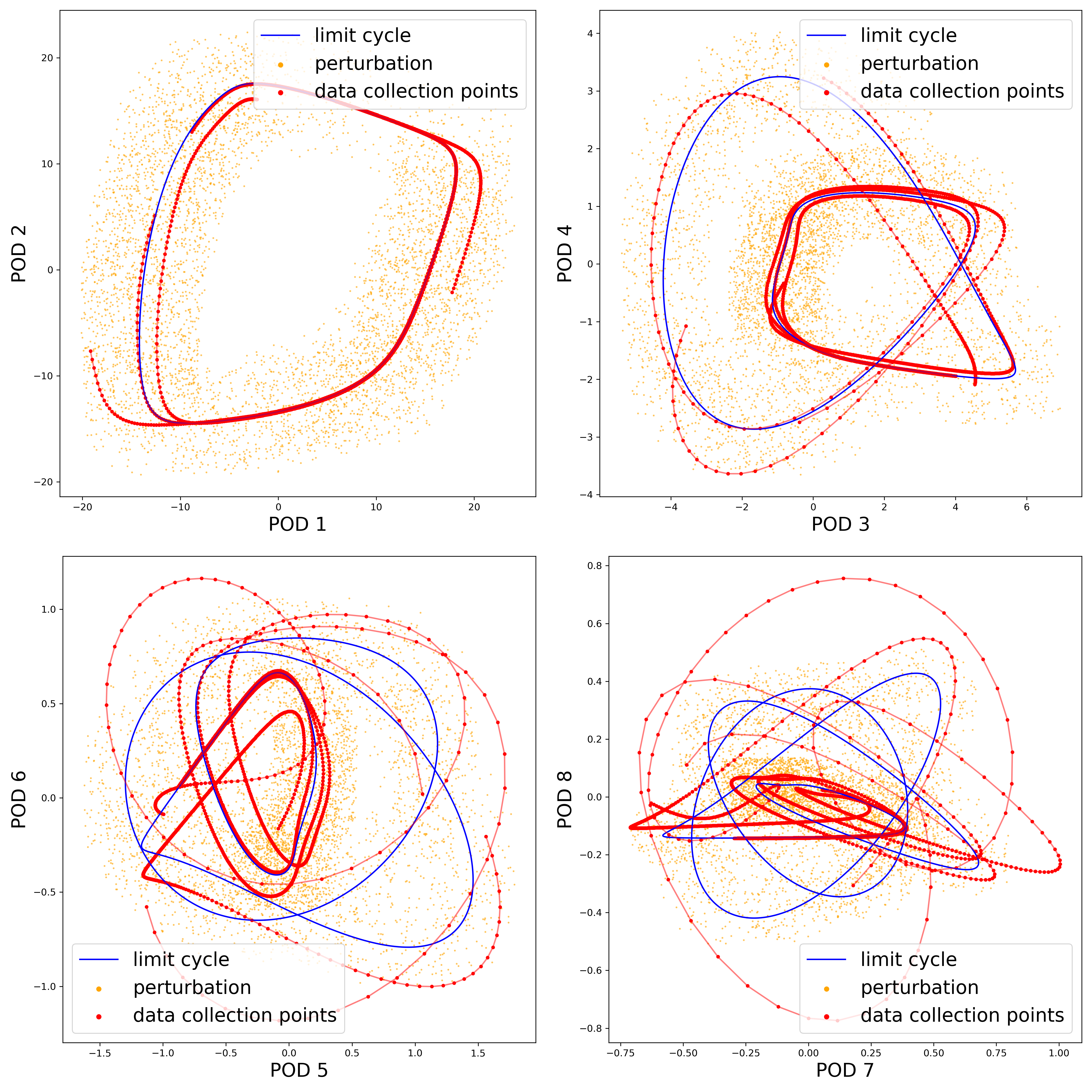}
    \caption{Depictions of phase portraits of the first 8 POD modes; the initial points corresponding to the limit cycle and their perturbations are shown in blue and orange respectively. The perturbed points were generated by adding uniformly randomly distributed noise over the interval $[-\mathrm{std}(\mathrm{POD}_{i})/2, \mathrm{std}(\mathrm{POD}_{i})/2]$ for each POD coordinate $i$. A subset of the data collection trajectories are illustrated in red, with the 500 sample points gathered along each of these trajectories marked by red dots. In the higher modes, the trajectories experience large excursions before being attracted back to the limit cycle.}
    \label{fig:POD_trajectories}
\end{figure}

Consider the set of 5000 snapshots used to compute the POD modes, depicted in Fig.~\ref{fig:starting_points}. We apply the POD transformation to these points to produce a set of initial points in POD space. In order to sample this space, we add uniformly randomly distributed perturbations over $[-\mathrm{std}(\mathrm{POD}_{i})/2, \mathrm{std}(\mathrm{POD}_{i})/2]$ to these points for each of the 8~POD modes that we plan to retain, where $\mathrm{std}(\mathrm{POD}_{i})$ is the standard deviation of POD coordinate $i$ over the set of 5000 initial points, yielding the set of perturbed points illustrated in Fig.~\ref{fig:POD_trajectories}. Each of these 5000 perturbed points is then transformed back to $(V, h)$ profiles and integrated forward in time from $t=0$ to $t=5$ to produce a trajectory which is sampled at 500 evenly spaced times (see Fig.~\ref{fig:POD_trajectories}). This process results in a set of 2,500,000 initial $(V, h)$ snapshots, which are each integrated for $\Delta t=10^{-4}$ to produce final snapshots. The time step, $\Delta t$, was selected small enough to ensure an accurate approximation of the time derivative from the flow data. Transforming these initial and final $(V, h)$ profiles to POD space and discarding all but the first 8 modes results in a set of initial points $\mathbf{c}(t)$ and final points $\mathbf{c}(t+\Delta t)$, which together makeup the training data set.

We use Tensorflow \cite{tensorflow2015-whitepaper} to construct our neural network, which consists of 11 hidden layers each comprised of a dense layer with 128 neurons followed by a batch normalization layer and a ReLU activation non-linearity. The dense layer kernels are initialized with a uniform Glorot initializer \cite{glorot2010understanding} and the biases are initialized to zero. A weighted mean squared error loss is used to compare the predicted output $\hat{\mathbf{c}}(t+\Delta t)$ to the ground truth $\mathbf{c}(t+\Delta t)$, where the weight accounts for differences in scale between the output coordinates. The weight is defined as $1/\mathrm{std}(\mathbf{c})$, where the standard deviation is computed over each dimension of the training data separately. The network is trained with Tensorflow's Adam optimizer \cite{kingma2014adam} for 250 epochs with 10\% of the training data (the training trajectories) set aside as validation data to verify that overfitting has not occurred. The hyperparameters of the network were tuned through trial and error.

\begin{figure}
    \centering
    \includegraphics[width=0.48\textwidth]{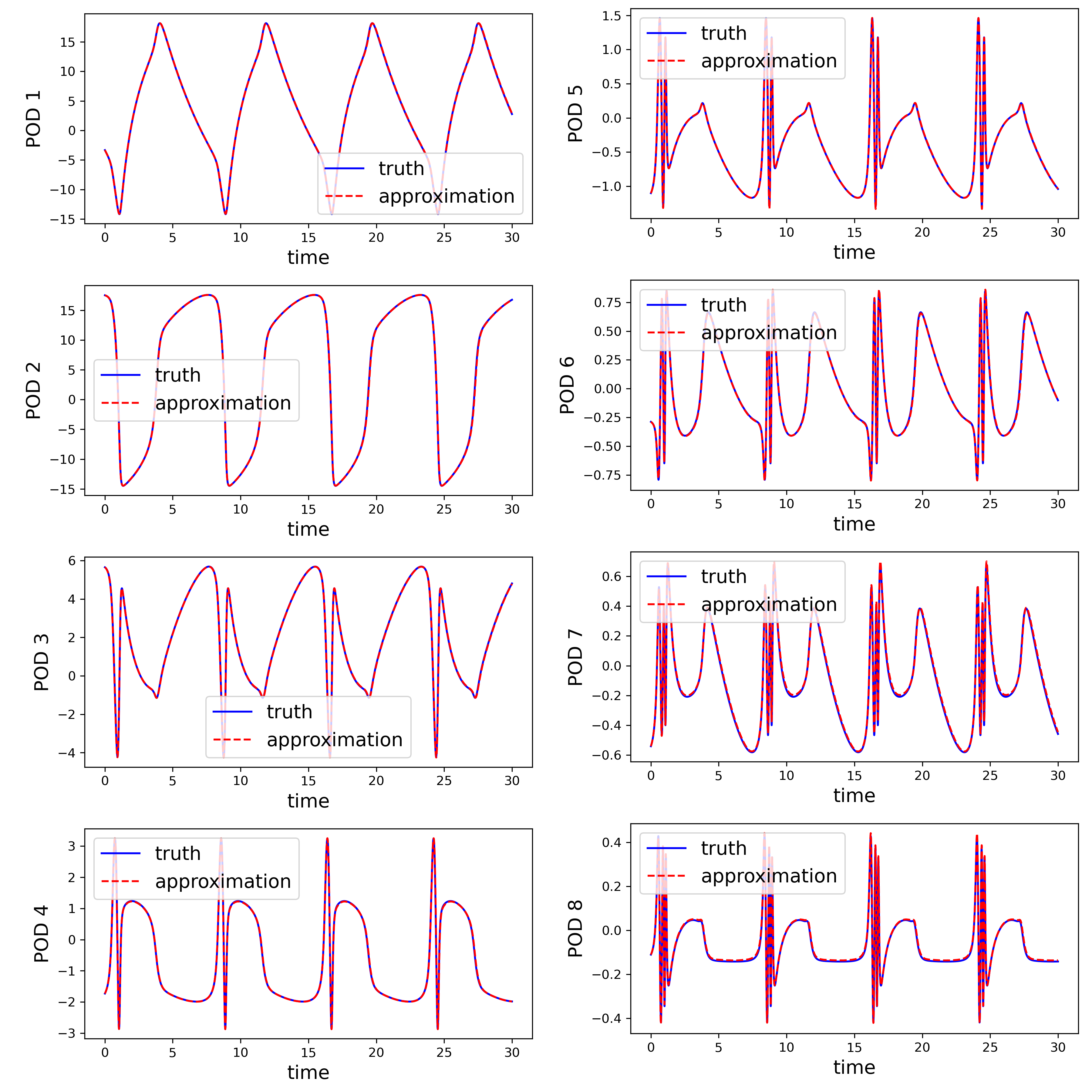}
    \caption{Time series of the 8 POD coordinates corresponding to the integration of a shared initial condition with both the full system (solid blue line) and the low-dimensional neural network derived approximation (dashed red line).}
    \label{fig:POD_timeseries}
\end{figure}

\begin{figure}
    \centering
    \includegraphics[width=0.48\textwidth]{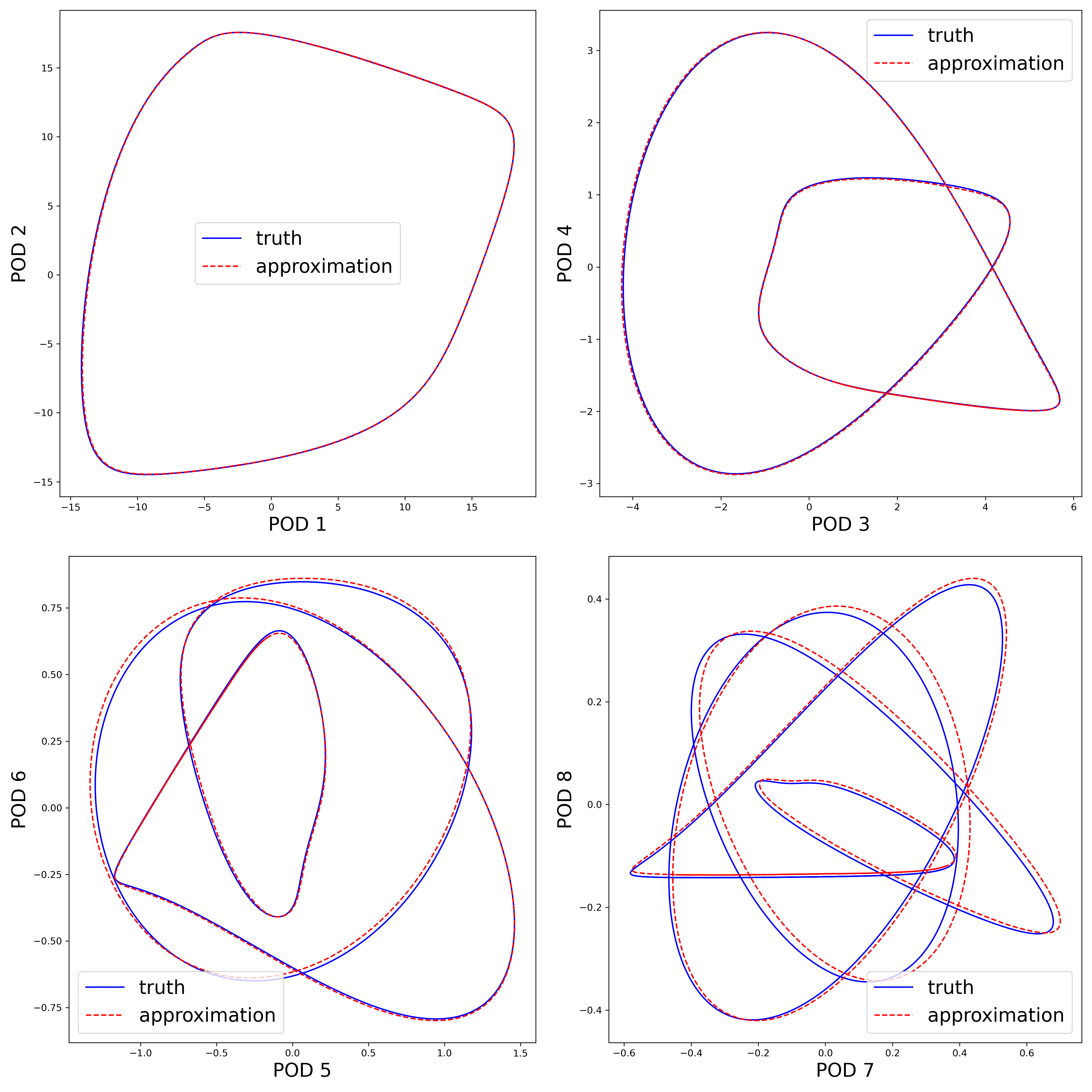}
    \caption{Phase portraits of the long-term behavior of the 8 POD modes corresponding to the neural network system (dashed red line) and the full system (solid blue line). These plots were created by integrating both systems for multiple cycles, to ensure convergence to their respective limit cycles, and plotting the final cycle of each.}
    \label{fig:POD_phase_portraits}
\end{figure}

As a comparison between the reduced system and the full system, we pick a point on the limit cycle and integrate forward in time with both the neural network approximation of the time derivative (in POD space) and the full model equations Eq.~\ref{eq:model_equations} (in $(V, h)$ space) using Scipy's Runge-Kutta integrator. In order to compare the full system to the reduced approximation meaningfully, we project the full system states onto the 8 POD modes used for the low-dimensional approximation to produce time series of each of the 8 POD coordinates. These time series are compared in Fig.~\ref{fig:POD_timeseries}, which depicts near perfect agreement between the coarse-grained approximation and the full system on these 8 modes. The phase portraits shown in Fig.~\ref{fig:POD_phase_portraits} similarly illustrate close agreement between the learned dynamics and the true dynamics on these 8 modes and furthermore suggest that the learned system exhibits a limit cycle that lies close to the limit cycle of the full system.

\begin{figure}
    \centering
    \includegraphics[width=0.48\textwidth]{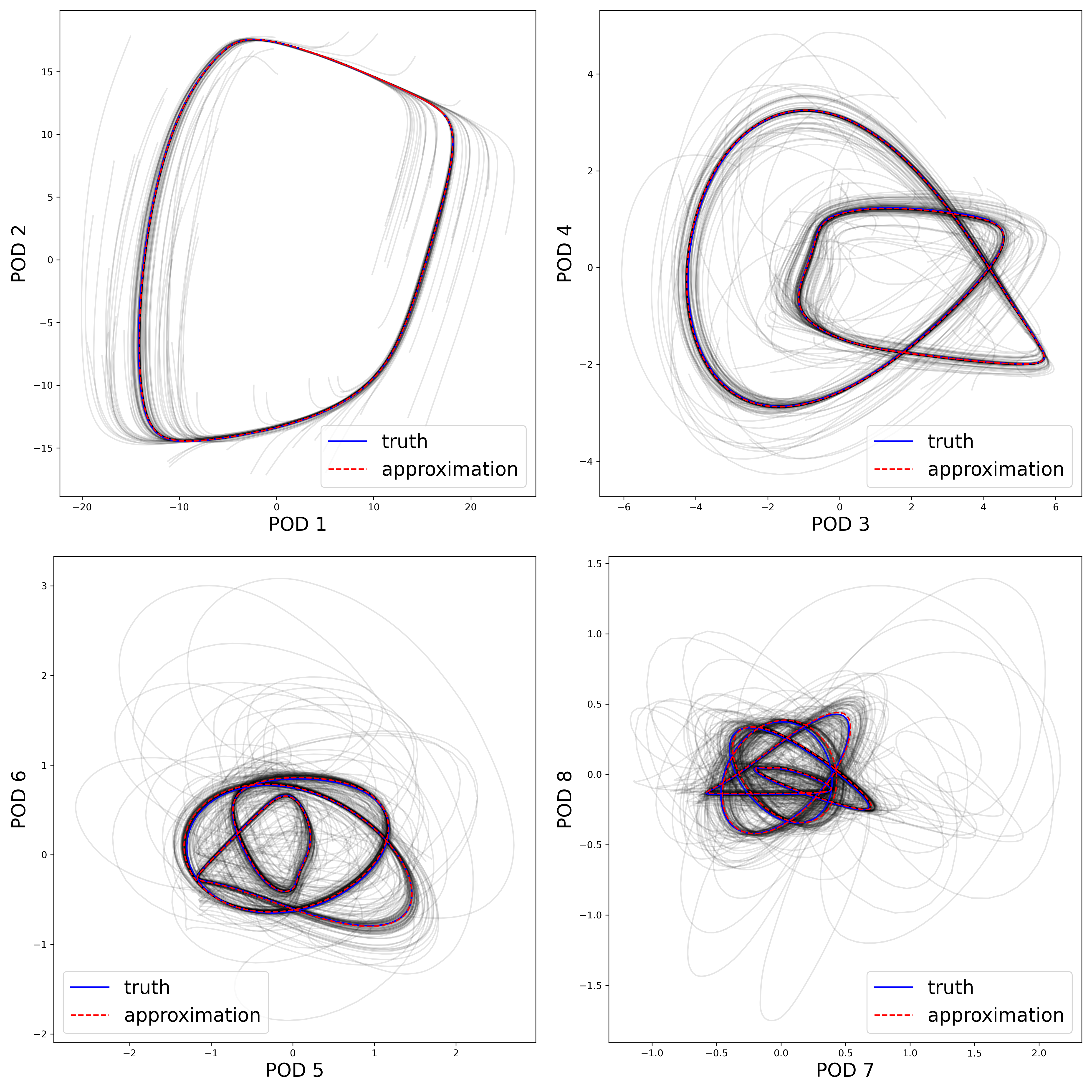}
    \caption{Phase portraits of the 8 POD modes for the approximate neural network system (dashed red line) and the full system (solid blue line). The 100 black curves correspond to trajectories of points that have been perturbed off of the limit cycle and integrated forward in time. The black curves experience large excursions, particularly for the higher POD modes, but are eventually attracted to the limit cycle (see the higher concentration of black curves near the limit cycle approximation in red).}
    \label{fig:POD_stability}
\end{figure}

Finally, we briefly address the stability of the POD approximation. Some degree of stability is already demonstrated by the fact that the POD system is stable under integration and exhibits limit cycle behavior that closely resembles the full system. However, we further illustrate the stability of the learned system by integrating 100 trajectories from initial points that have been randomly selected from the starting points of the data collection trajectories used in the generation of the training data set (the first point of one of the red trajectories seen in Fig.~\ref{fig:POD_trajectories}). The result of this experiment is depicted in Fig.~\ref{fig:POD_stability} where we plot the perturbed trajectories in POD space alongside the long-term behavior of the neural network approximation and the full system. The perturbed trajectories in black are observed to be attracted to the neural network approximation of the limit cycle in red, which closely matches the truth in blue.

\subsection{An Approximate Inertial Form: Reconstructing Higher POD Modes}\label{sec:reconstructing_POD_modes}

Even in cases where the number of POD modes retained is fewer than required to span the space of the spatiotemporal data, it may still be the case that the modes parametrize the low-dimensional dynamics. If this is the case, then the temporal dynamics of the higher modes can be learned as functions of the lower modes and only the behavior of the leading modes must be learned as part of a reduced model. Such a reduction is called an approximate inertial form. Due to the low-dimensional limit cycle behavior of the Hodgkin-Huxley system, we suspect that it admits an approximate inertial form and so retain only 4~POD modes, fewer than the 8 previously used for an accurate approximation of the system, and attempt to learn the behavior of modes 5-8 as a function of these 4.

\begin{figure}
    \centering
    \includegraphics[width=0.48\textwidth]{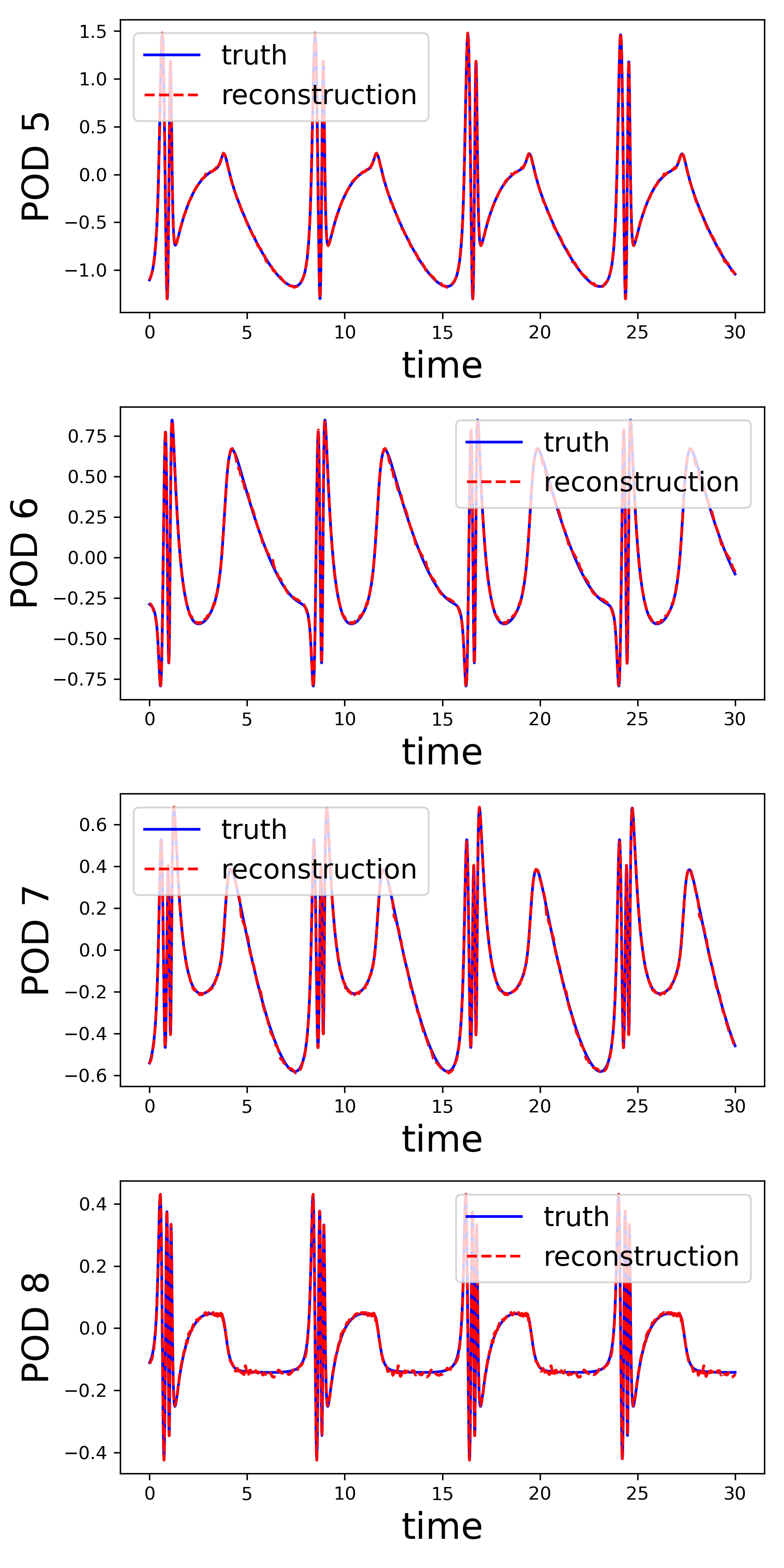}
    \caption{Time series of POD modes 5-8 predicted from time series of modes 1-4 via a dense neural network. The predicted time series are denoted with a dashed red line and compared to the true time series computed from the full system in blue.}
    \label{fig:POD_reconstruction}
\end{figure}

To do this, we use a dense neural network to approximate the function from the lower modes (1-4) to the higher modes (5-8). We train this network on limit cycle data of POD modes 1-4 and provide the corresponding 5-8 mode behavior as targets with a mean square error loss. After training, we check the accuracy of the fit by providing time series of POD modes 1-4 on the limit cycle as input, the first 4 modes of the truth shown in Fig.~\ref{fig:POD_timeseries}, and use the neural network to predict the corresponding time series for modes 5-8. The predicted time series for modes 5-8 are illustrated in Fig.~\ref{fig:POD_reconstruction} and are comparable to those pictured in Fig.~\ref{fig:POD_timeseries}, demonstrating that the dynamics of this system admit an approximate inertial form reduction and can be parametrized with only 4~POD modes.

\section{A Local ML Approach to Spatiotemporal Modeling} \label{sec:local_approximation_technique}

In this section we introduce a methodology for a locally-based model reduction technique that seeks to discover a PDE representation of the Hodgkin-Huxley coupled oscillator system. Specifically, we consider PDEs that depend on the system states and spatial partial derivatives of the states according to,
\begin{equation}
    \frac{\partial \textbf{u}}{\partial t}=\textbf{f}\left(\textbf{u},\frac{\partial \textbf{u}}{\partial x},\dots,\frac{\partial \textbf{u}^n}{\partial^n x}\right).
\label{eq:emergent_PDE}
\end{equation}
Where the spatial variable $x$ {\em is the heterogeneity of the model}, $I_{app}^{i}$ for the Hodgkin-Huxley system. Such a representation is inherently local as the time derivative at any given point only depends on points in a neighborhood around the point (i.e. through finite differences) and thus information in a neighborhood is sufficient to predict a future state. In the following section we demonstrate how such systems can be identified with an artificial neural network integrator in a data-driven way. At this point one may ask how we can hope to learn the behavior of a globally coupled system with only local information. The idea here is that much in the same way that time delays can be used to reconstruct dynamical systems, e.g. Takens' embedding theorem, the spatial derivatives provide observers with which to reconstruct the dynamics on the underlying low-dimensional manifold.

\subsection{Finding a Reduced Space}\label{sec:reduced_space}

We consider the same system studied in Section~\ref{sec:global_approximation_technique} and collect 2000 time snapshots of the dynamics on the low-dimensional limit cycle, each an element of $\mathbb{R}^{256}$. We now face the challenge of generating training data for this system. Unlike in the ODE example where each coordinate could be perturbed individually, our data for the PDE consists of profiles over space, which are in general infinite dimensional. However, because we know a priori that the dynamics of this system are low-dimensional, we use the proper orthogonal decomposition technique to filter the dynamics to a low-dimensional space. Following the same procedure as before, we normalize the snapshot data on a per dimension basis by subtracting the mean and dividing by the standard deviation and then applying the POD dimensionality reduction technique to the normalized data. As we remarked in Section~\ref{sec:global_approximation_technique}, we must keep 8~POD modes for a faithful reproduction of the original system dynamics. The POD modes provide a basis for the profiles in the reduced 8-dimensional function space and afford a set of coordinates with which to perturb the profiles.

We generate 2000 uniformly randomly distributed perturbations over $[-\mathrm{std}(\mathrm{POD}_{i})/2, \mathrm{std}(\mathrm{POD}_{i})/2]$ for each of the 8~POD coordinates $i$ to produce a set of perturbed points that sample the low-dimensional POD space, similar to the one illustrated in Fig.~\ref{fig:POD_trajectories}. We map our 2000 time snapshots into the full 256-dimensional POD space and then add these perturbations before mapping back to the $(V, h)$ space (where the Hodgkin-Huxley system is defined) and integrating forward in time from $t=0$ to $t=5$. By mapping to the full POD space instead of the reduced 8-dimensional space to add our perturbations we retain the higher mode information and ensure that the system is only perturbed in these 8 POD coordinates and not additionally by the loss of the higher modes. We sample these trajectories at 500 evenly spaced times to create a set of 1,000,000 initial points $(V(I_{app}^{i},t),h(I_{app}^{i},t))$ for our neural network integrator. We compute the final points $(V(I_{app}^{i},t+\Delta t),h(I_{app}^{i},t+\Delta t))$ by integrating each initial point for $\Delta t=10^{-4}$. Finally, we filter both the initial and final points down to the 8-dimensional POD space to produce the training data set.

\subsection{Learning a PDE}\label{sec:emergent_pde}

We use Tensorflow to build our neural network integrator, which consists of 4 hidden layers each with a one-dimensional convolutional layer with kernel width one and 128 filters followed by a batch normalization layer and a ReLU activation. The kernels of the convolutional layers are initialized with a uniform Glorot initializer and the biases are initialized to zero. We use a weighted mean squared error between the predicted states $(\hat{V}(I_{app}^{i},t+\Delta t),\hat{h}(I_{app}^{i},t+\Delta t))$ and the true states $(V(I_{app}^{i},t+\Delta t),h(I_{app}^{i},t+\Delta t))$ for the loss, where the weight accounts for the difference in scale between the two variables. A major difference between the neural ODE integrator and the neural PDE integrator is the inclusion of spatial partial derivatives in the PDE integrator. As part of the neural PDE integrator, we include a finite difference layer that computes the spatial partial derivatives of the input states using fourth order accurate centered difference formulas. The maximum order of spatial derivatives to include in the network is a hyperparameter of the model, however we found through experimentation that including up to fifth order derivatives for both $V$ and $h$ is sufficient.

A side effect of using centered differences is that their evaluation requires a neighborhood around each point; this in turn limits the time derivatives predicted by the neural network integrator to a reduced region lying inside the interior of the domain of the input. This restriction does not present an issue during training as the network can be trained with any profiles that sufficiently sample the space, even those corresponding to patches of data that are disjoint in both space and time. However, it does appear in the solution of the PDE and is suggestive of the requirement of additional information to define a well-posed PDE. Typically a set of both initial conditions and boundary conditions are required to define a well-posed PDE, so this naturally leads to the question: what should the boundary conditions of a coupled oscillator derived PDE be? Our answer to this question is to set aside portions of ground truth on the edges of the domain. These ``corridors of truth'' pin the edges of the profiles to prescribed values providing a set of boundary conditions that when combined with initial conditions serve to define a well-posed PDE. We select the corridors wide enough to provide a neighborhood sufficient for the computation of finite difference spatial partial derivatives in the interior region between the corridors.

\begin{figure}
    \centering
    \includegraphics[width=0.48\textwidth]{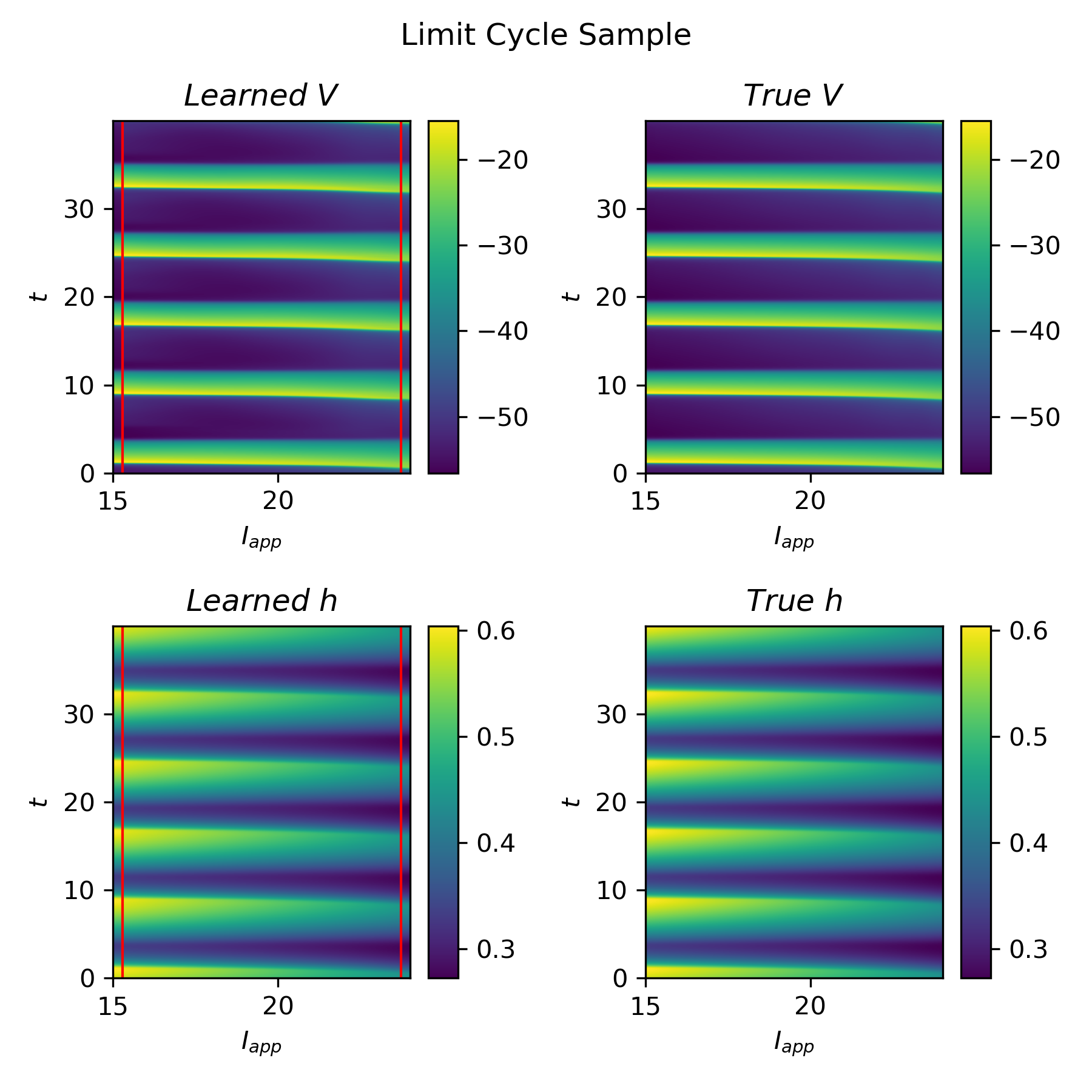}
    \caption{Spacetime plots of profiles generated by integrating a shared initial condition on the limit cycle with both the learned PDE approximation (left) and the full detailed system (right). The red vertical lines denote the extent of the corridors of truth provided to the PDE as boundary conditions. Both solutions have been filtered to 8~POD modes to provide a fair comparison.}
    \label{fig:PDE_spacetime_filtered}
\end{figure}

We train the neural PDE integrator with Tensorflow's Adam optimizer for 120 epochs, setting aside 10\% of the training data as validation to check for overfitting. To assess the quality of our PDE approximation, we select a common initial condition on the limit cycle and integrate both our learned PDE and the detailed ODEs forward in time. Because our neural network has only seen states in the reduced 8~POD mode space, and thus can only make reasonable predictions for these states, it is necessary to filter both the initial condition and the solution after each integration step to this reduced space. Performing the integration produces the spacetime plots illustrated in Fig.~\ref{fig:PDE_spacetime_filtered}, where the boundary conditions provided during integration (the ``corridors of truth'') are delineated with red lines. As depicted in Fig.~\ref{fig:PDE_spacetime_filtered} the dynamics of the learned model display close agreement with those from the detailed model.

\subsection{Subsampling the Domain}\label{sec:domain_subsampling}

In Section~\ref{sec:emergent_pde} we found a PDE representation of a simplified Hodgkin-Huxley system where we used the oscillators as the discretization points of the PDE. One of the benefits of a PDE description of a coupled agent-based system is that the discretization points used to solve the PDE do not have to correspond to the agents (here oscillators). Instead, one could subsample the system onto a coarse grid (where the space is defined by the heterogeneities of the system) and use a sparse set of points to both solve and learn a PDE with the only requirement being that the grid is fine enough to provide accurate approximations of the spatial and temporal partial derivatives. Frequently, it is difficult or impossible to reduce the populations of agent-based systems either due to the smaller systems not retaining the properties of the larger systems or because such a reduction is unknown/unavailable. In these cases, a coarse grid PDE surrogate model offers a comparatively straightforward and well understood alternative that could provide a significant reduction in computational complexity, especially for very large systems, and expedite simulation and analysis.

\begin{figure}
    \centering
    \includegraphics[width=0.48\textwidth]{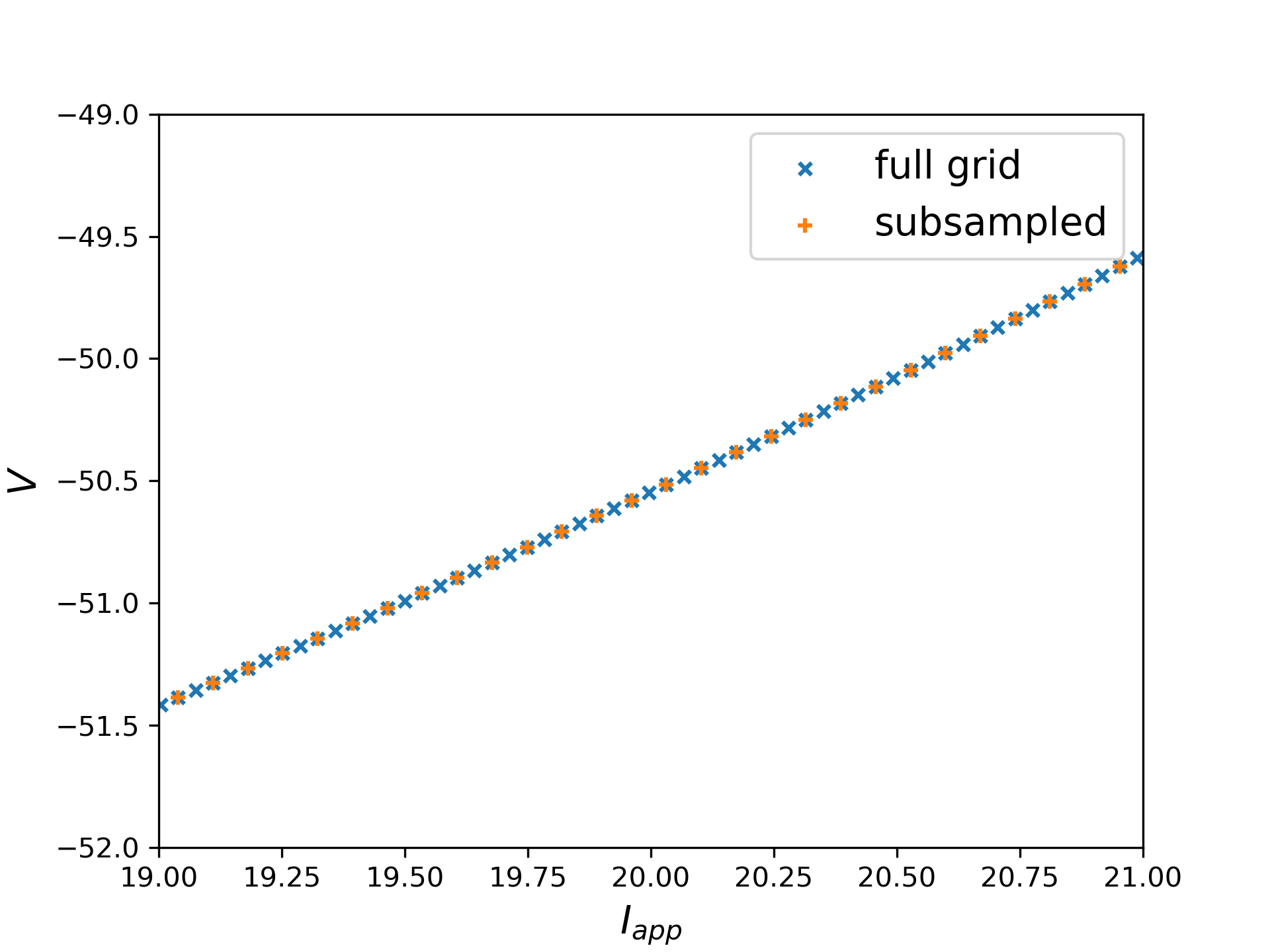}
    \caption{Close up of a subsampled profile of the $V$ state variable showing the use of every other agent for the coarse grid, denoted by the orange pluses, versus the original fine grid in blue. An identical subsampling was performed with the $h$ state variable.}
    \label{fig:subsample_PDE}
\end{figure}

Here we investigate such a subsampling approach and consider the system studied in Section~\ref{sec:reduced_space}, but with $N=255$ oscillators instead of the 128 studied earlier. We assign the oscillators equally spaced $I_{app}$ values over the $[15, 24]$ interval, and use model parameters identical to the smaller set. We generate the perturbations and training data with an equivalent process to the one used in Section~\ref{sec:reduced_space}, and then subsample the training data by selecting every other oscillator as a discretization point of our PDE, as depicted in Fig.~\ref{fig:subsample_PDE}. This process results in a set of 1,000,000 input-output pairs, each point a profile in $\mathbb{R}^{256}$, similar to the 128 oscillator system. We select this particular subsampling because we already know that 128 grid points is enough to compute accurate spatial partial derivatives for this system and the use of every other oscillator simplifies the subsampling procedure. In the case of resampled grids that do not align with the agents, interpolation techniques can be used to carry out the subsampling.

\begin{figure}
    \centering
    \includegraphics[width=0.48\textwidth]{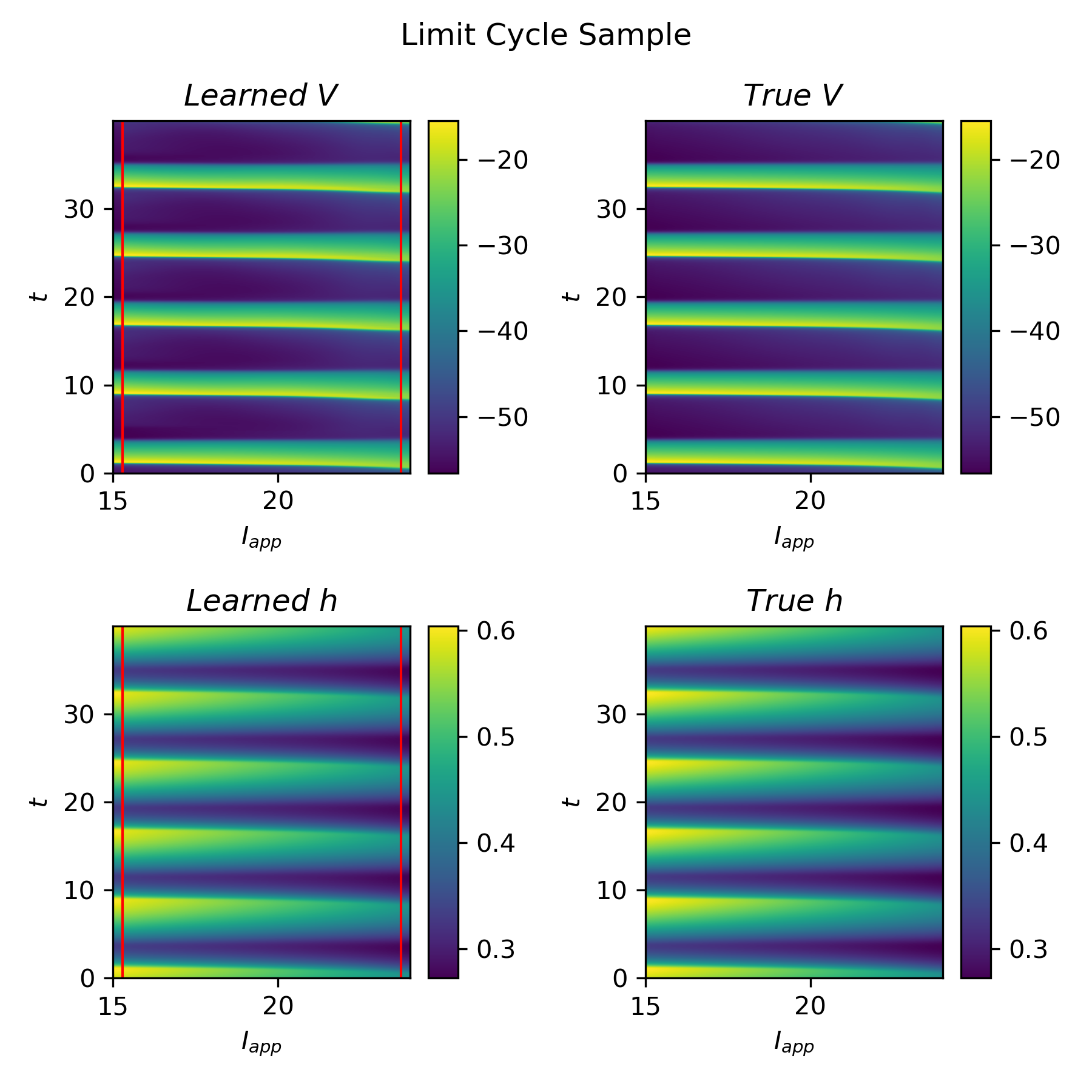}
    \caption{Spacetime plots of profiles generated by integrating a shared initial condition on the limit cycle with both the subsampled, learned PDE approximation (left) and the full detailed system (right). The red vertical lines denote the extent of the corridors of truth provided to the PDE as boundary conditions. Both solutions have been filtered to 8~POD modes to provide a fair comparison.}
    \label{fig:subsample_PDE_solution}
\end{figure}

We train a neural network with 4 hidden layers as described in Section~\ref{sec:emergent_pde} and compare the results of the subsampled PDE to the detailed system by integrating trajectories from each approach starting from a shared initial condition on the limit cycle. For the detailed system, we integrate the 255 oscillator system with the ODEs, subsample the results to every other oscillator, and apply the 8~POD mode filter. For the PDE system we use the same filtering and boundary conditions as in Section~\ref{sec:emergent_pde}, where we provide corridors of truth on the edges of the domain as boundary conditions and filter the initial condition as well as the states after each integration step. By filtering both approaches, we compare the dynamics on the 8~POD mode manifold and provide an equal comparison of the techniques. The results of this comparison are illustrated in Fig.~\ref{fig:subsample_PDE_solution}, which shows that the subsampled PDE can accurately model the behavior of the detailed system.

\section{Discussion}\label{sec:discussion}

The dynamics of large, coupled agent-based systems live in high-dimensional spaces and can produce complex dynamics that are computationally expensive to simulate and difficult to analyze. If the collective dynamics suggest the existence of low-dimensional behavior, then it may be possible to derive analytically coarse-grained surrogate models that are valid on the inertial manifold of the dynamics. However, such analytical model reductions are rarely available and when they are, they are often ad hoc in nature and require great insight and rumination. To address this deficiency we proposed a set of data-driven model reduction approaches for coupled systems including both a globally-based and a locally-based technique. We demonstrated the efficacy of our model reduction methodology through a series of examples based on a simplified Hodgkin-Huxley model.

For the globally-based method we showed how the proper orthogonal decomposition method could be leveraged to derive a set of coarse variables, the POD mode amplitudes, from time series of dynamical data. We then used a neural network integrator to successfully learn the ODEs governing the temporal evolution of the set of POD modes. We illustrated learning both the ``POD-Galerkin" version of the equations, and the ``nonlinear Galerkin" version; in the latter only the leading, ``determining POD mode" amplitude dynamics are learned -an approximate inertial form- and the remaining, higher POD mode amplitudes are recovered as {\em functions} of the lower, determining ones.  Comparisons between the reduced POD model(s) and the full system revealed an excellent reproduction of the Hodgkin-Huxley dynamics affirming the quality of the surrogate model. In our locally-based approach we sought to learn a PDE representation of the coupled Hodgkin-Huxley system based on the spatial partial derivatives of the oscillator states with respect to their heterogeneities. We showed how such a model could be learned from time series of data with a neural network integrator and used the POD method to filter the dynamics to a low-dimensional space to facilitate training as well as to generate the training data perturbations. Our resulting PDE exhibited excellent agreement with the dynamics of the detailed system. One advantage of a PDE description of an agent-based model over the detailed model is that the PDE can be solved with fewer discretization points than there are agents in the system. We showcased this approach with a larger system of Hodgkin-Huxley oscillators and found that the resulting PDE dynamics closely matched the detailed system.

In order to apply our PDE methodology we require that the states of the agents are smooth functions of the model heterogeneities and that the dynamics lie in a low-dimensional space. One of the limitations of this approach is that the heterogeneities provide the coordinates of the PDE and therefore must be known. This is not always possible and furthermore the known coordinates may not be minimal for the specific problem (i.e. there may be lower dimensional sets of coordinates that parametrize the dynamics and which are themselves functions/combinations of the higher dimensional heterogeneities). In related work we show how data-driven manifold learning techniques can be used to identify emergent, parsimonious coordinates from dynamical data that are suited to our PDE learning methodology \cite{kemeth2020learning}.

In Section~\ref{sec:emergent_pde} we learned a PDE representation of the coupled Hodgkin-Huxley system where we provided the states and their spatial partial derivatives as input features to the model. However, there is no reason why the input features of the model must be spatial partial derivatives: in principle we could have considered using arbitrary functions of the states as inputs. The concept of finding input features that are tailored to the specific system has been a topic of interest among the dynamical systems community with approaches such as the analytically derived holistic discretization technique \cite{mackenzie2000holistic, roberts2001holistic, roberts2001holisticburgers} able to provide bespoke finite difference stencils when the PDEs are known. In the case of data-driven system identification, recent work has explored the use of learnable finite difference stencils and made connections between the forms of the stencils and standard finite difference formulas \cite{long2018pde}. We believe that these techniques could be adapted to our PDE methodology enabling greater accuracy and improved generality.

There are several extensions to this work that would be of interest. Firstly, we only considered an all-to-all coupled network, so the neural network ``learned'' the structure of this network, which is trivial in this case. Consider a ring of identical oscillators each coupled to a fixed number of neighbors on either side. Such networks are known to support synchronized ``twisted'' states in which oscillators have the same frequency but not the same phases \cite{wiley2006size}. If the neural network was trained on these types of solutions, would it learn the underlying structure of the network that produced those dynamics, or some other structure, equally capable of fitting/reproducing the data?

Secondly, in the example considered the oscillators synchronized so that the {\em state} of an oscillator is a smooth function of the model heterogeneity. However, for many networks such synchrony does not occur, but the {\em distribution} of the state of an oscillator is a smooth function of the model heterogeneity \cite{laing2021dynamics,blasche20,laing2020effects}. In such cases it would be interesting to learn the dynamics governing the lowest few moments of these distributions. If the dynamics are such that the Ott-Antonsen ansatz is applicable \cite{ott2008low}, the distribution is completely specified by a single complex number, so it would suffice to learn the dynamics of this quantity, assuming it was a smooth function of the network's heterogeneity. In the works Refs.~\onlinecite{laing2021dynamics,blasche20,laing2020effects} it is {\em assumed} that the relevant heterogeneity is the degrees (both in-degree and out-degree) of an oscillator, but in some cases it can be shown analytically that only the in-degree is relevant, while in other cases both degrees are relevant. Our recently developed manifold learning techniques \cite{kemeth2020learning} should allow one to discover this automatically in these and more complex networks, and to determine whether other structural features of networks determine their dynamics.

\section*{Author Contributions}
All authors (TNT, FPK, TB, CRL, IGK) participated at different stages in planning the research, with IGK coordinating. TNT and FPK performed the research, supported by TB who had previous related experience. CRL contributed crucially in linking the mathematical aspects of the work with computational neuroscience modeling. TNT, FPK and IGK wrote the paper, with support from CRL and TB.

\section*{Acknowledgements}
This work was partially supported by an ARO MURI (Dr. M. Munson) and by the DARPA PAI program.

\section*{Source Code and Data Availability}
The data that supports the findings of this study are available within the article. The source code used to implement the framework and methodology presented in this paper is available at \url{https://github.com/TThiem/agent_model_reductions}.

\section*{References}

\bibliography{references}

\begin{thebibliography}{106}%
\makeatletter
\providecommand \@ifxundefined [1]{%
 \@ifx{#1\undefined}
}%
\providecommand \@ifnum [1]{%
 \ifnum #1\expandafter \@firstoftwo
 \else \expandafter \@secondoftwo
 \fi
}%
\providecommand \@ifx [1]{%
 \ifx #1\expandafter \@firstoftwo
 \else \expandafter \@secondoftwo
 \fi
}%
\providecommand \natexlab [1]{#1}%
\providecommand \enquote  [1]{``#1''}%
\providecommand \bibnamefont  [1]{#1}%
\providecommand \bibfnamefont [1]{#1}%
\providecommand \citenamefont [1]{#1}%
\providecommand \href@noop [0]{\@secondoftwo}%
\providecommand \href [0]{\begingroup \@sanitize@url \@href}%
\providecommand \@href[1]{\@@startlink{#1}\@@href}%
\providecommand \@@href[1]{\endgroup#1\@@endlink}%
\providecommand \@sanitize@url [0]{\catcode `\\12\catcode `\$12\catcode
  `\&12\catcode `\#12\catcode `\^12\catcode `\_12\catcode `\%12\relax}%
\providecommand \@@startlink[1]{}%
\providecommand \@@endlink[0]{}%
\providecommand \url  [0]{\begingroup\@sanitize@url \@url }%
\providecommand \@url [1]{\endgroup\@href {#1}{\urlprefix }}%
\providecommand \urlprefix  [0]{URL }%
\providecommand \Eprint [0]{\href }%
\providecommand \doibase [0]{http://dx.doi.org/}%
\providecommand \selectlanguage [0]{\@gobble}%
\providecommand \bibinfo  [0]{\@secondoftwo}%
\providecommand \bibfield  [0]{\@secondoftwo}%
\providecommand \translation [1]{[#1]}%
\providecommand \BibitemOpen [0]{}%
\providecommand \bibitemStop [0]{}%
\providecommand \bibitemNoStop [0]{.\EOS\space}%
\providecommand \EOS [0]{\spacefactor3000\relax}%
\providecommand \BibitemShut  [1]{\csname bibitem#1\endcsname}%
\let\auto@bib@innerbib\@empty
\bibitem [{\citenamefont {Carley}\ \emph {et~al.}(2006)\citenamefont {Carley},
  \citenamefont {Fridsma}, \citenamefont {Casman}, \citenamefont {Yahja},
  \citenamefont {Altman}, \citenamefont {Chen}, \citenamefont {Kaminsky},\ and\
  \citenamefont {Nave}}]{carley2006biowar}%
  \BibitemOpen
  \bibfield  {author} {\bibinfo {author} {\bibfnamefont {K.~M.}\ \bibnamefont
  {Carley}}, \bibinfo {author} {\bibfnamefont {D.~B.}\ \bibnamefont {Fridsma}},
  \bibinfo {author} {\bibfnamefont {E.}~\bibnamefont {Casman}}, \bibinfo
  {author} {\bibfnamefont {A.}~\bibnamefont {Yahja}}, \bibinfo {author}
  {\bibfnamefont {N.}~\bibnamefont {Altman}}, \bibinfo {author} {\bibfnamefont
  {L.-C.}\ \bibnamefont {Chen}}, \bibinfo {author} {\bibfnamefont
  {B.}~\bibnamefont {Kaminsky}}, \ and\ \bibinfo {author} {\bibfnamefont
  {D.}~\bibnamefont {Nave}},\ }\bibfield  {title} {\enquote {\bibinfo {title}
  {{BioWar: scalable agent-based model of bioattacks}},}\ }\href@noop {}
  {\bibfield  {journal} {\bibinfo  {journal} {IEEE Transactions on Systems,
  Man, and Cybernetics-Part A: Systems and Humans}\ }\textbf {\bibinfo {volume}
  {36}},\ \bibinfo {pages} {252--265} (\bibinfo {year} {2006})}\BibitemShut
  {NoStop}%
\bibitem [{\citenamefont {Kattis}\ \emph {et~al.}(2016)\citenamefont {Kattis},
  \citenamefont {Holiday}, \citenamefont {Stoica},\ and\ \citenamefont
  {Kevrekidis}}]{kattis2016modeling}%
  \BibitemOpen
  \bibfield  {author} {\bibinfo {author} {\bibfnamefont {A.~A.}\ \bibnamefont
  {Kattis}}, \bibinfo {author} {\bibfnamefont {A.}~\bibnamefont {Holiday}},
  \bibinfo {author} {\bibfnamefont {A.-A.}\ \bibnamefont {Stoica}}, \ and\
  \bibinfo {author} {\bibfnamefont {I.~G.}\ \bibnamefont {Kevrekidis}},\
  }\bibfield  {title} {\enquote {\bibinfo {title} {Modeling epidemics on
  adaptively evolving networks: a data-mining perspective},}\ }\href@noop {}
  {\bibfield  {journal} {\bibinfo  {journal} {Virulence}\ }\textbf {\bibinfo
  {volume} {7}},\ \bibinfo {pages} {153--162} (\bibinfo {year}
  {2016})}\BibitemShut {NoStop}%
\bibitem [{\citenamefont {Hoertel}\ \emph {et~al.}(2020)\citenamefont
  {Hoertel}, \citenamefont {Blachier}, \citenamefont {Blanco}, \citenamefont
  {Olfson}, \citenamefont {Massetti}, \citenamefont {Rico}, \citenamefont
  {Limosin},\ and\ \citenamefont {Leleu}}]{hoertel2020stochastic}%
  \BibitemOpen
  \bibfield  {author} {\bibinfo {author} {\bibfnamefont {N.}~\bibnamefont
  {Hoertel}}, \bibinfo {author} {\bibfnamefont {M.}~\bibnamefont {Blachier}},
  \bibinfo {author} {\bibfnamefont {C.}~\bibnamefont {Blanco}}, \bibinfo
  {author} {\bibfnamefont {M.}~\bibnamefont {Olfson}}, \bibinfo {author}
  {\bibfnamefont {M.}~\bibnamefont {Massetti}}, \bibinfo {author}
  {\bibfnamefont {M.~S.}\ \bibnamefont {Rico}}, \bibinfo {author}
  {\bibfnamefont {F.}~\bibnamefont {Limosin}}, \ and\ \bibinfo {author}
  {\bibfnamefont {H.}~\bibnamefont {Leleu}},\ }\bibfield  {title} {\enquote
  {\bibinfo {title} {{A stochastic agent-based model of the SARS-CoV-2 epidemic
  in France}},}\ }\href@noop {} {\bibfield  {journal} {\bibinfo  {journal}
  {Nature medicine}\ }\textbf {\bibinfo {volume} {26}},\ \bibinfo {pages}
  {1417--1421} (\bibinfo {year} {2020})}\BibitemShut {NoStop}%
\bibitem [{\citenamefont {Silva}\ \emph {et~al.}(2020)\citenamefont {Silva},
  \citenamefont {Batista}, \citenamefont {Lima}, \citenamefont {Alves},
  \citenamefont {Guimar{\~a}es},\ and\ \citenamefont {Silva}}]{silva2020covid}%
  \BibitemOpen
  \bibfield  {author} {\bibinfo {author} {\bibfnamefont {P.~C.}\ \bibnamefont
  {Silva}}, \bibinfo {author} {\bibfnamefont {P.~V.}\ \bibnamefont {Batista}},
  \bibinfo {author} {\bibfnamefont {H.~S.}\ \bibnamefont {Lima}}, \bibinfo
  {author} {\bibfnamefont {M.~A.}\ \bibnamefont {Alves}}, \bibinfo {author}
  {\bibfnamefont {F.~G.}\ \bibnamefont {Guimar{\~a}es}}, \ and\ \bibinfo
  {author} {\bibfnamefont {R.~C.}\ \bibnamefont {Silva}},\ }\bibfield  {title}
  {\enquote {\bibinfo {title} {{COVID-ABS: An agent-based model of COVID-19
  epidemic to simulate health and economic effects of social distancing
  interventions}},}\ }\href@noop {} {\bibfield  {journal} {\bibinfo  {journal}
  {Chaos, Solitons \& Fractals}\ }\textbf {\bibinfo {volume} {139}},\ \bibinfo
  {pages} {110088} (\bibinfo {year} {2020})}\BibitemShut {NoStop}%
\bibitem [{\citenamefont {Benenson}, \citenamefont {Martens},\ and\
  \citenamefont {Birfir}(2008)}]{benenson2008parkagent}%
  \BibitemOpen
  \bibfield  {author} {\bibinfo {author} {\bibfnamefont {I.}~\bibnamefont
  {Benenson}}, \bibinfo {author} {\bibfnamefont {K.}~\bibnamefont {Martens}}, \
  and\ \bibinfo {author} {\bibfnamefont {S.}~\bibnamefont {Birfir}},\
  }\bibfield  {title} {\enquote {\bibinfo {title} {{PARKAGENT: An agent-based
  model of parking in the city}},}\ }\href@noop {} {\bibfield  {journal}
  {\bibinfo  {journal} {Computers, Environment and Urban Systems}\ }\textbf
  {\bibinfo {volume} {32}},\ \bibinfo {pages} {431--439} (\bibinfo {year}
  {2008})}\BibitemShut {NoStop}%
\bibitem [{\citenamefont {Zou}\ \emph {et~al.}(2012{\natexlab{a}})\citenamefont
  {Zou}, \citenamefont {Torrens}, \citenamefont {Ghanem},\ and\ \citenamefont
  {Kevrekidis}}]{zou2012accelerating}%
  \BibitemOpen
  \bibfield  {author} {\bibinfo {author} {\bibfnamefont {Y.}~\bibnamefont
  {Zou}}, \bibinfo {author} {\bibfnamefont {P.~M.}\ \bibnamefont {Torrens}},
  \bibinfo {author} {\bibfnamefont {R.~G.}\ \bibnamefont {Ghanem}}, \ and\
  \bibinfo {author} {\bibfnamefont {I.~G.}\ \bibnamefont {Kevrekidis}},\
  }\bibfield  {title} {\enquote {\bibinfo {title} {Accelerating agent-based
  computation of complex urban systems},}\ }\href@noop {} {\bibfield  {journal}
  {\bibinfo  {journal} {International Journal of Geographical Information
  Science}\ }\textbf {\bibinfo {volume} {26}},\ \bibinfo {pages} {1917--1937}
  (\bibinfo {year} {2012}{\natexlab{a}})}\BibitemShut {NoStop}%
\bibitem [{\citenamefont {Torrens}\ \emph {et~al.}(2013)\citenamefont
  {Torrens}, \citenamefont {Kevrekidis}, \citenamefont {Ghanem},\ and\
  \citenamefont {Zou}}]{torrens2013simple}%
  \BibitemOpen
  \bibfield  {author} {\bibinfo {author} {\bibfnamefont {P.~M.}\ \bibnamefont
  {Torrens}}, \bibinfo {author} {\bibfnamefont {Y.}~\bibnamefont {Kevrekidis}},
  \bibinfo {author} {\bibfnamefont {R.}~\bibnamefont {Ghanem}}, \ and\ \bibinfo
  {author} {\bibfnamefont {Y.}~\bibnamefont {Zou}},\ }\bibfield  {title}
  {\enquote {\bibinfo {title} {Simple urban simulation atop complicated models:
  Multi-scale equation-free computing of sprawl using geographic automata},}\
  }\href@noop {} {\bibfield  {journal} {\bibinfo  {journal} {Entropy}\ }\textbf
  {\bibinfo {volume} {15}},\ \bibinfo {pages} {2606--2634} (\bibinfo {year}
  {2013})}\BibitemShut {NoStop}%
\bibitem [{\citenamefont {Fagnant}\ and\ \citenamefont
  {Kockelman}(2014)}]{fagnant2014travel}%
  \BibitemOpen
  \bibfield  {author} {\bibinfo {author} {\bibfnamefont {D.~J.}\ \bibnamefont
  {Fagnant}}\ and\ \bibinfo {author} {\bibfnamefont {K.~M.}\ \bibnamefont
  {Kockelman}},\ }\bibfield  {title} {\enquote {\bibinfo {title} {The travel
  and environmental implications of shared autonomous vehicles, using
  agent-based model scenarios},}\ }\href@noop {} {\bibfield  {journal}
  {\bibinfo  {journal} {Transportation Research Part C: Emerging Technologies}\
  }\textbf {\bibinfo {volume} {40}},\ \bibinfo {pages} {1--13} (\bibinfo {year}
  {2014})}\BibitemShut {NoStop}%
\bibitem [{\citenamefont {Martinez}\ and\ \citenamefont
  {Viegas}(2017)}]{martinez2017assessing}%
  \BibitemOpen
  \bibfield  {author} {\bibinfo {author} {\bibfnamefont {L.~M.}\ \bibnamefont
  {Martinez}}\ and\ \bibinfo {author} {\bibfnamefont {J.~M.}\ \bibnamefont
  {Viegas}},\ }\bibfield  {title} {\enquote {\bibinfo {title} {{Assessing the
  impacts of deploying a shared self-driving urban mobility system: An
  agent-based model applied to the city of Lisbon, Portugal}},}\ }\href@noop {}
  {\bibfield  {journal} {\bibinfo  {journal} {International Journal of
  Transportation Science and Technology}\ }\textbf {\bibinfo {volume} {6}},\
  \bibinfo {pages} {13--27} (\bibinfo {year} {2017})}\BibitemShut {NoStop}%
\bibitem [{\citenamefont {Siettos}, \citenamefont {Gear},\ and\ \citenamefont
  {Kevrekidis}(2012)}]{siettos2012equation}%
  \BibitemOpen
  \bibfield  {author} {\bibinfo {author} {\bibfnamefont {C.}~\bibnamefont
  {Siettos}}, \bibinfo {author} {\bibfnamefont {C.}~\bibnamefont {Gear}}, \
  and\ \bibinfo {author} {\bibfnamefont {I.}~\bibnamefont {Kevrekidis}},\
  }\bibfield  {title} {\enquote {\bibinfo {title} {An equation-free approach to
  agent-based computation: Bifurcation analysis and control of stationary
  states},}\ }\href@noop {} {\bibfield  {journal} {\bibinfo  {journal} {EPL
  (Europhysics Letters)}\ }\textbf {\bibinfo {volume} {99}},\ \bibinfo {pages}
  {48007} (\bibinfo {year} {2012})}\BibitemShut {NoStop}%
\bibitem [{\citenamefont {Liu}\ \emph {et~al.}(2015)\citenamefont {Liu},
  \citenamefont {Siettos}, \citenamefont {Gear},\ and\ \citenamefont
  {Kevrekidis}}]{liu2015equation}%
  \BibitemOpen
  \bibfield  {author} {\bibinfo {author} {\bibfnamefont {P.}~\bibnamefont
  {Liu}}, \bibinfo {author} {\bibfnamefont {C.}~\bibnamefont {Siettos}},
  \bibinfo {author} {\bibfnamefont {C.}~\bibnamefont {Gear}}, \ and\ \bibinfo
  {author} {\bibfnamefont {I.}~\bibnamefont {Kevrekidis}},\ }\bibfield  {title}
  {\enquote {\bibinfo {title} {Equation-free model reduction in agent-based
  computations: Coarse-grained bifurcation and variable-free rare event
  analysis},}\ }\href@noop {} {\bibfield  {journal} {\bibinfo  {journal}
  {Mathematical Modelling of Natural Phenomena}\ }\textbf {\bibinfo {volume}
  {10}},\ \bibinfo {pages} {71--90} (\bibinfo {year} {2015})}\BibitemShut
  {NoStop}%
\bibitem [{\citenamefont {Tsoumanis}\ \emph {et~al.}(2010)\citenamefont
  {Tsoumanis}, \citenamefont {Siettos}, \citenamefont {Bafas},\ and\
  \citenamefont {Kevrekidis}}]{tsoumanis2010equation}%
  \BibitemOpen
  \bibfield  {author} {\bibinfo {author} {\bibfnamefont {A.~C.}\ \bibnamefont
  {Tsoumanis}}, \bibinfo {author} {\bibfnamefont {C.~I.}\ \bibnamefont
  {Siettos}}, \bibinfo {author} {\bibfnamefont {G.~V.}\ \bibnamefont {Bafas}},
  \ and\ \bibinfo {author} {\bibfnamefont {I.~G.}\ \bibnamefont {Kevrekidis}},\
  }\bibfield  {title} {\enquote {\bibinfo {title} {Equation-free multiscale
  computations in social networks: from agent-based modeling to coarse-grained
  stability and bifurcation analysis},}\ }\href@noop {} {\bibfield  {journal}
  {\bibinfo  {journal} {International Journal of Bifurcation and Chaos}\
  }\textbf {\bibinfo {volume} {20}},\ \bibinfo {pages} {3673--3688} (\bibinfo
  {year} {2010})}\BibitemShut {NoStop}%
\bibitem [{\citenamefont {Zou}\ \emph {et~al.}(2012{\natexlab{b}})\citenamefont
  {Zou}, \citenamefont {Fonoberov}, \citenamefont {Fonoberova}, \citenamefont
  {Mezic},\ and\ \citenamefont {Kevrekidis}}]{zou2012model}%
  \BibitemOpen
  \bibfield  {author} {\bibinfo {author} {\bibfnamefont {Y.}~\bibnamefont
  {Zou}}, \bibinfo {author} {\bibfnamefont {V.~A.}\ \bibnamefont {Fonoberov}},
  \bibinfo {author} {\bibfnamefont {M.}~\bibnamefont {Fonoberova}}, \bibinfo
  {author} {\bibfnamefont {I.}~\bibnamefont {Mezic}}, \ and\ \bibinfo {author}
  {\bibfnamefont {I.~G.}\ \bibnamefont {Kevrekidis}},\ }\bibfield  {title}
  {\enquote {\bibinfo {title} {Model reduction for agent-based social
  simulation: coarse-graining a civil violence model},}\ }\href@noop {}
  {\bibfield  {journal} {\bibinfo  {journal} {Physical review E}\ }\textbf
  {\bibinfo {volume} {85}},\ \bibinfo {pages} {066106} (\bibinfo {year}
  {2012}{\natexlab{b}})}\BibitemShut {NoStop}%
\bibitem [{\citenamefont {Haer}\ \emph {et~al.}(2020)\citenamefont {Haer},
  \citenamefont {Husby}, \citenamefont {Botzen},\ and\ \citenamefont
  {Aerts}}]{haer2020safe}%
  \BibitemOpen
  \bibfield  {author} {\bibinfo {author} {\bibfnamefont {T.}~\bibnamefont
  {Haer}}, \bibinfo {author} {\bibfnamefont {T.~G.}\ \bibnamefont {Husby}},
  \bibinfo {author} {\bibfnamefont {W.~W.}\ \bibnamefont {Botzen}}, \ and\
  \bibinfo {author} {\bibfnamefont {J.~C.}\ \bibnamefont {Aerts}},\ }\bibfield
  {title} {\enquote {\bibinfo {title} {{The safe development paradox: An
  agent-based model for flood risk under climate change in the European
  Union}},}\ }\href@noop {} {\bibfield  {journal} {\bibinfo  {journal} {Global
  Environmental Change}\ }\textbf {\bibinfo {volume} {60}},\ \bibinfo {pages}
  {102009} (\bibinfo {year} {2020})}\BibitemShut {NoStop}%
\bibitem [{\citenamefont {Anderson}(1972)}]{anderson1972more}%
  \BibitemOpen
  \bibfield  {author} {\bibinfo {author} {\bibfnamefont {P.~W.}\ \bibnamefont
  {Anderson}},\ }\bibfield  {title} {\enquote {\bibinfo {title} {More is
  different},}\ }\href@noop {} {\bibfield  {journal} {\bibinfo  {journal}
  {Science}\ }\textbf {\bibinfo {volume} {177}},\ \bibinfo {pages} {393--396}
  (\bibinfo {year} {1972})}\BibitemShut {NoStop}%
\bibitem [{\citenamefont {Kuramoto}(1984)}]{kuramoto1984chemical}%
  \BibitemOpen
  \bibfield  {author} {\bibinfo {author} {\bibfnamefont {Y.}~\bibnamefont
  {Kuramoto}},\ }\bibfield  {title} {\enquote {\bibinfo {title} {{Chemical
  Oscillations, Waves and Turbulence}},}\ }\href@noop {} {\bibfield  {journal}
  {\bibinfo  {journal} {Synergetics}\ }\textbf {\bibinfo {volume} {19}}
  (\bibinfo {year} {1984})}\BibitemShut {NoStop}%
\bibitem [{\citenamefont {Berkooz}, \citenamefont {Holmes},\ and\ \citenamefont
  {Lumley}(1993)}]{berkooz1993proper}%
  \BibitemOpen
  \bibfield  {author} {\bibinfo {author} {\bibfnamefont {G.}~\bibnamefont
  {Berkooz}}, \bibinfo {author} {\bibfnamefont {P.}~\bibnamefont {Holmes}}, \
  and\ \bibinfo {author} {\bibfnamefont {J.~L.}\ \bibnamefont {Lumley}},\
  }\bibfield  {title} {\enquote {\bibinfo {title} {The proper orthogonal
  decomposition in the analysis of turbulent flows},}\ }\href@noop {}
  {\bibfield  {journal} {\bibinfo  {journal} {Annual review of fluid
  mechanics}\ }\textbf {\bibinfo {volume} {25}},\ \bibinfo {pages} {539--575}
  (\bibinfo {year} {1993})}\BibitemShut {NoStop}%
\bibitem [{\citenamefont {Kerschen}\ \emph {et~al.}(2005)\citenamefont
  {Kerschen}, \citenamefont {Golinval}, \citenamefont {Vakakis},\ and\
  \citenamefont {Bergman}}]{kerschen2005method}%
  \BibitemOpen
  \bibfield  {author} {\bibinfo {author} {\bibfnamefont {G.}~\bibnamefont
  {Kerschen}}, \bibinfo {author} {\bibfnamefont {J.-c.}\ \bibnamefont
  {Golinval}}, \bibinfo {author} {\bibfnamefont {A.~F.}\ \bibnamefont
  {Vakakis}}, \ and\ \bibinfo {author} {\bibfnamefont {L.~A.}\ \bibnamefont
  {Bergman}},\ }\bibfield  {title} {\enquote {\bibinfo {title} {The method of
  proper orthogonal decomposition for dynamical characterization and order
  reduction of mechanical systems: an overview},}\ }\href@noop {} {\bibfield
  {journal} {\bibinfo  {journal} {Nonlinear dynamics}\ }\textbf {\bibinfo
  {volume} {41}},\ \bibinfo {pages} {147--169} (\bibinfo {year}
  {2005})}\BibitemShut {NoStop}%
\bibitem [{\citenamefont {Hinze}\ and\ \citenamefont
  {Volkwein}(2005)}]{hinze2005proper}%
  \BibitemOpen
  \bibfield  {author} {\bibinfo {author} {\bibfnamefont {M.}~\bibnamefont
  {Hinze}}\ and\ \bibinfo {author} {\bibfnamefont {S.}~\bibnamefont
  {Volkwein}},\ }\bibfield  {title} {\enquote {\bibinfo {title} {Proper
  orthogonal decomposition surrogate models for nonlinear dynamical systems:
  Error estimates and suboptimal control},}\ }in\ \href@noop {} {\emph
  {\bibinfo {booktitle} {Dimension reduction of large-scale systems}}}\
  (\bibinfo  {publisher} {Springer},\ \bibinfo {year} {2005})\ pp.\ \bibinfo
  {pages} {261--306}\BibitemShut {NoStop}%
\bibitem [{\citenamefont {Kunisch}\ and\ \citenamefont
  {Volkwein}(2002)}]{kunisch2002galerkin}%
  \BibitemOpen
  \bibfield  {author} {\bibinfo {author} {\bibfnamefont {K.}~\bibnamefont
  {Kunisch}}\ and\ \bibinfo {author} {\bibfnamefont {S.}~\bibnamefont
  {Volkwein}},\ }\bibfield  {title} {\enquote {\bibinfo {title} {Galerkin
  proper orthogonal decomposition methods for a general equation in fluid
  dynamics},}\ }\href@noop {} {\bibfield  {journal} {\bibinfo  {journal} {SIAM
  Journal on Numerical analysis}\ }\textbf {\bibinfo {volume} {40}},\ \bibinfo
  {pages} {492--515} (\bibinfo {year} {2002})}\BibitemShut {NoStop}%
\bibitem [{\citenamefont {Sirovich}\ and\ \citenamefont
  {Rodriguez}(1987)}]{sirovich1987coherent}%
  \BibitemOpen
  \bibfield  {author} {\bibinfo {author} {\bibfnamefont {L.}~\bibnamefont
  {Sirovich}}\ and\ \bibinfo {author} {\bibfnamefont {J.}~\bibnamefont
  {Rodriguez}},\ }\bibfield  {title} {\enquote {\bibinfo {title} {Coherent
  structures and chaos: a model problem},}\ }\href@noop {} {\bibfield
  {journal} {\bibinfo  {journal} {Physics Letters A}\ }\textbf {\bibinfo
  {volume} {120}},\ \bibinfo {pages} {211--214} (\bibinfo {year}
  {1987})}\BibitemShut {NoStop}%
\bibitem [{\citenamefont {Deane}\ \emph {et~al.}(1991)\citenamefont {Deane},
  \citenamefont {Kevrekidis}, \citenamefont {Karniadakis},\ and\ \citenamefont
  {Orszag}}]{deane1991low}%
  \BibitemOpen
  \bibfield  {author} {\bibinfo {author} {\bibfnamefont {A.}~\bibnamefont
  {Deane}}, \bibinfo {author} {\bibfnamefont {I.}~\bibnamefont {Kevrekidis}},
  \bibinfo {author} {\bibfnamefont {G.~E.}\ \bibnamefont {Karniadakis}}, \ and\
  \bibinfo {author} {\bibfnamefont {S.}~\bibnamefont {Orszag}},\ }\bibfield
  {title} {\enquote {\bibinfo {title} {Low-dimensional models for complex
  geometry flows: application to grooved channels and circular cylinders},}\
  }\href@noop {} {\bibfield  {journal} {\bibinfo  {journal} {Physics of Fluids
  A: Fluid Dynamics}\ }\textbf {\bibinfo {volume} {3}},\ \bibinfo {pages}
  {2337--2354} (\bibinfo {year} {1991})}\BibitemShut {NoStop}%
\bibitem [{\citenamefont {Shvartsman}\ and\ \citenamefont
  {Kevrekidis}(1998{\natexlab{a}})}]{shvartsman1998low}%
  \BibitemOpen
  \bibfield  {author} {\bibinfo {author} {\bibfnamefont {S.~Y.}\ \bibnamefont
  {Shvartsman}}\ and\ \bibinfo {author} {\bibfnamefont {I.}~\bibnamefont
  {Kevrekidis}},\ }\bibfield  {title} {\enquote {\bibinfo {title}
  {Low-dimensional approximation and control of periodic solutions in spatially
  extended systems},}\ }\href@noop {} {\bibfield  {journal} {\bibinfo
  {journal} {Physical Review E}\ }\textbf {\bibinfo {volume} {58}},\ \bibinfo
  {pages} {361} (\bibinfo {year} {1998}{\natexlab{a}})}\BibitemShut {NoStop}%
\bibitem [{\citenamefont {Shvartsman}\ and\ \citenamefont
  {Kevrekidis}(1998{\natexlab{b}})}]{shvartsman1998nonlinear}%
  \BibitemOpen
  \bibfield  {author} {\bibinfo {author} {\bibfnamefont {S.~Y.}\ \bibnamefont
  {Shvartsman}}\ and\ \bibinfo {author} {\bibfnamefont {I.~G.}\ \bibnamefont
  {Kevrekidis}},\ }\bibfield  {title} {\enquote {\bibinfo {title} {Nonlinear
  model reduction for control of distributed systems: A computer-assisted
  study},}\ }\href@noop {} {\bibfield  {journal} {\bibinfo  {journal} {AIChE
  Journal}\ }\textbf {\bibinfo {volume} {44}},\ \bibinfo {pages} {1579--1595}
  (\bibinfo {year} {1998}{\natexlab{b}})}\BibitemShut {NoStop}%
\bibitem [{\citenamefont {Shvartsman}\ \emph {et~al.}(2000)\citenamefont
  {Shvartsman}, \citenamefont {Theodoropoulos}, \citenamefont
  {Rico-Mart{\i}{\'n}ez}, \citenamefont {Kevrekidis}, \citenamefont {Titi},\
  and\ \citenamefont {Mountziaris}}]{shvartsman2000order}%
  \BibitemOpen
  \bibfield  {author} {\bibinfo {author} {\bibfnamefont {S.~Y.}\ \bibnamefont
  {Shvartsman}}, \bibinfo {author} {\bibfnamefont {C.}~\bibnamefont
  {Theodoropoulos}}, \bibinfo {author} {\bibfnamefont {R.}~\bibnamefont
  {Rico-Mart{\i}{\'n}ez}}, \bibinfo {author} {\bibfnamefont {I.}~\bibnamefont
  {Kevrekidis}}, \bibinfo {author} {\bibfnamefont {E.~S.}\ \bibnamefont
  {Titi}}, \ and\ \bibinfo {author} {\bibfnamefont {T.}~\bibnamefont
  {Mountziaris}},\ }\bibfield  {title} {\enquote {\bibinfo {title} {Order
  reduction for nonlinear dynamic models of distributed reacting systems},}\
  }\href@noop {} {\bibfield  {journal} {\bibinfo  {journal} {Journal of Process
  Control}\ }\textbf {\bibinfo {volume} {10}},\ \bibinfo {pages} {177--184}
  (\bibinfo {year} {2000})}\BibitemShut {NoStop}%
\bibitem [{\citenamefont {Lee}\ and\ \citenamefont
  {Carlberg}(2020)}]{lee2020model}%
  \BibitemOpen
  \bibfield  {author} {\bibinfo {author} {\bibfnamefont {K.}~\bibnamefont
  {Lee}}\ and\ \bibinfo {author} {\bibfnamefont {K.~T.}\ \bibnamefont
  {Carlberg}},\ }\bibfield  {title} {\enquote {\bibinfo {title} {Model
  reduction of dynamical systems on nonlinear manifolds using deep
  convolutional autoencoders},}\ }\href@noop {} {\bibfield  {journal} {\bibinfo
   {journal} {Journal of Computational Physics}\ }\textbf {\bibinfo {volume}
  {404}},\ \bibinfo {pages} {108973} (\bibinfo {year} {2020})}\BibitemShut
  {NoStop}%
\bibitem [{\citenamefont {Kemeth}\ \emph {et~al.}(2018)\citenamefont {Kemeth},
  \citenamefont {Haugland}, \citenamefont {Dietrich}, \citenamefont {Bertalan},
  \citenamefont {H{\"o}hlein}, \citenamefont {Li}, \citenamefont {Bollt},
  \citenamefont {Talmon}, \citenamefont {Krischer},\ and\ \citenamefont
  {Kevrekidis}}]{kemeth2018emergent}%
  \BibitemOpen
  \bibfield  {author} {\bibinfo {author} {\bibfnamefont {F.~P.}\ \bibnamefont
  {Kemeth}}, \bibinfo {author} {\bibfnamefont {S.~W.}\ \bibnamefont
  {Haugland}}, \bibinfo {author} {\bibfnamefont {F.}~\bibnamefont {Dietrich}},
  \bibinfo {author} {\bibfnamefont {T.}~\bibnamefont {Bertalan}}, \bibinfo
  {author} {\bibfnamefont {K.}~\bibnamefont {H{\"o}hlein}}, \bibinfo {author}
  {\bibfnamefont {Q.}~\bibnamefont {Li}}, \bibinfo {author} {\bibfnamefont
  {E.~M.}\ \bibnamefont {Bollt}}, \bibinfo {author} {\bibfnamefont
  {R.}~\bibnamefont {Talmon}}, \bibinfo {author} {\bibfnamefont
  {K.}~\bibnamefont {Krischer}}, \ and\ \bibinfo {author} {\bibfnamefont
  {I.~G.}\ \bibnamefont {Kevrekidis}},\ }\bibfield  {title} {\enquote {\bibinfo
  {title} {An emergent space for distributed data with hidden internal order
  through manifold learning},}\ }\href@noop {} {\bibfield  {journal} {\bibinfo
  {journal} {IEEE Access}\ }\textbf {\bibinfo {volume} {6}},\ \bibinfo {pages}
  {77402--77413} (\bibinfo {year} {2018})}\BibitemShut {NoStop}%
\bibitem [{\citenamefont {Holiday}\ \emph {et~al.}(2019)\citenamefont
  {Holiday}, \citenamefont {Kooshkbaghi}, \citenamefont {Bello-Rivas},
  \citenamefont {Gear}, \citenamefont {Zagaris},\ and\ \citenamefont
  {Kevrekidis}}]{holiday2019manifold}%
  \BibitemOpen
  \bibfield  {author} {\bibinfo {author} {\bibfnamefont {A.}~\bibnamefont
  {Holiday}}, \bibinfo {author} {\bibfnamefont {M.}~\bibnamefont
  {Kooshkbaghi}}, \bibinfo {author} {\bibfnamefont {J.~M.}\ \bibnamefont
  {Bello-Rivas}}, \bibinfo {author} {\bibfnamefont {C.~W.}\ \bibnamefont
  {Gear}}, \bibinfo {author} {\bibfnamefont {A.}~\bibnamefont {Zagaris}}, \
  and\ \bibinfo {author} {\bibfnamefont {I.~G.}\ \bibnamefont {Kevrekidis}},\
  }\bibfield  {title} {\enquote {\bibinfo {title} {Manifold learning for
  parameter reduction},}\ }\href@noop {} {\bibfield  {journal} {\bibinfo
  {journal} {Journal of computational physics}\ }\textbf {\bibinfo {volume}
  {392}},\ \bibinfo {pages} {419--431} (\bibinfo {year} {2019})}\BibitemShut
  {NoStop}%
\bibitem [{\citenamefont {Thiem}\ \emph {et~al.}(2020)\citenamefont {Thiem},
  \citenamefont {Kooshkbaghi}, \citenamefont {Bertalan}, \citenamefont
  {Laing},\ and\ \citenamefont {Kevrekidis}}]{thiem2020emergent}%
  \BibitemOpen
  \bibfield  {author} {\bibinfo {author} {\bibfnamefont {T.~N.}\ \bibnamefont
  {Thiem}}, \bibinfo {author} {\bibfnamefont {M.}~\bibnamefont {Kooshkbaghi}},
  \bibinfo {author} {\bibfnamefont {T.}~\bibnamefont {Bertalan}}, \bibinfo
  {author} {\bibfnamefont {C.~R.}\ \bibnamefont {Laing}}, \ and\ \bibinfo
  {author} {\bibfnamefont {I.~G.}\ \bibnamefont {Kevrekidis}},\ }\bibfield
  {title} {\enquote {\bibinfo {title} {Emergent spaces for coupled
  oscillators},}\ }\href@noop {} {\bibfield  {journal} {\bibinfo  {journal}
  {Frontiers in Computational Neuroscience}\ }\textbf {\bibinfo {volume}
  {14}},\ \bibinfo {pages} {36} (\bibinfo {year} {2020})}\BibitemShut {NoStop}%
\bibitem [{\citenamefont {Watanabe}\ and\ \citenamefont
  {Strogatz}(1993)}]{watanabe1993integrability}%
  \BibitemOpen
  \bibfield  {author} {\bibinfo {author} {\bibfnamefont {S.}~\bibnamefont
  {Watanabe}}\ and\ \bibinfo {author} {\bibfnamefont {S.~H.}\ \bibnamefont
  {Strogatz}},\ }\bibfield  {title} {\enquote {\bibinfo {title} {Integrability
  of a globally coupled oscillator array},}\ }\href@noop {} {\bibfield
  {journal} {\bibinfo  {journal} {Physical review letters}\ }\textbf {\bibinfo
  {volume} {70}},\ \bibinfo {pages} {2391} (\bibinfo {year}
  {1993})}\BibitemShut {NoStop}%
\bibitem [{\citenamefont {O’Keeffe}, \citenamefont {Hong},\ and\
  \citenamefont {Strogatz}(2017)}]{o2017oscillators}%
  \BibitemOpen
  \bibfield  {author} {\bibinfo {author} {\bibfnamefont {K.~P.}\ \bibnamefont
  {O’Keeffe}}, \bibinfo {author} {\bibfnamefont {H.}~\bibnamefont {Hong}}, \
  and\ \bibinfo {author} {\bibfnamefont {S.~H.}\ \bibnamefont {Strogatz}},\
  }\bibfield  {title} {\enquote {\bibinfo {title} {Oscillators that sync and
  swarm},}\ }\href@noop {} {\bibfield  {journal} {\bibinfo  {journal} {Nature
  communications}\ }\textbf {\bibinfo {volume} {8}},\ \bibinfo {pages} {1--13}
  (\bibinfo {year} {2017})}\BibitemShut {NoStop}%
\bibitem [{\citenamefont {Strogatz}(2000)}]{strogatz2000kuramoto}%
  \BibitemOpen
  \bibfield  {author} {\bibinfo {author} {\bibfnamefont {S.~H.}\ \bibnamefont
  {Strogatz}},\ }\bibfield  {title} {\enquote {\bibinfo {title} {{From Kuramoto
  to Crawford: exploring the onset of synchronization in populations of coupled
  oscillators}},}\ }\href@noop {} {\bibfield  {journal} {\bibinfo  {journal}
  {Physica D: Nonlinear Phenomena}\ }\textbf {\bibinfo {volume} {143}},\
  \bibinfo {pages} {1--20} (\bibinfo {year} {2000})}\BibitemShut {NoStop}%
\bibitem [{\citenamefont {Ott}\ and\ \citenamefont
  {Antonsen}(2008)}]{ott2008low}%
  \BibitemOpen
  \bibfield  {author} {\bibinfo {author} {\bibfnamefont {E.}~\bibnamefont
  {Ott}}\ and\ \bibinfo {author} {\bibfnamefont {T.~M.}\ \bibnamefont
  {Antonsen}},\ }\bibfield  {title} {\enquote {\bibinfo {title} {Low
  dimensional behavior of large systems of globally coupled oscillators},}\
  }\href@noop {} {\bibfield  {journal} {\bibinfo  {journal} {Chaos: An
  Interdisciplinary Journal of Nonlinear Science}\ }\textbf {\bibinfo {volume}
  {18}},\ \bibinfo {pages} {037113} (\bibinfo {year} {2008})}\BibitemShut
  {NoStop}%
\bibitem [{\citenamefont {Tyulkina}\ \emph {et~al.}(2018)\citenamefont
  {Tyulkina}, \citenamefont {Goldobin}, \citenamefont {Klimenko},\ and\
  \citenamefont {Pikovsky}}]{tyulkina2018dynamics}%
  \BibitemOpen
  \bibfield  {author} {\bibinfo {author} {\bibfnamefont {I.~V.}\ \bibnamefont
  {Tyulkina}}, \bibinfo {author} {\bibfnamefont {D.~S.}\ \bibnamefont
  {Goldobin}}, \bibinfo {author} {\bibfnamefont {L.~S.}\ \bibnamefont
  {Klimenko}}, \ and\ \bibinfo {author} {\bibfnamefont {A.}~\bibnamefont
  {Pikovsky}},\ }\bibfield  {title} {\enquote {\bibinfo {title} {{Dynamics of
  noisy oscillator populations beyond the Ott-Antonsen ansatz}},}\ }\href@noop
  {} {\bibfield  {journal} {\bibinfo  {journal} {Physical review letters}\
  }\textbf {\bibinfo {volume} {120}},\ \bibinfo {pages} {264101} (\bibinfo
  {year} {2018})}\BibitemShut {NoStop}%
\bibitem [{\citenamefont {Bick}\ \emph {et~al.}(2020)\citenamefont {Bick},
  \citenamefont {Goodfellow}, \citenamefont {Laing},\ and\ \citenamefont
  {Martens}}]{bick2020understanding}%
  \BibitemOpen
  \bibfield  {author} {\bibinfo {author} {\bibfnamefont {C.}~\bibnamefont
  {Bick}}, \bibinfo {author} {\bibfnamefont {M.}~\bibnamefont {Goodfellow}},
  \bibinfo {author} {\bibfnamefont {C.~R.}\ \bibnamefont {Laing}}, \ and\
  \bibinfo {author} {\bibfnamefont {E.~A.}\ \bibnamefont {Martens}},\
  }\bibfield  {title} {\enquote {\bibinfo {title} {Understanding the dynamics
  of biological and neural oscillator networks through exact mean-field
  reductions: a review},}\ }\href@noop {} {\bibfield  {journal} {\bibinfo
  {journal} {The Journal of Mathematical Neuroscience}\ }\textbf {\bibinfo
  {volume} {10}},\ \bibinfo {pages} {1--43} (\bibinfo {year}
  {2020})}\BibitemShut {NoStop}%
\bibitem [{\citenamefont {Kumpati}, \citenamefont {Kannan}\ \emph
  {et~al.}(1990)\citenamefont {Kumpati}, \citenamefont {Kannan} \emph
  {et~al.}}]{kumpati1990identification}%
  \BibitemOpen
  \bibfield  {author} {\bibinfo {author} {\bibfnamefont {S.~N.}\ \bibnamefont
  {Kumpati}}, \bibinfo {author} {\bibfnamefont {P.}~\bibnamefont {Kannan}},
  \emph {et~al.},\ }\bibfield  {title} {\enquote {\bibinfo {title}
  {Identification and control of dynamical systems using neural networks},}\
  }\href@noop {} {\bibfield  {journal} {\bibinfo  {journal} {IEEE Transactions
  on neural networks}\ }\textbf {\bibinfo {volume} {1}},\ \bibinfo {pages}
  {4--27} (\bibinfo {year} {1990})}\BibitemShut {NoStop}%
\bibitem [{\citenamefont {Rico-Martinez}\ \emph {et~al.}(1992)\citenamefont
  {Rico-Martinez}, \citenamefont {Krischer}, \citenamefont {Kevrekidis},
  \citenamefont {Kube},\ and\ \citenamefont {Hudson}}]{rico1992discrete}%
  \BibitemOpen
  \bibfield  {author} {\bibinfo {author} {\bibfnamefont {R.}~\bibnamefont
  {Rico-Martinez}}, \bibinfo {author} {\bibfnamefont {K.}~\bibnamefont
  {Krischer}}, \bibinfo {author} {\bibfnamefont {I.}~\bibnamefont
  {Kevrekidis}}, \bibinfo {author} {\bibfnamefont {M.}~\bibnamefont {Kube}}, \
  and\ \bibinfo {author} {\bibfnamefont {J.}~\bibnamefont {Hudson}},\
  }\bibfield  {title} {\enquote {\bibinfo {title} {{Discrete-vs.
  continuous-time nonlinear signal processing of Cu electrodissolution
  data}},}\ }\href@noop {} {\bibfield  {journal} {\bibinfo  {journal} {Chemical
  Engineering Communications}\ }\textbf {\bibinfo {volume} {118}},\ \bibinfo
  {pages} {25--48} (\bibinfo {year} {1992})}\BibitemShut {NoStop}%
\bibitem [{\citenamefont {Rico-Martinez}, \citenamefont {Anderson},\ and\
  \citenamefont {Kevrekidis}(1994)}]{rico1994continuous}%
  \BibitemOpen
  \bibfield  {author} {\bibinfo {author} {\bibfnamefont {R.}~\bibnamefont
  {Rico-Martinez}}, \bibinfo {author} {\bibfnamefont {J.}~\bibnamefont
  {Anderson}}, \ and\ \bibinfo {author} {\bibfnamefont {I.}~\bibnamefont
  {Kevrekidis}},\ }\bibfield  {title} {\enquote {\bibinfo {title}
  {Continuous-time nonlinear signal processing: a neural network based approach
  for gray box identification},}\ }in\ \href@noop {} {\emph {\bibinfo
  {booktitle} {Proceedings of IEEE Workshop on Neural Networks for Signal
  Processing}}}\ (\bibinfo {organization} {IEEE},\ \bibinfo {year} {1994})\
  pp.\ \bibinfo {pages} {596--605}\BibitemShut {NoStop}%
\bibitem [{\citenamefont {Brunton}, \citenamefont {Noack},\ and\ \citenamefont
  {Koumoutsakos}(2020)}]{brunton2020machine}%
  \BibitemOpen
  \bibfield  {author} {\bibinfo {author} {\bibfnamefont {S.~L.}\ \bibnamefont
  {Brunton}}, \bibinfo {author} {\bibfnamefont {B.~R.}\ \bibnamefont {Noack}},
  \ and\ \bibinfo {author} {\bibfnamefont {P.}~\bibnamefont {Koumoutsakos}},\
  }\bibfield  {title} {\enquote {\bibinfo {title} {Machine learning for fluid
  mechanics},}\ }\href@noop {} {\bibfield  {journal} {\bibinfo  {journal}
  {Annual Review of Fluid Mechanics}\ }\textbf {\bibinfo {volume} {52}},\
  \bibinfo {pages} {477--508} (\bibinfo {year} {2020})}\BibitemShut {NoStop}%
\bibitem [{\citenamefont {Raissi}, \citenamefont {Perdikaris},\ and\
  \citenamefont {Karniadakis}(2018)}]{raissi2018multistep}%
  \BibitemOpen
  \bibfield  {author} {\bibinfo {author} {\bibfnamefont {M.}~\bibnamefont
  {Raissi}}, \bibinfo {author} {\bibfnamefont {P.}~\bibnamefont {Perdikaris}},
  \ and\ \bibinfo {author} {\bibfnamefont {G.~E.}\ \bibnamefont
  {Karniadakis}},\ }\bibfield  {title} {\enquote {\bibinfo {title} {Multistep
  neural networks for data-driven discovery of nonlinear dynamical systems},}\
  }\href@noop {} {\bibfield  {journal} {\bibinfo  {journal} {arXiv preprint
  arXiv:1801.01236}\ } (\bibinfo {year} {2018})}\BibitemShut {NoStop}%
\bibitem [{\citenamefont {Chen}\ \emph {et~al.}(2018)\citenamefont {Chen},
  \citenamefont {Rubanova}, \citenamefont {Bettencourt},\ and\ \citenamefont
  {Duvenaud}}]{chen2018neural}%
  \BibitemOpen
  \bibfield  {author} {\bibinfo {author} {\bibfnamefont {R.~T.}\ \bibnamefont
  {Chen}}, \bibinfo {author} {\bibfnamefont {Y.}~\bibnamefont {Rubanova}},
  \bibinfo {author} {\bibfnamefont {J.}~\bibnamefont {Bettencourt}}, \ and\
  \bibinfo {author} {\bibfnamefont {D.~K.}\ \bibnamefont {Duvenaud}},\
  }\bibfield  {title} {\enquote {\bibinfo {title} {Neural ordinary differential
  equations},}\ }in\ \href@noop {} {\emph {\bibinfo {booktitle} {Advances in
  neural information processing systems}}}\ (\bibinfo {year} {2018})\ pp.\
  \bibinfo {pages} {6571--6583}\BibitemShut {NoStop}%
\bibitem [{\citenamefont {Vlachas}\ \emph {et~al.}(2018)\citenamefont
  {Vlachas}, \citenamefont {Byeon}, \citenamefont {Wan}, \citenamefont
  {Sapsis},\ and\ \citenamefont {Koumoutsakos}}]{vlachas2018data}%
  \BibitemOpen
  \bibfield  {author} {\bibinfo {author} {\bibfnamefont {P.~R.}\ \bibnamefont
  {Vlachas}}, \bibinfo {author} {\bibfnamefont {W.}~\bibnamefont {Byeon}},
  \bibinfo {author} {\bibfnamefont {Z.~Y.}\ \bibnamefont {Wan}}, \bibinfo
  {author} {\bibfnamefont {T.~P.}\ \bibnamefont {Sapsis}}, \ and\ \bibinfo
  {author} {\bibfnamefont {P.}~\bibnamefont {Koumoutsakos}},\ }\bibfield
  {title} {\enquote {\bibinfo {title} {Data-driven forecasting of
  high-dimensional chaotic systems with long short-term memory networks},}\
  }\href@noop {} {\bibfield  {journal} {\bibinfo  {journal} {Proceedings of the
  Royal Society A: Mathematical, Physical and Engineering Sciences}\ }\textbf
  {\bibinfo {volume} {474}},\ \bibinfo {pages} {20170844} (\bibinfo {year}
  {2018})}\BibitemShut {NoStop}%
\bibitem [{\citenamefont {Lu}, \citenamefont {Jin},\ and\ \citenamefont
  {Karniadakis}(2019)}]{lu2019deeponet}%
  \BibitemOpen
  \bibfield  {author} {\bibinfo {author} {\bibfnamefont {L.}~\bibnamefont
  {Lu}}, \bibinfo {author} {\bibfnamefont {P.}~\bibnamefont {Jin}}, \ and\
  \bibinfo {author} {\bibfnamefont {G.~E.}\ \bibnamefont {Karniadakis}},\
  }\bibfield  {title} {\enquote {\bibinfo {title} {{DeepONet: Learning
  nonlinear operators for identifying differential equations based on the
  universal approximation theorem of operators}},}\ }\href@noop {} {\bibfield
  {journal} {\bibinfo  {journal} {arXiv preprint arXiv:1910.03193}\ } (\bibinfo
  {year} {2019})}\BibitemShut {NoStop}%
\bibitem [{\citenamefont {Nardini}\ \emph {et~al.}(2020)\citenamefont
  {Nardini}, \citenamefont {Baker}, \citenamefont {Simpson},\ and\
  \citenamefont {Flores}}]{nardini2020learning}%
  \BibitemOpen
  \bibfield  {author} {\bibinfo {author} {\bibfnamefont {J.~T.}\ \bibnamefont
  {Nardini}}, \bibinfo {author} {\bibfnamefont {R.~E.}\ \bibnamefont {Baker}},
  \bibinfo {author} {\bibfnamefont {M.~J.}\ \bibnamefont {Simpson}}, \ and\
  \bibinfo {author} {\bibfnamefont {K.~B.}\ \bibnamefont {Flores}},\ }\bibfield
   {title} {\enquote {\bibinfo {title} {Learning differential equation models
  from stochastic agent-based model simulations},}\ }\href@noop {} {\bibfield
  {journal} {\bibinfo  {journal} {arXiv preprint arXiv:2011.08255}\ } (\bibinfo
  {year} {2020})}\BibitemShut {NoStop}%
\bibitem [{\citenamefont {Long}\ \emph {et~al.}(2018)\citenamefont {Long},
  \citenamefont {Lu}, \citenamefont {Ma},\ and\ \citenamefont
  {Dong}}]{long2018pde}%
  \BibitemOpen
  \bibfield  {author} {\bibinfo {author} {\bibfnamefont {Z.}~\bibnamefont
  {Long}}, \bibinfo {author} {\bibfnamefont {Y.}~\bibnamefont {Lu}}, \bibinfo
  {author} {\bibfnamefont {X.}~\bibnamefont {Ma}}, \ and\ \bibinfo {author}
  {\bibfnamefont {B.}~\bibnamefont {Dong}},\ }\bibfield  {title} {\enquote
  {\bibinfo {title} {{PDE-net: Learning PDEs from data}},}\ }in\ \href@noop {}
  {\emph {\bibinfo {booktitle} {International Conference on Machine
  Learning}}}\ (\bibinfo {organization} {PMLR},\ \bibinfo {year} {2018})\ pp.\
  \bibinfo {pages} {3208--3216}\BibitemShut {NoStop}%
\bibitem [{\citenamefont {Raissi}(2018)}]{raissi2018forward}%
  \BibitemOpen
  \bibfield  {author} {\bibinfo {author} {\bibfnamefont {M.}~\bibnamefont
  {Raissi}},\ }\bibfield  {title} {\enquote {\bibinfo {title} {Forward-backward
  stochastic neural networks: Deep learning of high-dimensional partial
  differential equations},}\ }\href@noop {} {\bibfield  {journal} {\bibinfo
  {journal} {arXiv preprint arXiv:1804.07010}\ } (\bibinfo {year}
  {2018})}\BibitemShut {NoStop}%
\bibitem [{\citenamefont {Arbabi}\ \emph {et~al.}(2020)\citenamefont {Arbabi},
  \citenamefont {Bunder}, \citenamefont {Samaey}, \citenamefont {Roberts},\
  and\ \citenamefont {Kevrekidis}}]{arbabi2020linking}%
  \BibitemOpen
  \bibfield  {author} {\bibinfo {author} {\bibfnamefont {H.}~\bibnamefont
  {Arbabi}}, \bibinfo {author} {\bibfnamefont {J.~E.}\ \bibnamefont {Bunder}},
  \bibinfo {author} {\bibfnamefont {G.}~\bibnamefont {Samaey}}, \bibinfo
  {author} {\bibfnamefont {A.~J.}\ \bibnamefont {Roberts}}, \ and\ \bibinfo
  {author} {\bibfnamefont {I.~G.}\ \bibnamefont {Kevrekidis}},\ }\bibfield
  {title} {\enquote {\bibinfo {title} {{Linking Machine Learning with
  Multiscale Numerics: Data-Driven Discovery of Homogenized Equations}},}\
  }\href@noop {} {\bibfield  {journal} {\bibinfo  {journal} {Jom}\ }\textbf
  {\bibinfo {volume} {72}},\ \bibinfo {pages} {4444--4457} (\bibinfo {year}
  {2020})}\BibitemShut {NoStop}%
\bibitem [{\citenamefont {Arbabi}\ and\ \citenamefont
  {Kevrekidis}(2020)}]{arbabi2020particles}%
  \BibitemOpen
  \bibfield  {author} {\bibinfo {author} {\bibfnamefont {H.}~\bibnamefont
  {Arbabi}}\ and\ \bibinfo {author} {\bibfnamefont {I.}~\bibnamefont
  {Kevrekidis}},\ }\bibfield  {title} {\enquote {\bibinfo {title} {{Particles
  to Partial Differential Equations Parsimoniously}},}\ }\href@noop {}
  {\bibfield  {journal} {\bibinfo  {journal} {arXiv preprint arXiv:2011.04517}\
  } (\bibinfo {year} {2020})}\BibitemShut {NoStop}%
\bibitem [{\citenamefont {Linot}\ and\ \citenamefont
  {Graham}(2020)}]{linot2020deep}%
  \BibitemOpen
  \bibfield  {author} {\bibinfo {author} {\bibfnamefont {A.~J.}\ \bibnamefont
  {Linot}}\ and\ \bibinfo {author} {\bibfnamefont {M.~D.}\ \bibnamefont
  {Graham}},\ }\bibfield  {title} {\enquote {\bibinfo {title} {Deep learning to
  discover and predict dynamics on an inertial manifold},}\ }\href@noop {}
  {\bibfield  {journal} {\bibinfo  {journal} {Physical Review E}\ }\textbf
  {\bibinfo {volume} {101}},\ \bibinfo {pages} {062209} (\bibinfo {year}
  {2020})}\BibitemShut {NoStop}%
\bibitem [{\citenamefont {Lu}\ \emph {et~al.}(2019{\natexlab{a}})\citenamefont
  {Lu}, \citenamefont {Meng}, \citenamefont {Mao},\ and\ \citenamefont
  {Karniadakis}}]{lu2019deepxde}%
  \BibitemOpen
  \bibfield  {author} {\bibinfo {author} {\bibfnamefont {L.}~\bibnamefont
  {Lu}}, \bibinfo {author} {\bibfnamefont {X.}~\bibnamefont {Meng}}, \bibinfo
  {author} {\bibfnamefont {Z.}~\bibnamefont {Mao}}, \ and\ \bibinfo {author}
  {\bibfnamefont {G.~E.}\ \bibnamefont {Karniadakis}},\ }\bibfield  {title}
  {\enquote {\bibinfo {title} {{DeepXDE: A deep learning library for solving
  differential equations}},}\ }\href@noop {} {\bibfield  {journal} {\bibinfo
  {journal} {arXiv preprint arXiv:1907.04502}\ } (\bibinfo {year}
  {2019}{\natexlab{a}})}\BibitemShut {NoStop}%
\bibitem [{\citenamefont {Raissi}, \citenamefont {Perdikaris},\ and\
  \citenamefont {Karniadakis}(2019)}]{raissi2019physics}%
  \BibitemOpen
  \bibfield  {author} {\bibinfo {author} {\bibfnamefont {M.}~\bibnamefont
  {Raissi}}, \bibinfo {author} {\bibfnamefont {P.}~\bibnamefont {Perdikaris}},
  \ and\ \bibinfo {author} {\bibfnamefont {G.~E.}\ \bibnamefont
  {Karniadakis}},\ }\bibfield  {title} {\enquote {\bibinfo {title}
  {Physics-informed neural networks: A deep learning framework for solving
  forward and inverse problems involving nonlinear partial differential
  equations},}\ }\href@noop {} {\bibfield  {journal} {\bibinfo  {journal}
  {Journal of Computational Physics}\ }\textbf {\bibinfo {volume} {378}},\
  \bibinfo {pages} {686--707} (\bibinfo {year} {2019})}\BibitemShut {NoStop}%
\bibitem [{\citenamefont {Bhattacharya}\ \emph {et~al.}(2020)\citenamefont
  {Bhattacharya}, \citenamefont {Hosseini}, \citenamefont {Kovachki},\ and\
  \citenamefont {Stuart}}]{bhattacharya2020model}%
  \BibitemOpen
  \bibfield  {author} {\bibinfo {author} {\bibfnamefont {K.}~\bibnamefont
  {Bhattacharya}}, \bibinfo {author} {\bibfnamefont {B.}~\bibnamefont
  {Hosseini}}, \bibinfo {author} {\bibfnamefont {N.~B.}\ \bibnamefont
  {Kovachki}}, \ and\ \bibinfo {author} {\bibfnamefont {A.~M.}\ \bibnamefont
  {Stuart}},\ }\bibfield  {title} {\enquote {\bibinfo {title} {{Model reduction
  and neural networks for parametric PDEs}},}\ }\href@noop {} {\bibfield
  {journal} {\bibinfo  {journal} {arXiv preprint arXiv:2005.03180}\ } (\bibinfo
  {year} {2020})}\BibitemShut {NoStop}%
\bibitem [{\citenamefont {Lu}\ \emph {et~al.}(2019{\natexlab{b}})\citenamefont
  {Lu}, \citenamefont {Zhong}, \citenamefont {Tang},\ and\ \citenamefont
  {Maggioni}}]{lu2019nonparametric}%
  \BibitemOpen
  \bibfield  {author} {\bibinfo {author} {\bibfnamefont {F.}~\bibnamefont
  {Lu}}, \bibinfo {author} {\bibfnamefont {M.}~\bibnamefont {Zhong}}, \bibinfo
  {author} {\bibfnamefont {S.}~\bibnamefont {Tang}}, \ and\ \bibinfo {author}
  {\bibfnamefont {M.}~\bibnamefont {Maggioni}},\ }\bibfield  {title} {\enquote
  {\bibinfo {title} {Nonparametric inference of interaction laws in systems of
  agents from trajectory data},}\ }\href@noop {} {\bibfield  {journal}
  {\bibinfo  {journal} {Proceedings of the National Academy of Sciences}\
  }\textbf {\bibinfo {volume} {116}},\ \bibinfo {pages} {14424--14433}
  (\bibinfo {year} {2019}{\natexlab{b}})}\BibitemShut {NoStop}%
\bibitem [{\citenamefont {Lu}, \citenamefont {Maggioni},\ and\ \citenamefont
  {Tang}(2021)}]{lu2021learning}%
  \BibitemOpen
  \bibfield  {author} {\bibinfo {author} {\bibfnamefont {F.}~\bibnamefont
  {Lu}}, \bibinfo {author} {\bibfnamefont {M.}~\bibnamefont {Maggioni}}, \ and\
  \bibinfo {author} {\bibfnamefont {S.}~\bibnamefont {Tang}},\ }\bibfield
  {title} {\enquote {\bibinfo {title} {Learning interaction kernels in
  heterogeneous systems of agents from multiple trajectories},}\ }\href@noop {}
  {\bibfield  {journal} {\bibinfo  {journal} {Journal of Machine Learning
  Research}\ }\textbf {\bibinfo {volume} {22}},\ \bibinfo {pages} {1--67}
  (\bibinfo {year} {2021})}\BibitemShut {NoStop}%
\bibitem [{\citenamefont {He}\ \emph {et~al.}(2016)\citenamefont {He},
  \citenamefont {Zhang}, \citenamefont {Ren},\ and\ \citenamefont
  {Sun}}]{he2016deep}%
  \BibitemOpen
  \bibfield  {author} {\bibinfo {author} {\bibfnamefont {K.}~\bibnamefont
  {He}}, \bibinfo {author} {\bibfnamefont {X.}~\bibnamefont {Zhang}}, \bibinfo
  {author} {\bibfnamefont {S.}~\bibnamefont {Ren}}, \ and\ \bibinfo {author}
  {\bibfnamefont {J.}~\bibnamefont {Sun}},\ }\bibfield  {title} {\enquote
  {\bibinfo {title} {Deep residual learning for image recognition},}\ }in\
  \href@noop {} {\emph {\bibinfo {booktitle} {Proceedings of the IEEE
  conference on computer vision and pattern recognition}}}\ (\bibinfo {year}
  {2016})\ pp.\ \bibinfo {pages} {770--778}\BibitemShut {NoStop}%
\bibitem [{\citenamefont {Choi}\ \emph {et~al.}(2016)\citenamefont {Choi},
  \citenamefont {Bertalan}, \citenamefont {Laing},\ and\ \citenamefont
  {Kevrekidis}}]{choi2016dimension}%
  \BibitemOpen
  \bibfield  {author} {\bibinfo {author} {\bibfnamefont {M.}~\bibnamefont
  {Choi}}, \bibinfo {author} {\bibfnamefont {T.}~\bibnamefont {Bertalan}},
  \bibinfo {author} {\bibfnamefont {C.~R.}\ \bibnamefont {Laing}}, \ and\
  \bibinfo {author} {\bibfnamefont {I.~G.}\ \bibnamefont {Kevrekidis}},\
  }\bibfield  {title} {\enquote {\bibinfo {title} {{Dimension reduction in
  heterogeneous neural networks: generalized Polynomial Chaos (gPC) and
  ANalysis-Of-VAriance (ANOVA)}},}\ }\href@noop {} {\bibfield  {journal}
  {\bibinfo  {journal} {The European Physical Journal Special Topics}\ }\textbf
  {\bibinfo {volume} {225}},\ \bibinfo {pages} {1165--1180} (\bibinfo {year}
  {2016})}\BibitemShut {NoStop}%
\bibitem [{\citenamefont {Bertalan}\ \emph {et~al.}(2017)\citenamefont
  {Bertalan}, \citenamefont {Wu}, \citenamefont {Laing}, \citenamefont {Gear},\
  and\ \citenamefont {Kevrekidis}}]{bertalan2017coarse}%
  \BibitemOpen
  \bibfield  {author} {\bibinfo {author} {\bibfnamefont {T.}~\bibnamefont
  {Bertalan}}, \bibinfo {author} {\bibfnamefont {Y.}~\bibnamefont {Wu}},
  \bibinfo {author} {\bibfnamefont {C.}~\bibnamefont {Laing}}, \bibinfo
  {author} {\bibfnamefont {C.~W.}\ \bibnamefont {Gear}}, \ and\ \bibinfo
  {author} {\bibfnamefont {I.~G.}\ \bibnamefont {Kevrekidis}},\ }\bibfield
  {title} {\enquote {\bibinfo {title} {Coarse-grained descriptions of dynamics
  for networks with both intrinsic and structural heterogeneities},}\
  }\href@noop {} {\bibfield  {journal} {\bibinfo  {journal} {Frontiers in
  computational neuroscience}\ }\textbf {\bibinfo {volume} {11}},\ \bibinfo
  {pages} {43} (\bibinfo {year} {2017})}\BibitemShut {NoStop}%
\bibitem [{\citenamefont {Rajendran}\ \emph {et~al.}(2016)\citenamefont
  {Rajendran}, \citenamefont {Tsoumanis}, \citenamefont {Siettos},
  \citenamefont {Laing},\ and\ \citenamefont
  {Kevrekidis}}]{rajendran2016modeling}%
  \BibitemOpen
  \bibfield  {author} {\bibinfo {author} {\bibfnamefont {K.}~\bibnamefont
  {Rajendran}}, \bibinfo {author} {\bibfnamefont {A.~C.}\ \bibnamefont
  {Tsoumanis}}, \bibinfo {author} {\bibfnamefont {C.~I.}\ \bibnamefont
  {Siettos}}, \bibinfo {author} {\bibfnamefont {C.~R.}\ \bibnamefont {Laing}},
  \ and\ \bibinfo {author} {\bibfnamefont {I.~G.}\ \bibnamefont {Kevrekidis}},\
  }\bibfield  {title} {\enquote {\bibinfo {title} {{Modeling Heterogeneity in
  Networks using Polynomial Chaos}},}\ }\href@noop {} {\bibfield  {journal}
  {\bibinfo  {journal} {International Journal for Multiscale Computational
  Engineering}\ }\textbf {\bibinfo {volume} {14}} (\bibinfo {year}
  {2016})}\BibitemShut {NoStop}%
\bibitem [{\citenamefont {Smith}, \citenamefont {Chasnov},\ and\ \citenamefont
  {Waleffe}(1996)}]{smith1996crossover}%
  \BibitemOpen
  \bibfield  {author} {\bibinfo {author} {\bibfnamefont {L.~M.}\ \bibnamefont
  {Smith}}, \bibinfo {author} {\bibfnamefont {J.~R.}\ \bibnamefont {Chasnov}},
  \ and\ \bibinfo {author} {\bibfnamefont {F.}~\bibnamefont {Waleffe}},\
  }\bibfield  {title} {\enquote {\bibinfo {title} {Crossover from two-to
  three-dimensional turbulence},}\ }\href@noop {} {\bibfield  {journal}
  {\bibinfo  {journal} {Physical review letters}\ }\textbf {\bibinfo {volume}
  {77}},\ \bibinfo {pages} {2467} (\bibinfo {year} {1996})}\BibitemShut
  {NoStop}%
\bibitem [{\citenamefont {Lamorgese}, \citenamefont {Caughey},\ and\
  \citenamefont {Pope}(2005)}]{lamorgese2005direct}%
  \BibitemOpen
  \bibfield  {author} {\bibinfo {author} {\bibfnamefont {A.}~\bibnamefont
  {Lamorgese}}, \bibinfo {author} {\bibfnamefont {D.}~\bibnamefont {Caughey}},
  \ and\ \bibinfo {author} {\bibfnamefont {S.}~\bibnamefont {Pope}},\
  }\bibfield  {title} {\enquote {\bibinfo {title} {Direct numerical simulation
  of homogeneous turbulence with hyperviscosity},}\ }\href@noop {} {\bibfield
  {journal} {\bibinfo  {journal} {Physics of Fluids}\ }\textbf {\bibinfo
  {volume} {17}},\ \bibinfo {pages} {015106} (\bibinfo {year}
  {2005})}\BibitemShut {NoStop}%
\bibitem [{\citenamefont {Cook}\ and\ \citenamefont
  {Cabot}(2005)}]{cook2005hyperviscosity}%
  \BibitemOpen
  \bibfield  {author} {\bibinfo {author} {\bibfnamefont {A.~W.}\ \bibnamefont
  {Cook}}\ and\ \bibinfo {author} {\bibfnamefont {W.~H.}\ \bibnamefont
  {Cabot}},\ }\bibfield  {title} {\enquote {\bibinfo {title} {Hyperviscosity
  for shock-turbulence interactions},}\ }\href@noop {} {\bibfield  {journal}
  {\bibinfo  {journal} {Journal of Computational Physics}\ }\textbf {\bibinfo
  {volume} {203}},\ \bibinfo {pages} {379--385} (\bibinfo {year}
  {2005})}\BibitemShut {NoStop}%
\bibitem [{\citenamefont {Frisch}\ \emph {et~al.}(2008)\citenamefont {Frisch},
  \citenamefont {Kurien}, \citenamefont {Pandit}, \citenamefont {Pauls},
  \citenamefont {Ray}, \citenamefont {Wirth},\ and\ \citenamefont
  {Zhu}}]{frisch2008hyperviscosity}%
  \BibitemOpen
  \bibfield  {author} {\bibinfo {author} {\bibfnamefont {U.}~\bibnamefont
  {Frisch}}, \bibinfo {author} {\bibfnamefont {S.}~\bibnamefont {Kurien}},
  \bibinfo {author} {\bibfnamefont {R.}~\bibnamefont {Pandit}}, \bibinfo
  {author} {\bibfnamefont {W.}~\bibnamefont {Pauls}}, \bibinfo {author}
  {\bibfnamefont {S.~S.}\ \bibnamefont {Ray}}, \bibinfo {author} {\bibfnamefont
  {A.}~\bibnamefont {Wirth}}, \ and\ \bibinfo {author} {\bibfnamefont {J.-Z.}\
  \bibnamefont {Zhu}},\ }\bibfield  {title} {\enquote {\bibinfo {title}
  {{Hyperviscosity, Galerkin truncation, and bottlenecks in turbulence}},}\
  }\href@noop {} {\bibfield  {journal} {\bibinfo  {journal} {Physical review
  letters}\ }\textbf {\bibinfo {volume} {101}},\ \bibinfo {pages} {144501}
  (\bibinfo {year} {2008})}\BibitemShut {NoStop}%
\bibitem [{\citenamefont {Butera~Jr}, \citenamefont {Rinzel},\ and\
  \citenamefont {Smith}(1999)}]{butera1999models}%
  \BibitemOpen
  \bibfield  {author} {\bibinfo {author} {\bibfnamefont {R.~J.}\ \bibnamefont
  {Butera~Jr}}, \bibinfo {author} {\bibfnamefont {J.}~\bibnamefont {Rinzel}}, \
  and\ \bibinfo {author} {\bibfnamefont {J.~C.}\ \bibnamefont {Smith}},\
  }\bibfield  {title} {\enquote {\bibinfo {title} {{Models of Respiratory
  Rhythm Generation in the Pre-B\"{o}tzinger Complex. I. Bursting Pacemaker
  Neurons}},}\ }\href@noop {} {\bibfield  {journal} {\bibinfo  {journal}
  {Journal of neurophysiology}\ }\textbf {\bibinfo {volume} {82}},\ \bibinfo
  {pages} {382--397} (\bibinfo {year} {1999})}\BibitemShut {NoStop}%
\bibitem [{\citenamefont {Laing}\ \emph {et~al.}(2012)\citenamefont {Laing},
  \citenamefont {Zou}, \citenamefont {Smith},\ and\ \citenamefont
  {Kevrekidis}}]{laing2012managing}%
  \BibitemOpen
  \bibfield  {author} {\bibinfo {author} {\bibfnamefont {C.~R.}\ \bibnamefont
  {Laing}}, \bibinfo {author} {\bibfnamefont {Y.}~\bibnamefont {Zou}}, \bibinfo
  {author} {\bibfnamefont {B.}~\bibnamefont {Smith}}, \ and\ \bibinfo {author}
  {\bibfnamefont {I.~G.}\ \bibnamefont {Kevrekidis}},\ }\bibfield  {title}
  {\enquote {\bibinfo {title} {Managing heterogeneity in the study of neural
  oscillator dynamics},}\ }\href@noop {} {\bibfield  {journal} {\bibinfo
  {journal} {The Journal of Mathematical Neuroscience}\ }\textbf {\bibinfo
  {volume} {2}},\ \bibinfo {pages} {5} (\bibinfo {year} {2012})}\BibitemShut
  {NoStop}%
\bibitem [{\citenamefont {Rubin}\ and\ \citenamefont
  {Terman}(2002)}]{rubin2002synchronized}%
  \BibitemOpen
  \bibfield  {author} {\bibinfo {author} {\bibfnamefont {J.}~\bibnamefont
  {Rubin}}\ and\ \bibinfo {author} {\bibfnamefont {D.}~\bibnamefont {Terman}},\
  }\bibfield  {title} {\enquote {\bibinfo {title} {Synchronized activity and
  loss of synchrony among heterogeneous conditional oscillators},}\ }\href@noop
  {} {\bibfield  {journal} {\bibinfo  {journal} {SIAM Journal on Applied
  Dynamical Systems}\ }\textbf {\bibinfo {volume} {1}},\ \bibinfo {pages}
  {146--174} (\bibinfo {year} {2002})}\BibitemShut {NoStop}%
\bibitem [{\citenamefont {Li}\ \emph {et~al.}(2014)\citenamefont {Li},
  \citenamefont {Cai}, \citenamefont {Wang}, \citenamefont {Zhou},
  \citenamefont {Feng},\ and\ \citenamefont {Chen}}]{li2014medical}%
  \BibitemOpen
  \bibfield  {author} {\bibinfo {author} {\bibfnamefont {Q.}~\bibnamefont
  {Li}}, \bibinfo {author} {\bibfnamefont {W.}~\bibnamefont {Cai}}, \bibinfo
  {author} {\bibfnamefont {X.}~\bibnamefont {Wang}}, \bibinfo {author}
  {\bibfnamefont {Y.}~\bibnamefont {Zhou}}, \bibinfo {author} {\bibfnamefont
  {D.~D.}\ \bibnamefont {Feng}}, \ and\ \bibinfo {author} {\bibfnamefont
  {M.}~\bibnamefont {Chen}},\ }\bibfield  {title} {\enquote {\bibinfo {title}
  {Medical image classification with convolutional neural network},}\ }in\
  \href@noop {} {\emph {\bibinfo {booktitle} {2014 13th International
  Conference on Control Automation Robotics \& Vision (ICARCV)}}}\ (\bibinfo
  {organization} {IEEE},\ \bibinfo {year} {2014})\ pp.\ \bibinfo {pages}
  {844--848}\BibitemShut {NoStop}%
\bibitem [{\citenamefont {Liang}\ \emph {et~al.}(2018)\citenamefont {Liang},
  \citenamefont {Hong}, \citenamefont {Xie},\ and\ \citenamefont
  {Zheng}}]{liang2018combining}%
  \BibitemOpen
  \bibfield  {author} {\bibinfo {author} {\bibfnamefont {G.}~\bibnamefont
  {Liang}}, \bibinfo {author} {\bibfnamefont {H.}~\bibnamefont {Hong}},
  \bibinfo {author} {\bibfnamefont {W.}~\bibnamefont {Xie}}, \ and\ \bibinfo
  {author} {\bibfnamefont {L.}~\bibnamefont {Zheng}},\ }\bibfield  {title}
  {\enquote {\bibinfo {title} {Combining convolutional neural network with
  recursive neural network for blood cell image classification},}\ }\href@noop
  {} {\bibfield  {journal} {\bibinfo  {journal} {IEEE Access}\ }\textbf
  {\bibinfo {volume} {6}},\ \bibinfo {pages} {36188--36197} (\bibinfo {year}
  {2018})}\BibitemShut {NoStop}%
\bibitem [{\citenamefont {Hou}\ \emph {et~al.}(2016)\citenamefont {Hou},
  \citenamefont {Samaras}, \citenamefont {Kurc}, \citenamefont {Gao},
  \citenamefont {Davis},\ and\ \citenamefont {Saltz}}]{hou2016patch}%
  \BibitemOpen
  \bibfield  {author} {\bibinfo {author} {\bibfnamefont {L.}~\bibnamefont
  {Hou}}, \bibinfo {author} {\bibfnamefont {D.}~\bibnamefont {Samaras}},
  \bibinfo {author} {\bibfnamefont {T.~M.}\ \bibnamefont {Kurc}}, \bibinfo
  {author} {\bibfnamefont {Y.}~\bibnamefont {Gao}}, \bibinfo {author}
  {\bibfnamefont {J.~E.}\ \bibnamefont {Davis}}, \ and\ \bibinfo {author}
  {\bibfnamefont {J.~H.}\ \bibnamefont {Saltz}},\ }\bibfield  {title} {\enquote
  {\bibinfo {title} {Patch-based convolutional neural network for whole slide
  tissue image classification},}\ }in\ \href@noop {} {\emph {\bibinfo
  {booktitle} {Proceedings of the IEEE conference on computer vision and
  pattern recognition}}}\ (\bibinfo {year} {2016})\ pp.\ \bibinfo {pages}
  {2424--2433}\BibitemShut {NoStop}%
\bibitem [{\citenamefont {Mou}, \citenamefont {Ghamisi},\ and\ \citenamefont
  {Zhu}(2017)}]{mou2017deep}%
  \BibitemOpen
  \bibfield  {author} {\bibinfo {author} {\bibfnamefont {L.}~\bibnamefont
  {Mou}}, \bibinfo {author} {\bibfnamefont {P.}~\bibnamefont {Ghamisi}}, \ and\
  \bibinfo {author} {\bibfnamefont {X.~X.}\ \bibnamefont {Zhu}},\ }\bibfield
  {title} {\enquote {\bibinfo {title} {Deep recurrent neural networks for
  hyperspectral image classification},}\ }\href@noop {} {\bibfield  {journal}
  {\bibinfo  {journal} {IEEE Transactions on Geoscience and Remote Sensing}\
  }\textbf {\bibinfo {volume} {55}},\ \bibinfo {pages} {3639--3655} (\bibinfo
  {year} {2017})}\BibitemShut {NoStop}%
\bibitem [{\citenamefont {Ciregan}, \citenamefont {Meier},\ and\ \citenamefont
  {Schmidhuber}(2012)}]{ciregan2012multi}%
  \BibitemOpen
  \bibfield  {author} {\bibinfo {author} {\bibfnamefont {D.}~\bibnamefont
  {Ciregan}}, \bibinfo {author} {\bibfnamefont {U.}~\bibnamefont {Meier}}, \
  and\ \bibinfo {author} {\bibfnamefont {J.}~\bibnamefont {Schmidhuber}},\
  }\bibfield  {title} {\enquote {\bibinfo {title} {Multi-column deep neural
  networks for image classification},}\ }in\ \href@noop {} {\emph {\bibinfo
  {booktitle} {2012 IEEE conference on computer vision and pattern
  recognition}}}\ (\bibinfo {organization} {IEEE},\ \bibinfo {year} {2012})\
  pp.\ \bibinfo {pages} {3642--3649}\BibitemShut {NoStop}%
\bibitem [{\citenamefont {Specht}\ \emph {et~al.}(1991)\citenamefont {Specht}
  \emph {et~al.}}]{specht1991general}%
  \BibitemOpen
  \bibfield  {author} {\bibinfo {author} {\bibfnamefont {D.~F.}\ \bibnamefont
  {Specht}} \emph {et~al.},\ }\bibfield  {title} {\enquote {\bibinfo {title} {A
  general regression neural network},}\ }\href@noop {} {\bibfield  {journal}
  {\bibinfo  {journal} {IEEE transactions on neural networks}\ }\textbf
  {\bibinfo {volume} {2}},\ \bibinfo {pages} {568--576} (\bibinfo {year}
  {1991})}\BibitemShut {NoStop}%
\bibitem [{\citenamefont {Kolehmainen}, \citenamefont {Martikainen},\ and\
  \citenamefont {Ruuskanen}(2001)}]{kolehmainen2001neural}%
  \BibitemOpen
  \bibfield  {author} {\bibinfo {author} {\bibfnamefont {M.}~\bibnamefont
  {Kolehmainen}}, \bibinfo {author} {\bibfnamefont {H.}~\bibnamefont
  {Martikainen}}, \ and\ \bibinfo {author} {\bibfnamefont {J.}~\bibnamefont
  {Ruuskanen}},\ }\bibfield  {title} {\enquote {\bibinfo {title} {Neural
  networks and periodic components used in air quality forecasting},}\
  }\href@noop {} {\bibfield  {journal} {\bibinfo  {journal} {Atmospheric
  Environment}\ }\textbf {\bibinfo {volume} {35}},\ \bibinfo {pages} {815--825}
  (\bibinfo {year} {2001})}\BibitemShut {NoStop}%
\bibitem [{\citenamefont {Kalchbrenner}, \citenamefont {Grefenstette},\ and\
  \citenamefont {Blunsom}(2014)}]{kalchbrenner2014convolutional}%
  \BibitemOpen
  \bibfield  {author} {\bibinfo {author} {\bibfnamefont {N.}~\bibnamefont
  {Kalchbrenner}}, \bibinfo {author} {\bibfnamefont {E.}~\bibnamefont
  {Grefenstette}}, \ and\ \bibinfo {author} {\bibfnamefont {P.}~\bibnamefont
  {Blunsom}},\ }\bibfield  {title} {\enquote {\bibinfo {title} {A convolutional
  neural network for modelling sentences},}\ }\href@noop {} {\bibfield
  {journal} {\bibinfo  {journal} {arXiv preprint arXiv:1404.2188}\ } (\bibinfo
  {year} {2014})}\BibitemShut {NoStop}%
\bibitem [{\citenamefont {Kudugunta}\ and\ \citenamefont
  {Ferrara}(2018)}]{kudugunta2018deep}%
  \BibitemOpen
  \bibfield  {author} {\bibinfo {author} {\bibfnamefont {S.}~\bibnamefont
  {Kudugunta}}\ and\ \bibinfo {author} {\bibfnamefont {E.}~\bibnamefont
  {Ferrara}},\ }\bibfield  {title} {\enquote {\bibinfo {title} {Deep neural
  networks for bot detection},}\ }\href@noop {} {\bibfield  {journal} {\bibinfo
   {journal} {Information Sciences}\ }\textbf {\bibinfo {volume} {467}},\
  \bibinfo {pages} {312--322} (\bibinfo {year} {2018})}\BibitemShut {NoStop}%
\bibitem [{\citenamefont {Mohammad}\ \emph {et~al.}(2019)\citenamefont
  {Mohammad}, \citenamefont {Khan}, \citenamefont {Ali}, \citenamefont {Liu},
  \citenamefont {Shardlow},\ and\ \citenamefont {Nawaz}}]{mohammad2019bot}%
  \BibitemOpen
  \bibfield  {author} {\bibinfo {author} {\bibfnamefont {S.}~\bibnamefont
  {Mohammad}}, \bibinfo {author} {\bibfnamefont {M.~U.}\ \bibnamefont {Khan}},
  \bibinfo {author} {\bibfnamefont {M.}~\bibnamefont {Ali}}, \bibinfo {author}
  {\bibfnamefont {L.}~\bibnamefont {Liu}}, \bibinfo {author} {\bibfnamefont
  {M.}~\bibnamefont {Shardlow}}, \ and\ \bibinfo {author} {\bibfnamefont
  {R.}~\bibnamefont {Nawaz}},\ }\bibfield  {title} {\enquote {\bibinfo {title}
  {Bot detection using a single post on social media},}\ }in\ \href@noop {}
  {\emph {\bibinfo {booktitle} {2019 Third World Conference on Smart Trends in
  Systems Security and Sustainablity (WorldS4)}}}\ (\bibinfo {organization}
  {IEEE},\ \bibinfo {year} {2019})\ pp.\ \bibinfo {pages}
  {215--220}\BibitemShut {NoStop}%
\bibitem [{\citenamefont {Alemany}\ \emph {et~al.}(2019)\citenamefont
  {Alemany}, \citenamefont {Beltran}, \citenamefont {Perez},\ and\
  \citenamefont {Ganzfried}}]{alemany2019predicting}%
  \BibitemOpen
  \bibfield  {author} {\bibinfo {author} {\bibfnamefont {S.}~\bibnamefont
  {Alemany}}, \bibinfo {author} {\bibfnamefont {J.}~\bibnamefont {Beltran}},
  \bibinfo {author} {\bibfnamefont {A.}~\bibnamefont {Perez}}, \ and\ \bibinfo
  {author} {\bibfnamefont {S.}~\bibnamefont {Ganzfried}},\ }\bibfield  {title}
  {\enquote {\bibinfo {title} {Predicting hurricane trajectories using a
  recurrent neural network},}\ }in\ \href@noop {} {\emph {\bibinfo {booktitle}
  {Proceedings of the AAAI Conference on Artificial Intelligence}}},\
  Vol.~\bibinfo {volume} {33}\ (\bibinfo {year} {2019})\ pp.\ \bibinfo {pages}
  {468--475}\BibitemShut {NoStop}%
\bibitem [{\citenamefont {Ghosh}\ and\ \citenamefont
  {Krishnamurti}(2018)}]{ghosh2018improvements}%
  \BibitemOpen
  \bibfield  {author} {\bibinfo {author} {\bibfnamefont {T.}~\bibnamefont
  {Ghosh}}\ and\ \bibinfo {author} {\bibfnamefont {T.}~\bibnamefont
  {Krishnamurti}},\ }\bibfield  {title} {\enquote {\bibinfo {title}
  {Improvements in hurricane intensity forecasts from a multimodel
  superensemble utilizing a generalized neural network technique},}\
  }\href@noop {} {\bibfield  {journal} {\bibinfo  {journal} {Weather and
  Forecasting}\ }\textbf {\bibinfo {volume} {33}},\ \bibinfo {pages} {873--885}
  (\bibinfo {year} {2018})}\BibitemShut {NoStop}%
\bibitem [{\citenamefont {Cardaliaguet}\ and\ \citenamefont
  {Euvrard}(1992)}]{cardaliaguet1992approximation}%
  \BibitemOpen
  \bibfield  {author} {\bibinfo {author} {\bibfnamefont {P.}~\bibnamefont
  {Cardaliaguet}}\ and\ \bibinfo {author} {\bibfnamefont {G.}~\bibnamefont
  {Euvrard}},\ }\bibfield  {title} {\enquote {\bibinfo {title} {Approximation
  of a function and its derivative with a neural network},}\ }\href@noop {}
  {\bibfield  {journal} {\bibinfo  {journal} {Neural Networks}\ }\textbf
  {\bibinfo {volume} {5}},\ \bibinfo {pages} {207--220} (\bibinfo {year}
  {1992})}\BibitemShut {NoStop}%
\bibitem [{\citenamefont {Funahashi}\ and\ \citenamefont
  {Nakamura}(1993)}]{funahashi1993approximation}%
  \BibitemOpen
  \bibfield  {author} {\bibinfo {author} {\bibfnamefont {K.-i.}\ \bibnamefont
  {Funahashi}}\ and\ \bibinfo {author} {\bibfnamefont {Y.}~\bibnamefont
  {Nakamura}},\ }\bibfield  {title} {\enquote {\bibinfo {title} {Approximation
  of dynamical systems by continuous time recurrent neural networks},}\
  }\href@noop {} {\bibfield  {journal} {\bibinfo  {journal} {Neural networks}\
  }\textbf {\bibinfo {volume} {6}},\ \bibinfo {pages} {801--806} (\bibinfo
  {year} {1993})}\BibitemShut {NoStop}%
\bibitem [{\citenamefont {Wang}\ and\ \citenamefont
  {Lin}(1998)}]{wang1998runge}%
  \BibitemOpen
  \bibfield  {author} {\bibinfo {author} {\bibfnamefont {Y.-J.}\ \bibnamefont
  {Wang}}\ and\ \bibinfo {author} {\bibfnamefont {C.-T.}\ \bibnamefont {Lin}},\
  }\bibfield  {title} {\enquote {\bibinfo {title} {{Runge-Kutta neural network
  for identification of dynamical systems in high accuracy}},}\ }\href@noop {}
  {\bibfield  {journal} {\bibinfo  {journal} {IEEE Transactions on Neural
  Networks}\ }\textbf {\bibinfo {volume} {9}},\ \bibinfo {pages} {294--307}
  (\bibinfo {year} {1998})}\BibitemShut {NoStop}%
\bibitem [{\citenamefont {Karpatne}\ \emph {et~al.}(2017)\citenamefont
  {Karpatne}, \citenamefont {Watkins}, \citenamefont {Read},\ and\
  \citenamefont {Kumar}}]{karpatne2017physics}%
  \BibitemOpen
  \bibfield  {author} {\bibinfo {author} {\bibfnamefont {A.}~\bibnamefont
  {Karpatne}}, \bibinfo {author} {\bibfnamefont {W.}~\bibnamefont {Watkins}},
  \bibinfo {author} {\bibfnamefont {J.}~\bibnamefont {Read}}, \ and\ \bibinfo
  {author} {\bibfnamefont {V.}~\bibnamefont {Kumar}},\ }\bibfield  {title}
  {\enquote {\bibinfo {title} {Physics-guided neural networks (pgnn): An
  application in lake temperature modeling},}\ }\href@noop {} {\bibfield
  {journal} {\bibinfo  {journal} {arXiv preprint arXiv:1710.11431}\ } (\bibinfo
  {year} {2017})}\BibitemShut {NoStop}%
\bibitem [{\citenamefont {Pascanu}, \citenamefont {Mikolov},\ and\
  \citenamefont {Bengio}(2013)}]{pascanu2013difficulty}%
  \BibitemOpen
  \bibfield  {author} {\bibinfo {author} {\bibfnamefont {R.}~\bibnamefont
  {Pascanu}}, \bibinfo {author} {\bibfnamefont {T.}~\bibnamefont {Mikolov}}, \
  and\ \bibinfo {author} {\bibfnamefont {Y.}~\bibnamefont {Bengio}},\
  }\bibfield  {title} {\enquote {\bibinfo {title} {On the difficulty of
  training recurrent neural networks},}\ }in\ \href@noop {} {\emph {\bibinfo
  {booktitle} {International conference on machine learning}}}\ (\bibinfo
  {year} {2013})\ pp.\ \bibinfo {pages} {1310--1318}\BibitemShut {NoStop}%
\bibitem [{\citenamefont {Krischer}\ \emph {et~al.}(1993)\citenamefont
  {Krischer}, \citenamefont {Rico-Mart{\'\i}nez}, \citenamefont {Kevrekidis},
  \citenamefont {Rotermund}, \citenamefont {Ertl},\ and\ \citenamefont
  {Hudson}}]{krischer1993model}%
  \BibitemOpen
  \bibfield  {author} {\bibinfo {author} {\bibfnamefont {K.}~\bibnamefont
  {Krischer}}, \bibinfo {author} {\bibfnamefont {R.}~\bibnamefont
  {Rico-Mart{\'\i}nez}}, \bibinfo {author} {\bibfnamefont {I.}~\bibnamefont
  {Kevrekidis}}, \bibinfo {author} {\bibfnamefont {H.}~\bibnamefont
  {Rotermund}}, \bibinfo {author} {\bibfnamefont {G.}~\bibnamefont {Ertl}}, \
  and\ \bibinfo {author} {\bibfnamefont {J.}~\bibnamefont {Hudson}},\
  }\bibfield  {title} {\enquote {\bibinfo {title} {Model identification of a
  spatiotemporally varying catalytic reaction},}\ }\href@noop {} {\bibfield
  {journal} {\bibinfo  {journal} {AIChE Journal}\ }\textbf {\bibinfo {volume}
  {39}},\ \bibinfo {pages} {89--98} (\bibinfo {year} {1993})}\BibitemShut
  {NoStop}%
\bibitem [{\citenamefont {Rico-Martinez}\ and\ \citenamefont
  {Kevrekidis}(1993)}]{rico1993continuous}%
  \BibitemOpen
  \bibfield  {author} {\bibinfo {author} {\bibfnamefont {R.}~\bibnamefont
  {Rico-Martinez}}\ and\ \bibinfo {author} {\bibfnamefont {I.~G.}\ \bibnamefont
  {Kevrekidis}},\ }\bibfield  {title} {\enquote {\bibinfo {title} {Continuous
  time modeling of nonlinear systems: A neural network-based approach},}\ }in\
  \href@noop {} {\emph {\bibinfo {booktitle} {IEEE International Conference on
  Neural Networks}}}\ (\bibinfo {organization} {IEEE},\ \bibinfo {year}
  {1993})\ pp.\ \bibinfo {pages} {1522--1525}\BibitemShut {NoStop}%
\bibitem [{\citenamefont {Gonz{\'a}lez-Garc{\'\i}a}, \citenamefont
  {Rico-Mart{\'\i}nez},\ and\ \citenamefont
  {Kevrekidis}(1998)}]{gonzalez1998identification}%
  \BibitemOpen
  \bibfield  {author} {\bibinfo {author} {\bibfnamefont {R.}~\bibnamefont
  {Gonz{\'a}lez-Garc{\'\i}a}}, \bibinfo {author} {\bibfnamefont
  {R.}~\bibnamefont {Rico-Mart{\'\i}nez}}, \ and\ \bibinfo {author}
  {\bibfnamefont {I.~G.}\ \bibnamefont {Kevrekidis}},\ }\bibfield  {title}
  {\enquote {\bibinfo {title} {Identification of distributed parameter systems:
  A neural net based approach},}\ }\href@noop {} {\bibfield  {journal}
  {\bibinfo  {journal} {Computers \& chemical engineering}\ }\textbf {\bibinfo
  {volume} {22}},\ \bibinfo {pages} {S965--S968} (\bibinfo {year}
  {1998})}\BibitemShut {NoStop}%
\bibitem [{\citenamefont {Lee}\ \emph {et~al.}(2020)\citenamefont {Lee},
  \citenamefont {Kooshkbaghi}, \citenamefont {Spiliotis}, \citenamefont
  {Siettos},\ and\ \citenamefont {Kevrekidis}}]{lee2020coarse}%
  \BibitemOpen
  \bibfield  {author} {\bibinfo {author} {\bibfnamefont {S.}~\bibnamefont
  {Lee}}, \bibinfo {author} {\bibfnamefont {M.}~\bibnamefont {Kooshkbaghi}},
  \bibinfo {author} {\bibfnamefont {K.}~\bibnamefont {Spiliotis}}, \bibinfo
  {author} {\bibfnamefont {C.~I.}\ \bibnamefont {Siettos}}, \ and\ \bibinfo
  {author} {\bibfnamefont {I.~G.}\ \bibnamefont {Kevrekidis}},\ }\bibfield
  {title} {\enquote {\bibinfo {title} {{Coarse-scale PDEs from fine-scale
  observations via machine learning}},}\ }\href@noop {} {\bibfield  {journal}
  {\bibinfo  {journal} {Chaos: An Interdisciplinary Journal of Nonlinear
  Science}\ }\textbf {\bibinfo {volume} {30}},\ \bibinfo {pages} {013141}
  (\bibinfo {year} {2020})}\BibitemShut {NoStop}%
\bibitem [{\citenamefont {Zhang}\ \emph {et~al.}(1988)\citenamefont {Zhang}
  \emph {et~al.}}]{zhang1988shift}%
  \BibitemOpen
  \bibfield  {author} {\bibinfo {author} {\bibfnamefont {W.}~\bibnamefont
  {Zhang}} \emph {et~al.},\ }\bibfield  {title} {\enquote {\bibinfo {title}
  {Shift-invariant pattern recognition neural network and its optical
  architecture},}\ }in\ \href@noop {} {\emph {\bibinfo {booktitle} {Proceedings
  of annual conference of the Japan Society of Applied Physics}}}\ (\bibinfo
  {year} {1988})\BibitemShut {NoStop}%
\bibitem [{\citenamefont {LeCun}, \citenamefont {Bengio}\ \emph
  {et~al.}(1995)\citenamefont {LeCun}, \citenamefont {Bengio} \emph
  {et~al.}}]{lecun1995convolutional}%
  \BibitemOpen
  \bibfield  {author} {\bibinfo {author} {\bibfnamefont {Y.}~\bibnamefont
  {LeCun}}, \bibinfo {author} {\bibfnamefont {Y.}~\bibnamefont {Bengio}},
  \emph {et~al.},\ }\bibfield  {title} {\enquote {\bibinfo {title}
  {Convolutional networks for images, speech, and time series},}\ }\href@noop
  {} {\bibfield  {journal} {\bibinfo  {journal} {The handbook of brain theory
  and neural networks}\ }\textbf {\bibinfo {volume} {3361}},\ \bibinfo {pages}
  {1995} (\bibinfo {year} {1995})}\BibitemShut {NoStop}%
\bibitem [{\citenamefont {Pearson}(1901)}]{pearson1901liii}%
  \BibitemOpen
  \bibfield  {author} {\bibinfo {author} {\bibfnamefont {K.}~\bibnamefont
  {Pearson}},\ }\bibfield  {title} {\enquote {\bibinfo {title} {{LIII. On lines
  and planes of closest fit to systems of points in space}},}\ }\href@noop {}
  {\bibfield  {journal} {\bibinfo  {journal} {The London, Edinburgh, and Dublin
  Philosophical Magazine and Journal of Science}\ }\textbf {\bibinfo {volume}
  {2}},\ \bibinfo {pages} {559--572} (\bibinfo {year} {1901})}\BibitemShut
  {NoStop}%
\bibitem [{\citenamefont {Hotelling}(1933)}]{hotelling1933analysis}%
  \BibitemOpen
  \bibfield  {author} {\bibinfo {author} {\bibfnamefont {H.}~\bibnamefont
  {Hotelling}},\ }\bibfield  {title} {\enquote {\bibinfo {title} {Analysis of a
  complex of statistical variables into principal components.}}\ }\href@noop {}
  {\bibfield  {journal} {\bibinfo  {journal} {Journal of educational
  psychology}\ }\textbf {\bibinfo {volume} {24}},\ \bibinfo {pages} {417}
  (\bibinfo {year} {1933})}\BibitemShut {NoStop}%
\bibitem [{\citenamefont {Liang}\ \emph {et~al.}(2002)\citenamefont {Liang},
  \citenamefont {Lee}, \citenamefont {Lim}, \citenamefont {Lin}, \citenamefont
  {Lee},\ and\ \citenamefont {Wu}}]{liang2002proper}%
  \BibitemOpen
  \bibfield  {author} {\bibinfo {author} {\bibfnamefont {Y.}~\bibnamefont
  {Liang}}, \bibinfo {author} {\bibfnamefont {H.}~\bibnamefont {Lee}}, \bibinfo
  {author} {\bibfnamefont {S.}~\bibnamefont {Lim}}, \bibinfo {author}
  {\bibfnamefont {W.}~\bibnamefont {Lin}}, \bibinfo {author} {\bibfnamefont
  {K.}~\bibnamefont {Lee}}, \ and\ \bibinfo {author} {\bibfnamefont
  {C.}~\bibnamefont {Wu}},\ }\bibfield  {title} {\enquote {\bibinfo {title}
  {{Proper orthogonal decomposition and its applications—Part I: Theory}},}\
  }\href@noop {} {\bibfield  {journal} {\bibinfo  {journal} {Journal of Sound
  and vibration}\ }\textbf {\bibinfo {volume} {252}},\ \bibinfo {pages}
  {527--544} (\bibinfo {year} {2002})}\BibitemShut {NoStop}%
\bibitem [{\citenamefont {Van~Loan}\ and\ \citenamefont
  {Golub}(1983)}]{van1983matrix}%
  \BibitemOpen
  \bibfield  {author} {\bibinfo {author} {\bibfnamefont {C.~F.}\ \bibnamefont
  {Van~Loan}}\ and\ \bibinfo {author} {\bibfnamefont {G.~H.}\ \bibnamefont
  {Golub}},\ }\href@noop {} {\emph {\bibinfo {title} {Matrix computations}}}\
  (\bibinfo  {publisher} {Johns Hopkins University Press Baltimore},\ \bibinfo
  {year} {1983})\BibitemShut {NoStop}%
\bibitem [{\citenamefont {Rico-Martinez}, \citenamefont {Kevrekidis},\ and\
  \citenamefont {Krischer}(1995)}]{rico1995nonlinear}%
  \BibitemOpen
  \bibfield  {author} {\bibinfo {author} {\bibfnamefont {R.}~\bibnamefont
  {Rico-Martinez}}, \bibinfo {author} {\bibfnamefont {I.}~\bibnamefont
  {Kevrekidis}}, \ and\ \bibinfo {author} {\bibfnamefont {K.}~\bibnamefont
  {Krischer}},\ }\bibfield  {title} {\enquote {\bibinfo {title} {Nonlinear
  system identification using neural networks: dynamics and instabilities},}\
  }\href@noop {} {\bibfield  {journal} {\bibinfo  {journal} {Neural networks
  for chemical engineers}\ ,\ \bibinfo {pages} {409--442}} (\bibinfo {year}
  {1995})}\BibitemShut {NoStop}%
\bibitem [{\citenamefont {Foias}\ and\ \citenamefont
  {Titi}(1991)}]{foias1991determining}%
  \BibitemOpen
  \bibfield  {author} {\bibinfo {author} {\bibfnamefont {C.}~\bibnamefont
  {Foias}}\ and\ \bibinfo {author} {\bibfnamefont {E.~S.}\ \bibnamefont
  {Titi}},\ }\bibfield  {title} {\enquote {\bibinfo {title} {Determining nodes,
  finite difference schemes and inertial manifolds},}\ }\href@noop {}
  {\bibfield  {journal} {\bibinfo  {journal} {Nonlinearity}\ }\textbf {\bibinfo
  {volume} {4}},\ \bibinfo {pages} {135} (\bibinfo {year} {1991})}\BibitemShut
  {NoStop}%
\bibitem [{\citenamefont {Virtanen}\ \emph {et~al.}(2020)\citenamefont
  {Virtanen}, \citenamefont {Gommers}, \citenamefont {Oliphant}, \citenamefont
  {Haberland}, \citenamefont {Reddy}, \citenamefont {Cournapeau}, \citenamefont
  {Burovski}, \citenamefont {Peterson}, \citenamefont {Weckesser},
  \citenamefont {Bright}, \citenamefont {{van der Walt}}, \citenamefont
  {Brett}, \citenamefont {Wilson}, \citenamefont {Millman}, \citenamefont
  {Mayorov}, \citenamefont {Nelson}, \citenamefont {Jones}, \citenamefont
  {Kern}, \citenamefont {Larson}, \citenamefont {Carey}, \citenamefont {Polat},
  \citenamefont {Feng}, \citenamefont {Moore}, \citenamefont {{VanderPlas}},
  \citenamefont {Laxalde}, \citenamefont {Perktold}, \citenamefont {Cimrman},
  \citenamefont {Henriksen}, \citenamefont {Quintero}, \citenamefont {Harris},
  \citenamefont {Archibald}, \citenamefont {Ribeiro}, \citenamefont
  {Pedregosa}, \citenamefont {{van Mulbregt}},\ and\ \citenamefont {{SciPy 1.0
  Contributors}}}]{2020SciPy-NMeth}%
  \BibitemOpen
  \bibfield  {author} {\bibinfo {author} {\bibfnamefont {P.}~\bibnamefont
  {Virtanen}}, \bibinfo {author} {\bibfnamefont {R.}~\bibnamefont {Gommers}},
  \bibinfo {author} {\bibfnamefont {T.~E.}\ \bibnamefont {Oliphant}}, \bibinfo
  {author} {\bibfnamefont {M.}~\bibnamefont {Haberland}}, \bibinfo {author}
  {\bibfnamefont {T.}~\bibnamefont {Reddy}}, \bibinfo {author} {\bibfnamefont
  {D.}~\bibnamefont {Cournapeau}}, \bibinfo {author} {\bibfnamefont
  {E.}~\bibnamefont {Burovski}}, \bibinfo {author} {\bibfnamefont
  {P.}~\bibnamefont {Peterson}}, \bibinfo {author} {\bibfnamefont
  {W.}~\bibnamefont {Weckesser}}, \bibinfo {author} {\bibfnamefont
  {J.}~\bibnamefont {Bright}}, \bibinfo {author} {\bibfnamefont {S.~J.}\
  \bibnamefont {{van der Walt}}}, \bibinfo {author} {\bibfnamefont
  {M.}~\bibnamefont {Brett}}, \bibinfo {author} {\bibfnamefont
  {J.}~\bibnamefont {Wilson}}, \bibinfo {author} {\bibfnamefont {K.~J.}\
  \bibnamefont {Millman}}, \bibinfo {author} {\bibfnamefont {N.}~\bibnamefont
  {Mayorov}}, \bibinfo {author} {\bibfnamefont {A.~R.~J.}\ \bibnamefont
  {Nelson}}, \bibinfo {author} {\bibfnamefont {E.}~\bibnamefont {Jones}},
  \bibinfo {author} {\bibfnamefont {R.}~\bibnamefont {Kern}}, \bibinfo {author}
  {\bibfnamefont {E.}~\bibnamefont {Larson}}, \bibinfo {author} {\bibfnamefont
  {C.~J.}\ \bibnamefont {Carey}}, \bibinfo {author} {\bibfnamefont
  {{\.I}.}~\bibnamefont {Polat}}, \bibinfo {author} {\bibfnamefont
  {Y.}~\bibnamefont {Feng}}, \bibinfo {author} {\bibfnamefont {E.~W.}\
  \bibnamefont {Moore}}, \bibinfo {author} {\bibfnamefont {J.}~\bibnamefont
  {{VanderPlas}}}, \bibinfo {author} {\bibfnamefont {D.}~\bibnamefont
  {Laxalde}}, \bibinfo {author} {\bibfnamefont {J.}~\bibnamefont {Perktold}},
  \bibinfo {author} {\bibfnamefont {R.}~\bibnamefont {Cimrman}}, \bibinfo
  {author} {\bibfnamefont {I.}~\bibnamefont {Henriksen}}, \bibinfo {author}
  {\bibfnamefont {E.~A.}\ \bibnamefont {Quintero}}, \bibinfo {author}
  {\bibfnamefont {C.~R.}\ \bibnamefont {Harris}}, \bibinfo {author}
  {\bibfnamefont {A.~M.}\ \bibnamefont {Archibald}}, \bibinfo {author}
  {\bibfnamefont {A.~H.}\ \bibnamefont {Ribeiro}}, \bibinfo {author}
  {\bibfnamefont {F.}~\bibnamefont {Pedregosa}}, \bibinfo {author}
  {\bibfnamefont {P.}~\bibnamefont {{van Mulbregt}}}, \ and\ \bibinfo {author}
  {\bibnamefont {{SciPy 1.0 Contributors}}},\ }\bibfield  {title} {\enquote
  {\bibinfo {title} {{{SciPy} 1.0: Fundamental Algorithms for Scientific
  Computing in Python}},}\ }\href {\doibase 10.1038/s41592-019-0686-2}
  {\bibfield  {journal} {\bibinfo  {journal} {Nature Methods}\ }\textbf
  {\bibinfo {volume} {17}},\ \bibinfo {pages} {261--272} (\bibinfo {year}
  {2020})}\BibitemShut {NoStop}%
\bibitem [{\citenamefont {Abadi}\ \emph {et~al.}(2015)\citenamefont {Abadi},
  \citenamefont {Agarwal}, \citenamefont {Barham}, \citenamefont {Brevdo},
  \citenamefont {Chen}, \citenamefont {Citro}, \citenamefont {Corrado},
  \citenamefont {Davis}, \citenamefont {Dean}, \citenamefont {Devin},
  \citenamefont {Ghemawat}, \citenamefont {Goodfellow}, \citenamefont {Harp},
  \citenamefont {Irving}, \citenamefont {Isard}, \citenamefont {Jia},
  \citenamefont {Jozefowicz}, \citenamefont {Kaiser}, \citenamefont {Kudlur},
  \citenamefont {Levenberg}, \citenamefont {Man\'{e}}, \citenamefont {Monga},
  \citenamefont {Moore}, \citenamefont {Murray}, \citenamefont {Olah},
  \citenamefont {Schuster}, \citenamefont {Shlens}, \citenamefont {Steiner},
  \citenamefont {Sutskever}, \citenamefont {Talwar}, \citenamefont {Tucker},
  \citenamefont {Vanhoucke}, \citenamefont {Vasudevan}, \citenamefont
  {Vi\'{e}gas}, \citenamefont {Vinyals}, \citenamefont {Warden}, \citenamefont
  {Wattenberg}, \citenamefont {Wicke}, \citenamefont {Yu},\ and\ \citenamefont
  {Zheng}}]{tensorflow2015-whitepaper}%
  \BibitemOpen
  \bibfield  {author} {\bibinfo {author} {\bibfnamefont {M.}~\bibnamefont
  {Abadi}}, \bibinfo {author} {\bibfnamefont {A.}~\bibnamefont {Agarwal}},
  \bibinfo {author} {\bibfnamefont {P.}~\bibnamefont {Barham}}, \bibinfo
  {author} {\bibfnamefont {E.}~\bibnamefont {Brevdo}}, \bibinfo {author}
  {\bibfnamefont {Z.}~\bibnamefont {Chen}}, \bibinfo {author} {\bibfnamefont
  {C.}~\bibnamefont {Citro}}, \bibinfo {author} {\bibfnamefont {G.~S.}\
  \bibnamefont {Corrado}}, \bibinfo {author} {\bibfnamefont {A.}~\bibnamefont
  {Davis}}, \bibinfo {author} {\bibfnamefont {J.}~\bibnamefont {Dean}},
  \bibinfo {author} {\bibfnamefont {M.}~\bibnamefont {Devin}}, \bibinfo
  {author} {\bibfnamefont {S.}~\bibnamefont {Ghemawat}}, \bibinfo {author}
  {\bibfnamefont {I.}~\bibnamefont {Goodfellow}}, \bibinfo {author}
  {\bibfnamefont {A.}~\bibnamefont {Harp}}, \bibinfo {author} {\bibfnamefont
  {G.}~\bibnamefont {Irving}}, \bibinfo {author} {\bibfnamefont
  {M.}~\bibnamefont {Isard}}, \bibinfo {author} {\bibfnamefont
  {Y.}~\bibnamefont {Jia}}, \bibinfo {author} {\bibfnamefont {R.}~\bibnamefont
  {Jozefowicz}}, \bibinfo {author} {\bibfnamefont {L.}~\bibnamefont {Kaiser}},
  \bibinfo {author} {\bibfnamefont {M.}~\bibnamefont {Kudlur}}, \bibinfo
  {author} {\bibfnamefont {J.}~\bibnamefont {Levenberg}}, \bibinfo {author}
  {\bibfnamefont {D.}~\bibnamefont {Man\'{e}}}, \bibinfo {author}
  {\bibfnamefont {R.}~\bibnamefont {Monga}}, \bibinfo {author} {\bibfnamefont
  {S.}~\bibnamefont {Moore}}, \bibinfo {author} {\bibfnamefont
  {D.}~\bibnamefont {Murray}}, \bibinfo {author} {\bibfnamefont
  {C.}~\bibnamefont {Olah}}, \bibinfo {author} {\bibfnamefont {M.}~\bibnamefont
  {Schuster}}, \bibinfo {author} {\bibfnamefont {J.}~\bibnamefont {Shlens}},
  \bibinfo {author} {\bibfnamefont {B.}~\bibnamefont {Steiner}}, \bibinfo
  {author} {\bibfnamefont {I.}~\bibnamefont {Sutskever}}, \bibinfo {author}
  {\bibfnamefont {K.}~\bibnamefont {Talwar}}, \bibinfo {author} {\bibfnamefont
  {P.}~\bibnamefont {Tucker}}, \bibinfo {author} {\bibfnamefont
  {V.}~\bibnamefont {Vanhoucke}}, \bibinfo {author} {\bibfnamefont
  {V.}~\bibnamefont {Vasudevan}}, \bibinfo {author} {\bibfnamefont
  {F.}~\bibnamefont {Vi\'{e}gas}}, \bibinfo {author} {\bibfnamefont
  {O.}~\bibnamefont {Vinyals}}, \bibinfo {author} {\bibfnamefont
  {P.}~\bibnamefont {Warden}}, \bibinfo {author} {\bibfnamefont
  {M.}~\bibnamefont {Wattenberg}}, \bibinfo {author} {\bibfnamefont
  {M.}~\bibnamefont {Wicke}}, \bibinfo {author} {\bibfnamefont
  {Y.}~\bibnamefont {Yu}}, \ and\ \bibinfo {author} {\bibfnamefont
  {X.}~\bibnamefont {Zheng}},\ }\bibfield  {title} {\enquote {\bibinfo {title}
  {{TensorFlow}: Large-scale machine learning on heterogeneous systems.}}\
  }\href {https://www.tensorflow.org/} {\  (\bibinfo {year} {2015})},\ \bibinfo
  {note} {software available from https://www.tensorflow.org}\BibitemShut
  {NoStop}%
\bibitem [{\citenamefont {Glorot}\ and\ \citenamefont
  {Bengio}(2010)}]{glorot2010understanding}%
  \BibitemOpen
  \bibfield  {author} {\bibinfo {author} {\bibfnamefont {X.}~\bibnamefont
  {Glorot}}\ and\ \bibinfo {author} {\bibfnamefont {Y.}~\bibnamefont
  {Bengio}},\ }\bibfield  {title} {\enquote {\bibinfo {title} {Understanding
  the difficulty of training deep feedforward neural networks},}\ }in\
  \href@noop {} {\emph {\bibinfo {booktitle} {Proceedings of the thirteenth
  international conference on artificial intelligence and statistics}}}\
  (\bibinfo {year} {2010})\ pp.\ \bibinfo {pages} {249--256}\BibitemShut
  {NoStop}%
\bibitem [{\citenamefont {Kingma}\ and\ \citenamefont
  {Ba}(2014)}]{kingma2014adam}%
  \BibitemOpen
  \bibfield  {author} {\bibinfo {author} {\bibfnamefont {D.~P.}\ \bibnamefont
  {Kingma}}\ and\ \bibinfo {author} {\bibfnamefont {J.}~\bibnamefont {Ba}},\
  }\bibfield  {title} {\enquote {\bibinfo {title} {Adam: A method for
  stochastic optimization},}\ }\href@noop {} {\bibfield  {journal} {\bibinfo
  {journal} {arXiv preprint arXiv:1412.6980}\ } (\bibinfo {year}
  {2014})}\BibitemShut {NoStop}%
\bibitem [{\citenamefont {Kemeth}\ \emph {et~al.}(2020)\citenamefont {Kemeth},
  \citenamefont {Bertalan}, \citenamefont {Thiem}, \citenamefont {Dietrich},
  \citenamefont {Moon}, \citenamefont {Laing},\ and\ \citenamefont
  {Kevrekidis}}]{kemeth2020learning}%
  \BibitemOpen
  \bibfield  {author} {\bibinfo {author} {\bibfnamefont {F.~P.}\ \bibnamefont
  {Kemeth}}, \bibinfo {author} {\bibfnamefont {T.}~\bibnamefont {Bertalan}},
  \bibinfo {author} {\bibfnamefont {T.}~\bibnamefont {Thiem}}, \bibinfo
  {author} {\bibfnamefont {F.}~\bibnamefont {Dietrich}}, \bibinfo {author}
  {\bibfnamefont {S.~J.}\ \bibnamefont {Moon}}, \bibinfo {author}
  {\bibfnamefont {C.~R.}\ \bibnamefont {Laing}}, \ and\ \bibinfo {author}
  {\bibfnamefont {I.~G.}\ \bibnamefont {Kevrekidis}},\ }\bibfield  {title}
  {\enquote {\bibinfo {title} {{Learning emergent PDEs in a learned emergent
  space}},}\ }\href@noop {} {\bibfield  {journal} {\bibinfo  {journal} {arXiv
  preprint arXiv:2012.12738}\ } (\bibinfo {year} {2020})}\BibitemShut {NoStop}%
\bibitem [{\citenamefont {Mackenzie}\ and\ \citenamefont
  {Roberts}(2000)}]{mackenzie2000holistic}%
  \BibitemOpen
  \bibfield  {author} {\bibinfo {author} {\bibfnamefont {T.}~\bibnamefont
  {Mackenzie}}\ and\ \bibinfo {author} {\bibfnamefont {A.}~\bibnamefont
  {Roberts}},\ }\bibfield  {title} {\enquote {\bibinfo {title} {{Holistic
  finite differences accurately model the dynamics of the Kuramoto-Sivashinsky
  equation}},}\ }\href@noop {} {\bibfield  {journal} {\bibinfo  {journal}
  {ANZIAM Journal}\ }\textbf {\bibinfo {volume} {42}},\ \bibinfo {pages}
  {918--935} (\bibinfo {year} {2000})}\BibitemShut {NoStop}%
\bibitem [{\citenamefont {Roberts}(2001{\natexlab{a}})}]{roberts2001holistic}%
  \BibitemOpen
  \bibfield  {author} {\bibinfo {author} {\bibfnamefont {A.}~\bibnamefont
  {Roberts}},\ }\bibfield  {title} {\enquote {\bibinfo {title} {Holistic
  discretisation illuminates and enhances the numerical modelling of
  differential equations},}\ }\href@noop {} {\bibfield  {journal} {\bibinfo
  {journal} {Topics in Applied and Theoretical Mathematics and Computer
  Science}\ ,\ \bibinfo {pages} {81--89}} (\bibinfo {year}
  {2001}{\natexlab{a}})}\BibitemShut {NoStop}%
\bibitem [{\citenamefont
  {Roberts}(2001{\natexlab{b}})}]{roberts2001holisticburgers}%
  \BibitemOpen
  \bibfield  {author} {\bibinfo {author} {\bibfnamefont {A.}~\bibnamefont
  {Roberts}},\ }\bibfield  {title} {\enquote {\bibinfo {title} {{Holistic
  discretization ensures fidelity to Burgers' equation}},}\ }\href@noop {}
  {\bibfield  {journal} {\bibinfo  {journal} {Applied numerical mathematics}\
  }\textbf {\bibinfo {volume} {37}},\ \bibinfo {pages} {371--396} (\bibinfo
  {year} {2001}{\natexlab{b}})}\BibitemShut {NoStop}%
\bibitem [{\citenamefont {Wiley}, \citenamefont {Strogatz},\ and\ \citenamefont
  {Girvan}(2006)}]{wiley2006size}%
  \BibitemOpen
  \bibfield  {author} {\bibinfo {author} {\bibfnamefont {D.~A.}\ \bibnamefont
  {Wiley}}, \bibinfo {author} {\bibfnamefont {S.~H.}\ \bibnamefont {Strogatz}},
  \ and\ \bibinfo {author} {\bibfnamefont {M.}~\bibnamefont {Girvan}},\
  }\bibfield  {title} {\enquote {\bibinfo {title} {The size of the sync
  basin},}\ }\href@noop {} {\bibfield  {journal} {\bibinfo  {journal} {Chaos:
  An Interdisciplinary Journal of Nonlinear Science}\ }\textbf {\bibinfo
  {volume} {16}},\ \bibinfo {pages} {015103} (\bibinfo {year}
  {2006})}\BibitemShut {NoStop}%
\bibitem [{\citenamefont {Laing}, \citenamefont {Bl{\"a}sche},\ and\
  \citenamefont {Means}(2021)}]{laing2021dynamics}%
  \BibitemOpen
  \bibfield  {author} {\bibinfo {author} {\bibfnamefont {C.}~\bibnamefont
  {Laing}}, \bibinfo {author} {\bibfnamefont {C.}~\bibnamefont {Bl{\"a}sche}},
  \ and\ \bibinfo {author} {\bibfnamefont {S.}~\bibnamefont {Means}},\
  }\bibfield  {title} {\enquote {\bibinfo {title} {{Dynamics of structured
  networks of Winfree oscillators}},}\ }\href@noop {} {\bibfield  {journal}
  {\bibinfo  {journal} {Frontiers in Systems Neuroscience}\ }\textbf {\bibinfo
  {volume} {15}},\ \bibinfo {pages} {7} (\bibinfo {year} {2021})}\BibitemShut
  {NoStop}%
\bibitem [{\citenamefont {Bl{\"a}sche}, \citenamefont {Means},\ and\
  \citenamefont {Laing}(2020)}]{blasche20}%
  \BibitemOpen
  \bibfield  {author} {\bibinfo {author} {\bibfnamefont {C.}~\bibnamefont
  {Bl{\"a}sche}}, \bibinfo {author} {\bibfnamefont {S.}~\bibnamefont {Means}},
  \ and\ \bibinfo {author} {\bibfnamefont {C.~R.}\ \bibnamefont {Laing}},\
  }\bibfield  {title} {\enquote {\bibinfo {title} {Degree assortativity in
  networks of spiking neurons},}\ }\href@noop {} {\bibfield  {journal}
  {\bibinfo  {journal} {Journal of Computational Dynamics}\ }\textbf {\bibinfo
  {volume} {7}},\ \bibinfo {pages} {401} (\bibinfo {year} {2020})}\BibitemShut
  {NoStop}%
\bibitem [{\citenamefont {Laing}\ and\ \citenamefont
  {Bl{\"a}sche}(2020)}]{laing2020effects}%
  \BibitemOpen
  \bibfield  {author} {\bibinfo {author} {\bibfnamefont {C.~R.}\ \bibnamefont
  {Laing}}\ and\ \bibinfo {author} {\bibfnamefont {C.}~\bibnamefont
  {Bl{\"a}sche}},\ }\bibfield  {title} {\enquote {\bibinfo {title} {The effects
  of within-neuron degree correlations in networks of spiking neurons},}\
  }\href@noop {} {\bibfield  {journal} {\bibinfo  {journal} {Biological
  cybernetics}\ }\textbf {\bibinfo {volume} {114}},\ \bibinfo {pages}
  {337--347} (\bibinfo {year} {2020})}\BibitemShut {NoStop}%
\end{thebibliography}%

\end{document}